\documentclass[12pt]{iopart}

\eqnobysec

\usepackage{latexsym,amssymb}
\usepackage{graphics}
\usepackage{graphicx}
\usepackage{epsfig}
\usepackage{cite}

\newcommand{\Cee}{\mathcal{C}}
\newcommand{\Eff}{\mathcal{F}}
\newcommand{\Zed}{\mathcal{Z}}

\newcommand{\One}{\ensuremath{1\!\!1}}

\renewcommand{\th}{\ensuremath{{}^{\rm th}}}

\newcommand{\DD}{{D^\prime}}
\newcommand{\EE}{{E^\prime}}

\newcommand{\avg}[1]{\langle #1 \rangle}

\newcommand{\bra}[1]{\langle #1 |}
\newcommand{\ket}[1]{| #1 \rangle}
\newcommand{\braket}[2]{\langle #1 | #2 \rangle}

\newcommand{\Order}{{\rm O}}
\newcommand{\coeff}[1]{\{#1\}}

\newcommand{\RR}{(\ref{int:DE})--(\ref{int:EW})}
\newcommand{\RRR}{(\ref{ring:DE})--(\ref{ring:DA})}
\newcommand{\PRR}{(\ref{pasep:DE})--(\ref{pasep:EW})}

\newcommand{\D}{{\rm d}}
\newcommand{\ds}{\displaystyle}

\begin{document}

\topical{Nonequilibrium Steady States of Matrix Product Form: A Solver's Guide}

\author{R.\ A.\ Blythe and M.\ R.\ Evans}

\address{SUPA, School of Physics, University of Edinburgh,
  Mayfield Road, Edinburgh, EH9 3JZ, UK\\[3ex] \today} 

\begin{abstract}
We consider the general problem of determining the steady state of
stochastic nonequilibrium systems such as those that have been used to
model (among other things) biological transport and traffic flow.  We
begin with a broad overview of this class of driven diffusive
systems---which includes exclusion processes---focusing on interesting
physical properties, such as shocks and phase transitions.  We then
turn our attention specifically to those models for which the exact
distribution of microstates in the steady state can be expressed in a
matrix product form.  In addition to a gentle introduction to this
matrix product approach, how it works and how it relates to similar
constructions that arise in other physical contexts, we present a
unified, pedagogical account of the various means by which the
statistical mechanical calculations of macroscopic physical quantities
are actually performed.  We also review a number of more advanced
topics, including nonequilibrium free energy functionals, the
classification of exclusion processes involving multiple particle
species, existence proofs of a matrix product state for a given model
and more complicated variants of the matrix product state that allow
various types of parallel dynamics to be handled.  We conclude with a
brief discussion of open problems for future research.
\end{abstract}


\tableofcontents
\pagestyle{plain}

\maketitle


\section{Introduction}
\label{intro}

\subsection{What is a nonequilibrium steady state?}

\begin{figure}[b]
\begin{center}
\includegraphics[scale=0.5]{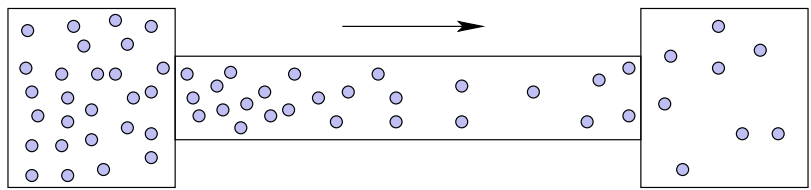}
\end{center}
\caption{\label{fig:chempotgrad} A channel held out of equilibrium
through the application of a chemical potential gradient.  It is
anticipated that a nonequilibrium steady state will be reached in
which particles flow from one reservoir to another at a constant
rate.}
\end{figure}

Consider a finite volume containing interacting particles and that is
coupled at opposite boundaries to particle reservoirs held at
\emph{different} chemical potentials, as shown in
Figure~\ref{fig:chempotgrad}.  In this situation one anticipates a net
flux of particles from the reservoir with the greater chemical
potential through the system and into the opposite reservoir.  After
some initial transients we expect a state to arise in which there is a
nonzero mean flux that is constant over space and time.  This flux
reveals the system to be in a \emph{nonequilibrium steady state}.
More generally, we have in mind any physical system all of whose
observables do not change with time, but nevertheless exhibits an
irreversible exchange of heat, particles, volume or some other
physical quantity with its environment.

When heat, particle or volume exchanges are reversible, the system is
at equilibrium with its environment and, as is well established, its
microstates have a Gibbs-Boltzmann distribution.  This knowledge
allows one then to predict the macroscopic physics of a many-body
system given a microscopic model couched in terms of energetic
interactions.  Furthermore, these Gibbs-Boltzmann statistics emerge
from the application of a very simple principle, namely that of equal
\textit{a priori} probabilities for the system combined with its
environment.

By contrast, when a system is out of equilibrium, very little is known
about the statistics of the microstates, not even in the steady state.
This is despite a wide range of approaches aimed at addressing this
deficiency in our understanding of nonequilibrium physics.  These
approaches fall roughly into two broad categories.

First, one has macroscopic theories, such as Onsager and Machlup's
pioneering work on near-equilibrium fluctuations \cite{OM53a,OM53b}.
In this regime, it is assumed that the thermodynamic forces that
restore an equilibrium are linear, and it is possible to determine the
probability of witnessing certain fluctuations in the equilibrium
steady state, the most probable trajectory given by a principle of
minimum energy dissipation.  In this work, we are interested in steady
states that are far from equilibrium, i.e., where the system is driven
beyond the linear response regime.  Recently there has been some
success in extending the Onsager-Machlup theory of equilibrium
fluctuations to such nonequilibrium steady states, as long as one has
been able to derive macroscopic (hydrodynamic) equations of motion for
the system \cite{BDSGJLL01}.  Furthermore, over the past few years
fluctuation theorems have been derived under a wide range of
conditions.  Typically, these relate the probabilities of entropy
changes of equal magnitude but opposite sign occurring in a system
driven arbitrarily far from equilibrium (see \cite{Kurchan06} for a
brief introductory overview).  Although some connections between these
various topics have been established (see, e.g., \cite{TC06}), it is
fair to say that a complete coherent picture of the macroscopic
properties of nonequilibrium steady states is still lacking.

Somewhere between this macroscopic approach and a truly microscopic
approach (i.e., one that would follow directly from some microscopic
equations of motion) lie a range of mesoscopic models that are
intended to capture the essential features of nonequilibrium
dynamics---irreversibility, currents, dissipation of heat and so
on---but are nevertheless simple enough that analytical treatment is
possible.  In particular, it is possible to explore the macroscopic
consequences of the underlying dynamics, in the process identifying
both similarities with equilibrium states of matter and the novel
features peculiar to nonequilibrium systems.

It is this approach that is the focus of this review article.
Specifically, we will discuss a set of models that have a steady-state
distribution of microstates that can be expressed mathematically in
the form of a matrix product.  We will explain in detail how these
expressions can be used to calculate macroscopic steady-state
properties exactly.  These will include particle currents, density
profiles, correlation functions, the distribution of macroscopic
fluctuations and so on.  These calculations will reveal phenomena that
occur purely because of the far-from-equilibrium conditions: for
example, boundary induced phase transitions and shock fronts.
Meanwhile, we will also find conceptual connections with equilibrium
statistical physics: in some of the models we discuss, for example, it
is possible to construct partition functions and free-energy-like
quantities that are meaningful both for equilibrium and nonequilibrium
systems.  We begin by explaining this modelling approach in more
detail.

\subsection{Modelling nonequilibrium dynamics}

\begin{figure}
\begin{center}
\includegraphics[scale=0.5]{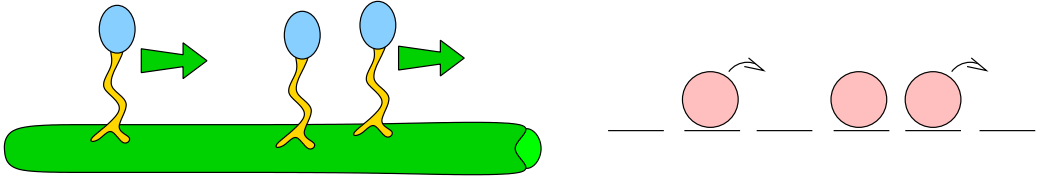}
\end{center}
\caption{\label{fig:microtubule} Kinesins attached to a microtubule
that move in a preferred direction by extracting chemical energy from
the environment.  At a mesoscopic level, this can be modelled as a
stochastic process in which particles hop along a one-dimensional
lattice as shown.}
\end{figure}

In Figure~\ref{fig:microtubule} we have sketched some molecular motors
(kinesins) that are attached to a microtubule.  This is essentially a
track along which the motors can walk by using packets of energy
carried by ATP molecules present in the surroundings.  Clearly, at
the most microscopic level, there are a number chemical processes at
play that combine to give rise to the motor's progress along the
track.  At a more coarse-grained, mesoscopic level, we can simply
observe that the motor takes steps of a well-defined size
(approximately 8nm for a kinesin \cite{Bray00}) at a time, and thus
model the system as a set of particles that hop
\textit{stochastically} between sites of a lattice.  This stochastic
prescription is in part intended to reflect the fact that the internal
degrees of freedom that govern when a particle hops are not explicitly
included in the model.

The simplest such stochastic model for the particle hops is a Poisson
process that occurs at some prescribed rate.  Let $\Cee$ and $\Cee'$
be two configurations of the lattice that differ by a single particle
hop.  We then define $W(\Cee\to\Cee')$ as the rate at which this hop
occurs, such that in an infinitesimal time interval $\rmd t$ the
probability that that hop takes place is $W(\Cee\to\Cee') \rmd t$.
The rate of change of the probability $P(\Cee, t)$ for the system to
be in configuration $\Cee$ at time $t$ is then the solution of the
master equation
\begin{equation} 
\label{int:me}
\frac{\partial}{\partial t} P(\Cee, t) = \sum_{\Cee^\prime \ne \Cee}
P(\Cee^\prime, t) W(\Cee^\prime \to \Cee) - \sum_{\Cee^\prime \ne
\Cee} P(\Cee, t) W(\Cee \to \Cee^\prime)
\end{equation}
subject to some initial condition $P(\Cee, 0)$.  The first term on the
right-hand side of this equation gives contributions from all possible
hops (transitions) into the configuration $\Cee$ from other
configurations $\Cee'$; the second term gives contributions from hops
out of $\Cee$ into other configuration $\Cee'$.  We are concerned here
with steady states, where these gain and loss terms exactly balance
causing the time derivative to vanish.  Such solutions of the master
equation we denote $P(\Cee)$.  In this work, nearly all the models we
discuss have the property that a single steady state is reached from
any initial condition: i.e., they are ergodic \cite{EB02,Durrett01}.
Given then that the distribution $P(\Cee)$ is unique for these models,
we concentrate almost exclusively on how it is obtained analytically,
and how expectation values of observables are calculated.  The
remaining models that are not ergodic over the full space of
configurations turn out to have a unique steady-state distribution
over some subspace of configurations.  We will see that such a
distributions of this type can also be handled using the methods
described in this work.

\subsection{A catalogue of steady states}

Solutions of the master equation (\ref{int:me}) can assume a number of
structures.  Some general classes of steady state---not intended to be
mutually exclusive---can be summarised as follows:

\paragraph{Equilibrium (Gibbs-Boltzmann) steady state}
If the internal energy of a microstate $\Cee$ is $E(\Cee)$ and the
system as a whole is in equilibrium with a heat bath with inverse
temperature $\beta$, the microstates have a Gibbs-Boltzmann
distribution $P(\Cee) \propto \rme^{-\beta E(\Cee)}$.  
The dynamics is reversible in that 
the probability of witnessing any particular trajectory
through phase space is equal to that of its time reversal.  
This then implies that stochastic dynamics expressed
in terms of transition rates $W(\Cee\to\Cee')$ must satisfy the
detailed balance condition \cite{Kelly79,vanKampen92}
\begin{equation}
\label{int:db}
P(\Cee) W(\Cee \to \Cee^\prime) = P(\Cee^\prime) W(\Cee^\prime \to
\Cee)
\end{equation}
where here $P(\Cee)$ is the Gibbs-Boltzmann distribution.  A simple
consequence of (\ref{int:db}) is that there are no fluxes in the
steady state.  Hence any system that does exhibit currents in the
steady state, i.e., nonequilibrium systems, generally do not have a
stationary distribution that satisfies detailed balance.

\paragraph{Factorised steady state}
Sometimes, whether in or out of equilibrium, one has a factorised
steady state.  Typically one has in mind a lattice (or graph) with $N$
sites with configurations specified by occupancy variables $n_i$
(i.e., site $i$ contains $n_i$ particles).  A factorised steady
state then takes the form $P(\Cee) \propto \prod_{i=1}^{N}
f_i(n_i)$.  For such a structure to arise, certain constraints on
the transition rates $W(\Cee\to\Cee')$ must be satisfied.  For
example, in an equilibrium system, one would need the energy of the
system to be the sum of single-site energies.  Out of equilibrium, one
finds factorised steady states in zero-range processes, in which
particles hop at a rate that depends only on the occupation number at
the departure site.  Moreover, necessary and sufficient conditions for
factorisation in a broad class of models which includes the ZRP have
been established \cite{EMZ04}.  Zero-range processes and various
generalisations have been reviewed recently in a companion paper
\cite{EH05} and so we do not discuss them further here.

\paragraph{Matrix product steady state}
A matrix product steady state is an extension to a factorised steady
state that is of particular utility for one-dimensional models.  The
rough idea is to replace the scalar factors $f_i(n_i)$ with
matrices $X_{n_i}$, the steady state probability then being given
by an element of the resulting matrix product $\prod_{i=1}^{N}
X_{n_i}$.  Since the matrices $X_{n}$ for different occupancy
numbers $n$ need not commute, one opens up the possibility for
correlations between the occupancy of different sites (above those
that emerge from global constraints, such as a fixed particle number,
that one sees in factorised states).  It turns out that quite a number
of nonequilibrium models have a matrix product steady state, and we
will encounter them all in the course of this review.

\paragraph{Most general steady state}
In principle, the (assumed unique) stationary solution of the master
equation (\ref{int:me}) can always be found if the number of
configurations is finite, since (\ref{int:me}) is a system of linear
equations in the probabilities $P(\Cee)$.  One way to express this
solution is in terms of \textit{statistical weights} $f(\Cee)$, each
of which is given by the determinant of the matrix of transition rates
obtained by removing the row and column corresponding to the
configuration $\Cee$ (see e.g. \cite{BlythePhD,SZ07}).  To arrive at the probability
distribution $P(\Cee)$ one requires the \textit{normalisation}
$Z=\sum_{\Cee} f(\Cee)$, so that then $P(\Cee) = f(\Cee)/Z$.  This
normalisation has some interesting properties: it is uniquely defined
for any ergodic Markov process with a finite number of configurations;
it can be shown to be equal to the product of all nonzero eigenvalues
of the matrix of transition rates; and it is a partition function of a
set of trees on the graph of transitions between different microscopic
configurations.  In this latter case, the densities of particular
edges in the ensemble of trees are controlled by tuning the transition
rates corresponding to those edges, so in this interpretation these
transition rates are equivalent to equilibrium fugacities.  Although
this interpretation of the normalisation is rather abstract, concrete
connections between transition rates in certain nonequilibrium models
that have a matrix product steady state and fugacities in an
equilibrium ensemble have been made, and these we shall discuss later
in this review.  We also remark that the equilibrium theory of
partition function zeros to characterise phase transitions also
carries over to the normalisation as just defined, at least for those
models that have been tested \cite{Arndt00,DDH02,BE02}.

\subsection{Purpose of this review}

We have three principal aims in this work.  First, we wish to
illustrate the insights into nonequilibrium statistical mechanics that
have been gained from exactly solvable models, particularly those that
have a steady state of the matrix product form.  Secondly, we seek to
provide a self-contained pedagogical account of the various analytical
methods and calculational tools that can be used to go from the matrix
product expressions to predictions for the macroscopic physics.
Finally, we wish to present a thorough review of the progress that has
been made using the matrix product approach over the last few years.
As far as we are aware, the matrix product approach has not been the
focus of a review for nearly a decade \cite{DE96,Derrida98}, and we
feel that it is high time that the significant developments that have
occurred in the meantime should be collected together in one place.
In order to prevent this review from becoming unmanageably long, we
focus purely on static physical properties exhibited in the steady
states of models solvable by the matrix product method.  This means we
have unfortunately had to omit discussion of some very interesting
topics, for example how some dynamical properties, such as fluctuation
phenomena, have been elucidated through the use of the Bethe ansatz
\cite{GS92,Kim95,dGE05},
determinental solutions and their connection to
random matrix theory \cite{Schutz97,PS00,NS04} and dynamical matrix products\cite{SS95,Schutz98}.  
Some of these
topics, however, have recently been reviewed elsewhere
\cite{GM06b,Sasamoto07}.  We also direct the reader to established
reviews, such as
\cite{SZ95,Krug97,Derrida98,Schutz00,CSS00,Stinchcombe01,Schutz03},
for any other background that we have been forced to leave out here.

To realise the aims stated above, we provide in the next section a
general account of the physics one expects to see in nonequilibrium
dynamical systems, and outline the essential ideas underlying the
matrix product approach.  Thereafter, we go into the details of how
the simplest models are solved, and show a number of contrasting (but
equivalent) approaches that have found application to more complex
problems.  These cases then form the material of the remainder of the
review, which we round off by posing some open problems for future
research.

\section{An overview of driven diffusive systems}
\label{asep}

\subsection{A paradigmatic example: the asymmetric
simple exclusion process (ASEP)}

The asymmetric simple exclusion process (ASEP) is a very simple model
of a driven system in one dimension that has biased diffusion of
hard-core particles.  We shall discuss two versions: the periodic
system and the open system.  The latter is coupled to particle
reservoirs at either end so that, as described in the previous
section, there is a steady state with a constant particle flux.  The
open system has a long history, having first appeared in the literature
to our knowledge as a model of biopolymerisation \cite{MGP68} and
transport across membranes \cite{Heckmann72}.  In the mathematical
literature meanwhile diffusion with collisions between particles was
first studied by Harris \cite{Harris65} and the terminology simple
exclusion was first defined by Spitzer \cite{Spitzer70}.  Over the
years, applications to other transport processes have appeared, e.g.,
as a general model for traffic flow \cite{Schadschneider01} and
various other theoretical and experimental studies of biophysical
transport \cite{CL99,KL03,PFF03,Leduc04,CSN05}.  Whilst interesting
and important, these many applications are not our primary concern
here.  Also we shall not do justice to the many rigorous mathematical
results which have been summarised in the books by
Liggett\cite{Liggett85,Liggett99}.  Rather, our interest in the ASEP
lies in its having acquired the status of a fundamental model of
nonequilibrium statistical physics in its own right in much the same
way that the Ising model has become a paradigm for equilibrium
critical phenomena.  In particular, the ASEP and related models
show---despite their superficial simplicity---a range of nontrivial
macroscopic phenomena, such as phase transitions, spontaneous symmetry
breaking, shock fronts, condensation and jamming.

In common with the vast majority of the models we discuss in this
review, the ASEP is defined as a stochastic process taking place in
continuous time on a discrete one-dimensional lattice of $N$ sites.
Hopping is totally asymmetric: a particle sitting to the left of an
occupied site hops forwards one site (as in
Figure~\ref{fig:microtubule}), each hop being a Poisson process with
rate $p$.  That is, in an infinitesimal time interval $\rmd t$, there
is a probability $p \rmd t$ that one of the particles that can hop,
does hop, the identity of that particle being chosen at random.  In
simulation terms, this type of dynamics corresponds to a
\emph{random-sequential} updating scheme, in which bonds between
particles are chosen at random and then, if there is a particle at the
left-hand end of the bond and a vacancy at the right, the particle is
moved forwards.  In this scheme, each update corresponds (on average)
to $1/(Np)$ units of time; in the following we take $p=1$ without loss
of generality.  The definition of the model is completed by specifying
the boundary conditions.

\begin{figure}
\begin{center}
\includegraphics[scale=0.6]{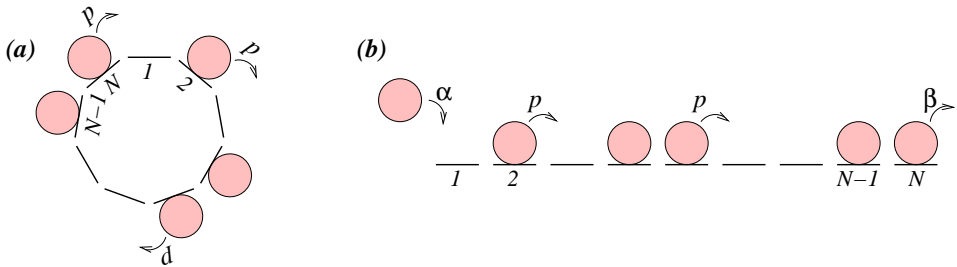}
\end{center}
\caption{\label{fig:asep} Asymmetric exclusion process with (a)
  periodic and (b) open boundary conditions.  The labels indicate the
  rates at which the various particle moves can occur.}
\end{figure}

\subsubsection{Periodic boundary conditions}

The simplest version of the model has a ring of $N$ sites as depicted
in Figure~\ref{fig:asep}(a): a particle hopping to the right on site $N$
lands on site $1$ if the receiving site is empty.  Under these
dynamics, the total particle number $M$ is conserved.  Furthermore, as
we will show in Section~\ref{RRproof}, the steady state is very
simple: all configurations (with the allowed number of particles) are
equally likely. As there are ${N \choose M}$ allowed configurations
the probability of any one is $1/{N\choose M}$.  The steady-state
current of particles, $J$, through a bond $i, i+1$ is given by $p$ (here
equal to 1) multiplied by the probability that there is a particle at site $i$ and
site $i+1$ is vacant. One finds
\begin{equation}
J = \frac{{N-2 \choose M-1}}{{N \choose M}}
= \frac{M}{N}\frac{(N-M)}{(N-1)} \to  \bar\rho(1-\bar\rho)
\end{equation}
in the thermodynamic limit where $N,M \to \infty$ at a constant ratio
$M/N=\bar\rho$.

One may also ask, starting from a known configuration, how long the
system takes to relax to the steady state.  It turns out that this
timescale $T$ depends on the system size $N$ via the relation $T \sim N^{z}$
which defines a \emph{dynamic exponent} $z$.  One can show using the
Bethe ansatz (not covered here, but see, e.g.,
\cite{GS92,GM05a,GM05b}) that for the ASEP on a ring, $z=3/2$.  Also
in the symmetric case where particles hop both to the left and to the
right with equal rates, but still with exclusion, one obtains $z=2$
which is the value one would find for a purely diffusive process.

\subsubsection{Open boundary conditions}

To model the interaction of an open system with reservoirs at
different densities, Poisson processes acting at the boundary sites
are added.  Specifically, a particle may be inserted onto the leftmost
site with rate $\alpha$ (if it is vacant), and may leave the system
from the rightmost site at rate $\beta$: see Figure~\ref{fig:asep}(b).
Here we will consider only $\alpha$, $\beta<1$ although there is no
problem in considering rates outwith this range.
Then we may think of  reservoirs at site 0 with density $\alpha$
and a reservoir at site $N+1$ with density $1-\beta$.
To include these moves in the simulation scheme described above, one
would additionally allow the bonds between the system and the
reservoirs to be chosen, and perform the particle updates with
probability $\alpha$ (entry) or $\beta$ (exit) according to the bond
chosen (assuming $\alpha$ and $\beta$ both less than $1$).  These
moves admit the possibility of a nonzero current in the steady state.

The open system is more interesting in that phase transitions may
occur in the steady state.  In the following we shall explore the
origins of these phase transitions, both through approximate
treatments and an exact solution.  In this context we recognise---by
direct analogy with equilibrium phase transitions---a phase transition
as a sudden change in form of macroscopic quantities such as the
particle current across a bond, or the density at a site.  It turns
out that the particle current plays a central role, analogous to the
free energy of an equilibrium system, in determining the nature of the
phase transition i.e. the order of the transition is determined by
which derivative of the current with respect to some external
parameter exhibits a discontinuity.

\medskip

We remark that other types of updating scheme can also be considered.
For example, in applications to traffic flow or pedestrian dynamics
\cite{CSS00,Helbing01}, it is more natural to consider parallel
dynamics where many particles can hop in concert.  The ASEP remains
solvable under certain classes of parallel updating schemes which will
be discussed towards the end of this review (Section~\ref{discretetime}).

\subsection{Mean-field and hydrodynamic approaches}
\label{sec:mfh}

\subsubsection{Lighthill-Whitham theory of kinematic waves}

We begin by reviewing a classical phenomenological theory, first applied
to traffic flow \cite{LW55b}, which serves as a first recourse in our
understanding of the phase diagrams of driven diffusive systems.  The
idea is to postulate a continuity equation for the local density
$\rho$
\begin{equation}
\frac{\partial \rho}{\partial t} + \frac{\partial J}{\partial x}  =0
\label{cty}
\end{equation}
where $J$ is the current of particles.
(Note  we use here a different
nomenclature and notation to \cite{LW55b} which discusses
concentration and flow rather density and current.)

The analysis rests on the key assumption that there is a unique
relation, $J(\rho)$, between the current and local density.
Also, it is assumed that there is a maximum current $J_m$ (the
capacity of the road) at density $\rho_m$.  The first assumption
implies that (\ref{cty}) becomes
\begin{equation}
\frac{\partial \rho}{\partial t} + v_g
\frac{\partial \rho}{\partial x} =0
\label{cty2}
\end{equation}
where 
\begin{equation}
v_g(\rho) = \frac{\D J(\rho)}{\D \rho}\; .
\label{vg}
\end{equation}
Now, an implicit  solution of (\ref{cty2}) is of travelling-wave  form
\begin{equation}
\rho(x,t) = f(x-v_g(\rho)t)
\end{equation}
as can readily be checked by substitution into (\ref{cty2}). The
arbitrary function $f$ is determined by the initial density profile
$\rho(x,0) = f(x)$.  The interpretation of the solution is that a
patch with local density $\rho$ propagates with velocity $v_g(\rho)$.
The propagation of such a patch is known as a collective phenomenon
known as a {\em kinematic wave} and $v_g$ is a group velocity.  The
velocity $c$ of a single particle in an environment of density $\rho$,
on the other hand, is defined through
\begin{equation}
J(\rho) = c(\rho) \rho\;.
\end{equation}
One sees that
\begin{equation}
v_g= \frac{\D}{\D \rho}(\rho c) = c + \rho\frac{\D c}{\D \rho}\;.
\end{equation}
Thus, if the single particle velocity is a decreasing function of
density, which is what we expect, we find $v_g <c$ and kinematic waves
travel backwards in the frame of a moving particle.

\begin{figure}
\begin{center}
\includegraphics[scale=0.7]{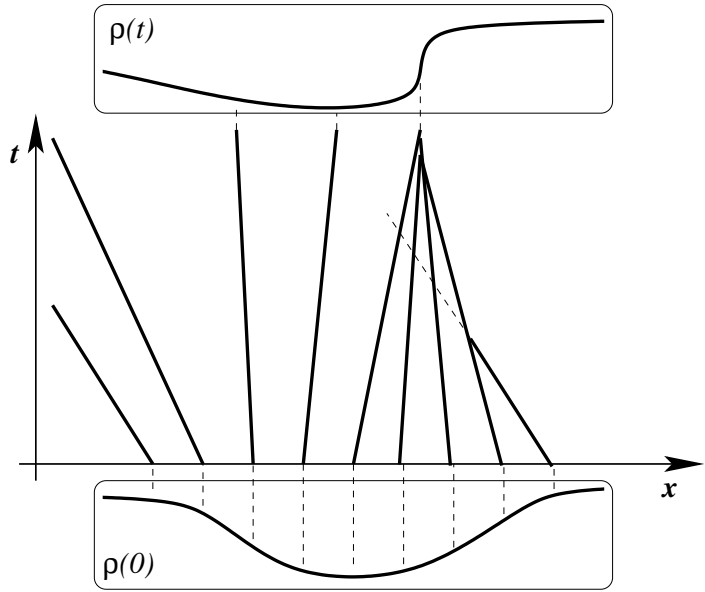}
\end{center}
\caption{\label{fig:kinematic} Kinematic waves and sharpening of a shock:
the figure sketches the evolution of a density profile $\rho(t)$
from an initial  profile $\rho(0)$ according to (\ref{cty})
and illustrates how sharp discontinuities in the profile may develop (see text for discussion).}
\end{figure}

If $v_g$ is a decreasing function of $\rho$ we have the phenomenon of
shock formation (see Figure~\ref{fig:kinematic}). That is, since the
patches of an initial density profile travel with different speeds,
the low-density regions catch up with higher density regions and
discontinuities in the density profile, known as shocks, develop.
Strictly, after the formation of a shock the description of the
density profile by the first order equation (\ref{cty}) breaks downs
and one has to supplement (\ref{cty}) with second-order spatial
derivatives to describe the shock profile.

To deduce the velocity of a shock consider two regions of density
$\rho_1$ to the left and $\rho_2$ to the right separated by a
shock. Mass conservation implies that the velocity (positive to the
right) of the shock is given by
\begin{equation}
v_s = \frac{J(\rho_1)- J(\rho_2)}{\rho_2-\rho_1}\;.
\label{vs}
\end{equation}
Note that if the current density relation has a maximum then one can have a
stationary shock, $v_s=0$.

A particular form for the current density relationship studied in
\cite{LW55b} and which is relevant to the ASEP is
\begin{equation}
J(\rho) = \rho(1-\rho) \;.
\label{jasep}
\end{equation}
In this case $J_m = 1/4$ at $\rho_m = 1/2$ and $J$ is symmetric about the maximum.
Expression (\ref{jasep}) may be easily understood for the ASEP
by noting that in order for a particle to hop across
a bond, the site to the left must be occupied and that to the right
empty.  For a bond at position $x$, the probability of the former is
$\rho(x)$ and that of the latter is $1-\rho(x)$.  Assuming these
events are uncorrelated (which they are not---but we shall come to
this later), we have $J(\rho) = \rho(1-\rho)$.  

In the case (\ref{jasep}) one can compute the kinematic wave and shock
velocities from (\ref{vg},\ref{vs})
\begin{equation}
v_g = 1-2\rho \qquad v_s = 1-\rho_1-\rho_2
\end{equation}
where, as above, $\rho_1$ and $\rho_2$ are the densities on either
side of the shock.  Thus the kinematic wave velocity is negative when
$\rho>1/2$ and the shock velocity is negative when $\rho_1 >
1-\rho_2$.

\begin{figure}
\begin{center}
\includegraphics[scale=0.8]{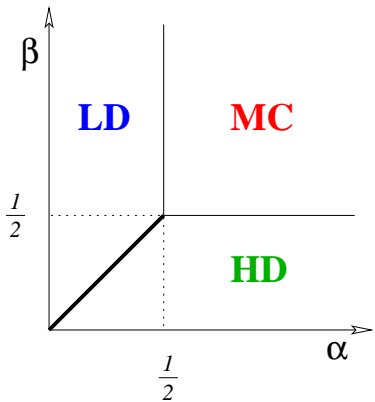}
\end{center}
\caption{\label{fig:asepmfpd} Mean field phase diagram for the ASEP
  showing the range of $\alpha$ and $\beta$ in which the low-density
  (LD), high-density (HD) and maximal-current (MC) phases are seen.}
\end{figure}

The kinematic wave theory can be used to predict the phase diagram of
the open boundary ASEP, in which distinct steady-state
behaviours are demarcated (Figure~\ref{fig:asepmfpd}).  The left hand boundary $x=0$ is considered as a reservoir
of particle of density $\rho_l = \alpha$ and the right hand
boundary ($x=N$) as a reservoir of density $\rho_r = 1-\beta$.
Associated with these boundary densities are kinematic waves with
velocities
\begin{equation}
v_l = 1-2\alpha \quad ; \quad
v_r = 2\beta -1 \;.
\end{equation}
In the case where $\alpha< 1/2$ and $\beta < 1/2$ both kinematic waves
propagate into the system.  So, for example, from an initially empty
system the kinematic waves of densities $\alpha$ and $1-\beta$ will
enter from the left and right of the system and meet somewhere in the
middle forming a shock which then moves with velocity
\begin{equation}
v_s = \beta - \alpha\;.
\end{equation}
If $\beta >\alpha$ the shock moves to the right hand boundary and
density associated with the left hand boundary, $\rho_l = \alpha$, is
adopted throughout the bulk of the system. On the other hand if
$\alpha <\beta$ the shock moves to the left hand boundary and density
associated with the right hand boundary, $\rho_r = 1-\beta$, is
adopted throughout the bulk of the system.

In the case $\alpha= \beta < 1/2$, $v_s=0$ and the shock 
is stationary. In the stochastic system  the shock,
although on average stationary,
diffuses around the system and effectively  reflects off the boundaries.
The result is that the shock is equally likely to be anywhere in the system.

In the case where one of $\alpha$ or $\beta > 1/2$, the kinematic wave
associated with that boundary does not propagate into the system and
the kinematic wave which does propagate from the other boundary
controls the bulk density. Thus, the boundary with $\alpha$ or $\beta
<1/2$ controls the bulk density

Finally if both $\alpha$, $\beta>1/2$. The kinematic waves from both
boundaries do not penetrate.  To describe this phase one
needs to add a diffusive contribution to the current (\ref{jasep})
i.e. to  consider  second-order spatial derivative of $\rho$
which we shall do in the next section. The result is that
steady state of the system
has density $1/2$; the system adopts the maximal current
density $\rho_m = 1/2$ which is the density associated with kinematic
wave velocity zero.

The resulting bulk
densities and currents are then as shown in Table~\ref{tab:asepmf}
which corresponds to the phase diagram given in
Figure~\ref{fig:asepmfpd}.  Here we have adopted what has become the
standard nomenclature for the three phases---the terminology should
hopefully be self-explanatory.

\begin{table}
\begin{center}
\begin{tabular}{c|c|c|c}
Region & Phase & Current $J$ & Bulk density $\bar{\rho}$ \\\hline
$\alpha < \beta, \alpha < \frac{1}{2}$ & Low-density (LD) & $\alpha(1-\alpha)$
& $\alpha$ \\
$\beta < \alpha, \beta < \frac{1}{2}$ & High-density (HD) &
$\beta(1-\beta)$ & $1 - \beta$ \\
$\alpha > \frac{1}{2}, \beta > \frac{1}{2}$ & Maximal current (MC) &
$\frac{1}{4}$ & $\frac{1}{2}$ 
\end{tabular}
\end{center}
\caption{\label{tab:asepmf} Properties of the ASEP obtained using the
  extremal-current principle under a mean-field approximation for the
  density-current relationship.}
\end{table}

This simple kinematic wave theory using the first order equation
(\ref{cty}) correctly predicts the phase diagram, as we shall see
from the exact solution.  The theory also
implies two sub-phases of the high and low-density phases according to
whether the boundary which does not control the bulk density has
$\alpha$ or $\beta$ greater than or equal to 1/2.

A more refined theory is the domain wall theory
\cite{KSKS98,PS99,HKPS01,SA02} which is able to make more general
predictions (such as for multi-species problems) and is reviewed in
\cite{Schutz03}.  We remark that kinematic waves are in fact the
characteristic curves of the differential equation (\ref{cty2}).  For
more general first-order differential equations modelling driven
diffusive systems one can use the method of characteristics to predict
the phase diagram as we briefly discuss in Appendix~\ref{char}.

\subsubsection{Comparison of mean-field and hydrodynamic
approaches}
In the previous subsection we showed how a simple phenomenological
description by a first-order differential equation can predict the
correct phase diagram of the ASEP.  
Generally the assumption that the current is
solely a function of the local density is an assumption which is {\em
mean-field} in nature.  Mean-field theories are usually considered as
a general class of approximations where correlations within the
configuration of the system are ignored at some level. Thus a very
simple mean-field theory is to only keep information about the density
profile. A more sophisticated mean-field
theory  might  keep  information about two  point
correlation functions but will still ignore correlations at some level. 
In some cases or limits a mean-field
description may  become an  exact description.

At this point, it is useful to compare the idea of a mean-field
description with the hydrodynamic limit usually considered
in the mathematical literature. In the latter case one defines, for
example, a local density by averaging or coarse graining over some
scale. Then one seeks to define a limit where this scale is allowed to
become large but the ratio of this scale to the system size is small.
The hydrodynamic limit is then proven by showing that the variables
coarse-grained on this scale satisfy a deterministic equation where
the noise has been scaled out.  The procedure involves determining the
local stationarity of the system on the coarse-grained scale. As in mean-field theory, the simplest case results in describing the system by
the density profile alone without correlations.

Therefore, at a practical level, the results of a mean-field
description often coincide with those obtained in a hydrodynamic
limit---both approaches result in the same differential equations
involving the density or some small number of variables.  But, it is
important to stress that the approaches differ significantly in
spirit: the hydrodynamic limit specifies a limit or scale where
an exact equation is proven.
On the other hand,  mean-field theory is often used
as a rough and ready approximation regardless
of whether the hydrodynamic limit exists.
In addition, the hydrodynamic limit introduces
coarse grained densities whereas the equivalent mean-field equations
are often obtained without such an explicit procedure as we show
below.  Finally mean-field or hydrodynamic equations can often be
written down phenomenologically as with the Lighthill-Witham theory.

\subsubsection{Mean-field treatment of the density profile}
We start  our considerations of mean-field approximations
by keeping space discrete,
and implementing  a  simple procedure 
which is to 
to assume a factorised form for the
stationary distribution.    That is, in terms of
the indicator variables $\tau_i$ which take values
$\tau_i=1$ if site $i$ is
occupied and $\tau_i=0$ otherwise, we suppose
\begin{equation}
\label{int:fac}
P(\tau_1, \tau_2, \ldots, \tau_N) = \prod_{i=1}^{N} \mu_i(\tau_i)
\quad\mbox{where}\quad \mu_i(\tau) = \left\{ \begin{array}{ccc} \rho_i &
  \mbox{if} & \tau=1 \\ 1-\rho_i & \mbox{if} & \tau=0 \end{array} \right.
\end{equation}
and  $\rho_i = \avg{\tau_i}$.  
Note that all higher order correlations vanish
i.e.
$\langle \tau_i \tau_j\rangle$ is approximated in this mean-field
theory by
$\rho_i \rho_j$.

Using the master equation (\ref{int:me}), one can derive an exact
equation for the evolution of the density at site $i$:
\begin{equation}
\frac{\partial \langle \tau_i\rangle }{\partial t} = 
\langle \tau_{i-1}(1-\tau_{i})\rangle
- \langle \tau_{i}(1-\tau_{i+1})\rangle
\label{currex} \;.
\end{equation}
A full derivation is given, for example, in \cite{DE96}.  To
understand this equation note that the right-hand side contains the
particle current from site $i-1$ to $i$ minus the particle current
from $i$ to $i+1$, which by particle conservation gives the rate of
change of density at site $i$.  In the steady state, where the density
is independent of time, the right hand side of (\ref{currex}) must
vanish which implies the exact steady state result
\begin{equation}
J= \langle \tau_{i}(1-\tau_{i+1})\rangle
\end{equation}
where the current $J$ between two neighbouring sites
is independent of position.

In the mean-field approximation equation (\ref{currex}) becomes
\begin{equation}
 \frac{\partial \rho_i}{\partial t} = \rho_{i-1}(1-\rho_{i}) -
\rho_{i}(1-\rho_{i+1})
\label{currmf}
\end{equation}
and the steady-state mean-field current becomes
\begin{equation} J= \rho_{i}(1-\rho_{i+1})\;. \end{equation}
The density profile and phase diagram can be obtained in the
mean-field approximation by considering the mapping
\begin{equation} 
\rho_{i+1} = 1- \frac{J}{\rho_i}\;.
\label{rhomap}
 \end{equation}
The mapping has fixed points at
\begin{equation}
\rho_{\pm}=  \frac{1}{2} \left[ 1 \pm \sqrt{1-4J} \right]\;.
\label{rhopm}
\end{equation}

\begin{figure}
\begin{center}
\includegraphics[scale=0.7]{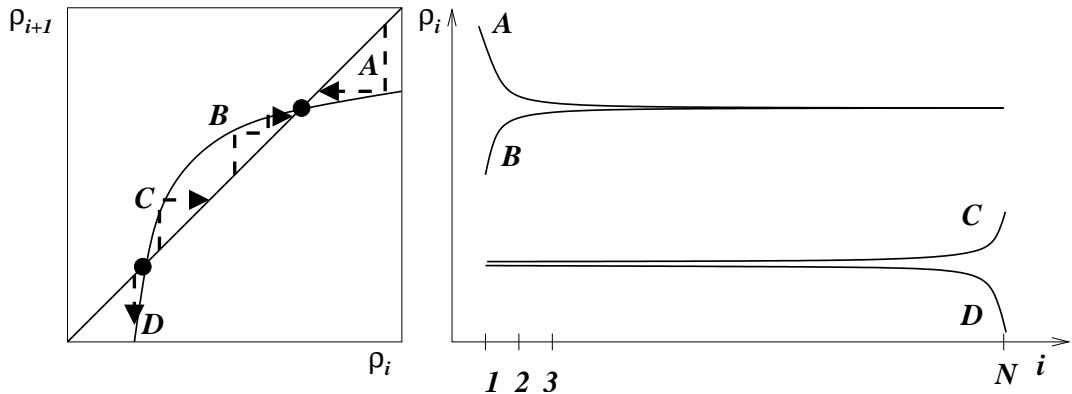}
\end{center}
\caption{\label{fig:HD-staircase}Iteration of the map (\ref{rhomap})
in the high and low-density phases. Profiles A and B iterate towards
the high-density fixed point; profiles C and D begin infinitesimally
close to the low-density fixed point and iterate away from it as the
right boundary is approached.}
\end{figure}

\begin{figure}
\begin{center}
\includegraphics[scale=0.7]{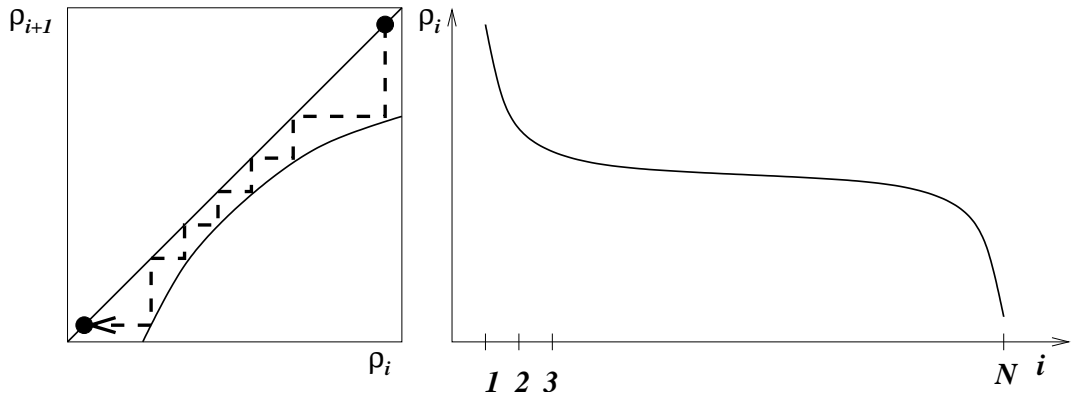}
\end{center}
\caption{\label{fig:MC-staircase} Iteration of the map (\ref{rhomap})
to give the density profile in the maximal-current phase.}
\end{figure}

The structure of possible solutions is illustrated in
Figures~\ref{fig:HD-staircase} and \ref{fig:MC-staircase} and
correspond to the three phases. In the high-density phase the map
iterates to a value of $\rho_i$ arbitrarily close to the stable
$\rho_+$ fixed point in some finite number of steps.  In the
low-density phase the map starts close to the unstable $\rho_-$ fixed
point. In the maximal current phase $J = 1/4 + O(N^{-2})$ so that
there are no real fixed points and the density profile takes on its
characteristic maximal current shape: the gradient of the density
profile is always negative and far away from both boundaries the
density approaches the value 1/2 (see Figure~\ref{fig:MC-staircase}).

\begin{figure}
\begin{center}
\includegraphics[scale=0.7]{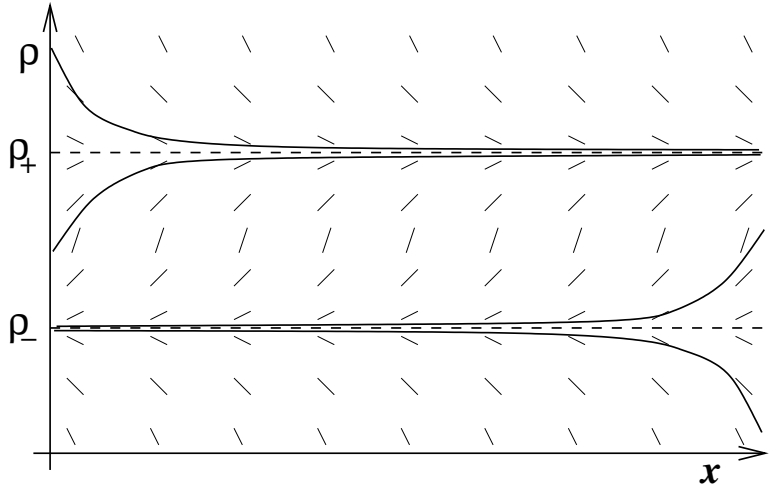}
\end{center}
\caption{\label{fig:LDHD-profiles} Sketches of the possible
solutions
of equation (\ref{rhoss}) with $\rho_\pm$ real, which corresponds
to  $J$ less than the maximum of $J_0(\rho)$.
}
\end{figure}

\begin{figure}
\begin{center}
\includegraphics[scale=0.7]{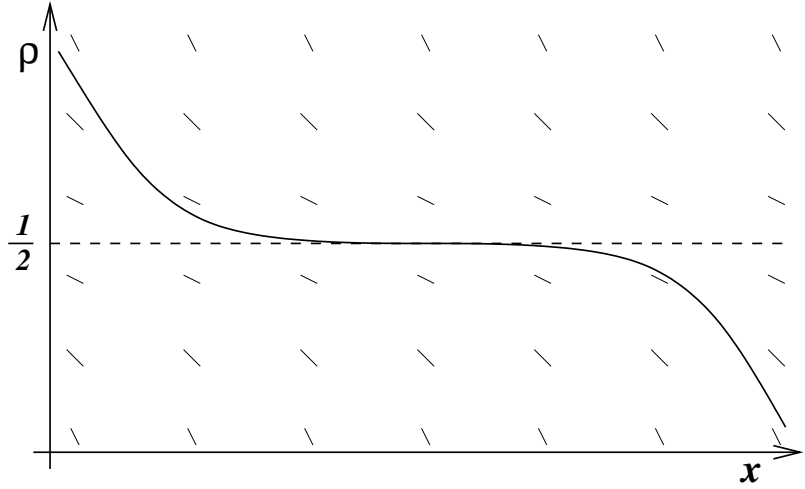}
\end{center}
\caption{\label{fig:MC-profile}
Sketches of the 
solutions
of equation (\ref{rhoss}) with $\rho_\pm$ complex, which corresponds
to  $J$ greater than the maximum of $J_0(\rho)$.}
\end{figure}

The phase diagram can be fully analysed within this discrete
mean-field approximation \cite{DDM92}.  Here we choose to make a
further continuum approximation as this approach can be easily
generalised to other systems as we shall see in the next subsection.
One can take a continuum limit of (\ref{currmf}) in a simple way by
replacing $\rho_i(t)$ by $\rho(x,t)$ where $x=ai$ and $a$ is
introduced as the lattice spacing.  Expanding to second order in $a$
\begin{equation}
\rho_{i\pm 1} = \rho  \pm a \frac{\partial \rho}{\partial x}
+ \frac{a^2}{2} \frac{\partial^2 \rho}{\partial x^2}
\end{equation}
yields
\begin{equation}
 \frac{\partial \rho}{\partial t} = - a \frac{\partial}{\partial x} 
\left[
J_0(\rho)   -  D \frac{\partial \rho}{\partial x}\right]
\label{cty4}
\end{equation}
where
\begin{equation}
J_0(\rho) = \rho(1-\rho ) \quad\mbox{and}\quad D = \frac{a}{2}\;.
\label{Jasep}
\end{equation}
$J_0(\rho)$ represents a contribution to the current due to the asymmetry of
the dynamics i.e. the external drive and $-D \frac{\partial}{\partial
x}\rho(x)$ is the diffusive contribution to the current (down the concentration
gradient).  An equation of this form was postulated by Krug \cite{Krug91}.
Ignoring the
diffusive part recovers the Lighthill-Whitham phenomenological equation
(\ref{cty}) or more formally rescaling time to $t' = t/a$ and letting
$a\to 0$ recovers the hydrodynamic limit on the Euler scale.
Retaining the diffusive term in (\ref{cty4})
allows one to calculate
the (approximate) density profiles explicitly. To see this
one considers the steady state of (\ref{cty4})
\begin{equation}
J= J_0(\rho)   -  D \frac{\partial \rho}{\partial x}
\label{Jss}
\end{equation}
where $J$ is a constant (the current) to be determined.
Rearranging (\ref{Jss},\ref{Jasep}) yields
\begin{equation}
\frac{\partial \rho}{\partial x} 
=-\frac{(\rho-\rho_+)(\rho-\rho_-)}{D}
\label{rhoss}
\end{equation}
where $\rho_\pm$ are  given by (\ref{rhopm}).
One can separate
variables of this first order differential equation and
integrate, for example, from the left boundary
\begin{equation}
-\int_{\rho(0)}^{\rho(x)} \rmd \rho\,
\frac{D}{(\rho-\rho_+)(\rho-\rho_-)} = x \;,
\label{int}
\end{equation}
which yields
\begin{equation}
\ln \left[ \frac{(\rho-\rho_+)(\rho(0)-\rho_-)}{(\rho-\rho_-)(\rho(0)-\rho_+)}
\right]
=  -\frac{(\rho_+-\rho_-)x}{D} \;.
\label{prof}
\end{equation}
The boundary conditions to be imposed on (\ref{Jss}) are
implied by the reservoir densities:
\begin{equation}
\rho(0)= \alpha \qquad  \rho(N+1) = 1-\beta
\end{equation}
the second of these will fix $J$ from (\ref{prof}):
\begin{equation}
 \left[ \frac{(1-\beta-\rho_+)(\alpha-\rho_-)}{(1-\beta-\rho_-)(\alpha-\rho_+)}
\right]
= \exp\left( -\frac{(\rho_+-\rho_-)(N+1)}{D}\right) \;.
\label{Jfix}
\end{equation}
One can rearrange  (\ref{prof}) to give an explicit expression for the profile,
but to deduce the phase diagram it is simplest to return to
(\ref{rhoss}).  One can easily sketch the possible solution curves to this equation
as is done in Figures~\ref{fig:LDHD-profiles} and
\ref{fig:MC-profile}.
The phase diagram in the limit of a large system can be deduced 
as follows:\\

\noindent {\em High-density phase}: The bulk density is $\rho_+= \rho(N+1)=1-\beta$
which yields $J= \beta(1-\beta)$ and $\rho_-=\beta$. Then the condition
$\rho_+ > \rho_-$  implies $\beta <1/2$ and the condition that $\rho(0) > \rho_-$
yields $\alpha<\beta$.\\

\noindent {\em Low-density phase}: The bulk density is $\rho_-= \rho(0)=\alpha$
which yields $J= \alpha(1-\alpha)$ and $\rho_+=\alpha(1-\alpha)$. 
Then the condition
$\rho_+ > \rho_-$  implies $\alpha <1/2$ and the condition that $\rho(N+1) < \rho_+$
yields $\alpha>\beta$.\\

\noindent {\em Co-existence line}:
When $\alpha = \beta <1/2$ the profile begins infinitesimally close
to $\rho_-=\alpha$ and finishes infinitesimally close to $\rho_+ = 1-\beta$.
In the bulk of the system there is a shock front in which the density switches
over. Thus, on this line the High-density and Low-density phases coexist.\\

\noindent {\em Maximal current phase}: In this case $J > 1/4$ and the
profile has the shape shown in Figure~\ref{fig:MC-profile}.  One can
show that to satisfy (\ref{Jfix}) one must take
\begin{equation}
J\approx \frac{1}{4} + \frac{(D\pi)^2}{N^2}
\end{equation}
Then setting $\rho(x)=1/2 + \eta(x)$, (\ref{rhoss}) becomes
\begin{equation}
\frac{\partial \eta}{\partial x} = - \frac{\eta^2}{D} - O(1/N^2)
\end{equation}
which implies  
 that in the large $N\to \infty$ limit the decay 
of the density away from the left boundary follows, for large $x$,
\begin{equation}
\rho(x)-\frac{1}{2} \sim \frac{D}{x}\;.
\end{equation}

\subsubsection{Extremal current principle}
\label{sec:PD}

Let us consider more general equations of the form (\ref{Jss}) where
the current of particles $J$ across each bond is the sum of a
systematic current $J_0(x)$ and a diffusive current $J_D$.  The former is
taken to be a known function $J_0(\rho)$ of the local density
$\rho(x)$ arising from the nature of the particle interactions whilst
the latter is taken to cause motion down a density gradient.  Thus the
diffusive current should depend only on the local gradient, take
opposite sign to its argument and vanish in regions of constant
density: e.g.  $J_D = - D(\rho) \rho'$ where $D(\rho)$ is some
(positive) diffusion constant.

The steady-state profile is given by (\ref{rhoss}) where $J$ is a
constant and $D$ may now depend on $\rho$.  Generally, if $J_0$ has a
single maximum $J_0^{max}$ as a function of $\rho$ we obtain density
profiles with the same qualitative structure as just deduced for the
ASEP.  To see this we can again sketch the solution curves as in
Figs.~ \ref{fig:LDHD-profiles} and \ref{fig:MC-profile}.  If $J<
J_0^{max}$ there are two fixed points $\rho_\pm$ where $\frac{\partial
\rho(x)}{\partial x}=0$ and the high and low-density profiles iterate
to and away from these fixed points.  If $J> J_0^{max}$ there are no
real fixed points and $\frac{\partial \rho(x)}{\partial x}=0$ yielding
a maximal current profile. As the boundary reservoir densities $\alpha$
and $1-\beta$ are varied, the system undergoes transitions, of the
same type as those exhibited by the ASEP, between regimes
characterised by different functional forms of the particle current
and density.

These phase transitions can be understood qualitatively using a
macroscopic extremal current principle proposed by Krug \cite{Krug91}
and later developed into a more general description of driven systems
\cite{PS99,HKPS01}.  The argument runs as follows.  From
Figs.~\ref{fig:LDHD-profiles} and \ref{fig:MC-profile} one sees that
the density profile has the following properties: it is monotonic in
$x$; it is bounded from above and below due to the finite values of
the boundary reservoir densities, and in the limit of an infinite
system there is a bulk region in which the density is roughly constant
$\rho(x)=\bar\rho$.  In the bulk region, the diffusive current $J_D$
vanishes, and so here $J=J_0(\bar\rho)$.  If the boundary conditions
are such that the density at the left boundary $\rho_L$ is greater
than that at the right $\rho_R$, we have regions near the boundaries
where the density is a decreasing function of $x$ (as a consequence of
monotonicity).  In these regions, the diffusive current $J_D(\rho)$ is
positive, and hence the systematic current is reduced compared to that
in the bulk.  This implies that
\begin{equation}
\label{int:Jmax}
J = J_0(\bar\rho) = \max_{\rho\in[\rho_R,\rho_L]} J_0(\rho)
\quad\mbox{when}\quad \rho_R<\rho_L
\end{equation}
since every possible density $\rho$ in the range $[\rho_R,\rho_L]$ is
represented somewhere in the system, and as we just argued, the
systematic current in the bulk must be greater than that at any other
point.  A similar argument goes through for the case $\rho_L <
\rho_R$, except here the diffusive current is negative in the
boundary regions and hence
\begin{equation}
\label{int:Jmin}
J = J_0(\bar\rho) = \min_{\rho\in[\rho_L,\rho_R]} J_0(\rho)
\quad\mbox{when}\quad \rho_L<\rho_R \;.
\end{equation}
The pair of equations (\ref{int:Jmax}) and (\ref{int:Jmin})
encapsulate the extremal current principle.

\subsection{The exact solution of open-boundary ASEP}
So far we have discussed mean-field  approaches to
deducing the phase diagram of the ASEP.  These approaches yield the
correct phase diagram but do not correctly predict correlation
functions.  For example, the density profiles discussed in the
previous subsections are only qualitatively correct.  The time has now
come to summarise the main ideas behind the exact solution of the ASEP
that uses the matrix product formulation briefly introduced in
Section~\ref{intro} \cite{DEHP93}.  (We note that an exact solution
using recursion relations for the stationary weights was found
independently \cite{SD93}, building on earlier work \cite{DDM92,DE93}
for the case $\alpha=\beta=1$.)

\subsubsection{Nonequilibrium matrix product steady states}
\label{mpa}

To recap, a matrix product state is constructed as a product of
matrices, one for each site, chosen according to the state of the
site.  Then, a scalar probability is obtained by combining in some way
the elements of this product.  The precise prescription for the ASEP
with open boundaries was determined in \cite{DEHP93}.  It is
summarised as follows.

\begin{figure}
\begin{center}
\includegraphics[scale=0.66]{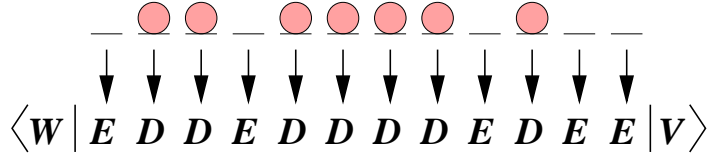}
\end{center}
\caption{\label{fig:visual} The procedure by which a particle
  configuration is transliterated to a string of matrices.}
\end{figure}

First, it is convenient to work with an unnormalised weights
$f(\tau_1, \ldots, \tau_N)$ 
so that
\begin{equation}
\label{int:PfZ}
P(\tau_1, \tau_2, \ldots, \tau_N) = \frac{f(\tau_1, \tau_2, \ldots,
  \tau_N)}{Z_N}
\end{equation}
in which $Z_N$ is the normalisation obtained by summing the weights
$f$ over all $2^N$ possible configurations of the $N$-site lattice.
It is these weights that take the matrix-product form
\begin{equation}
\label{int:mp}
f_N(\tau_1, \tau_2, \ldots, \tau_N) = \bra{W} X_{\tau_1} X_{\tau_2}
\cdots X_{\tau_N} \ket{V}
\end{equation}
in which $X_{\tau_i}$ is a matrix $D$ if site $i$ is occupied by a
particle ($\tau_i=1$) or a matrix $E$ otherwise.  Observe that this is
a very visual way of representing a particle configuration---see
Figure~\ref{fig:visual}.  Note further that the matrix appearing at a
particular point in the product does not depend explicitly on the site
label $i$, only the state of that site, and that the vectors $\bra{W}$
and $\ket{V}$ perform the necessary conversion of the matrix product
to a scalar.

With the notation established, expressions for macroscopic quantities
quickly follow.  In all of these, one requires the normalisation which
can be written as
\begin{equation}
\label{int:Z}
Z_N = \bra{W} (D+E)^N \ket{V} \equiv \bra{W} C^N \ket{V}
\end{equation}
in which the matrix $C=D+E$ has been introduced for convenience.  The
quantity $Z_N$ is simply the sum of the weights of all possible
configurations.  The density at site $i$ is then
\begin{equation}
\label{int:rho}
\rho_i = \avg{\tau_i} = \frac{\bra{W} C^{i-1} D C^{N-i} \ket{V}}{Z_N}
\end{equation}
and the current between bonds $i$ and $i+1$ is
\begin{equation}
\label{int:J}
J_{i,i+1} = \avg{\tau_i (1-\tau_{i+1})} =
  \frac{\bra{W} C^{i-1} DE C^{N-i-1} \ket{V}}{Z_N} \;.
\end{equation}
In order actually to evaluate these expressions, one needs to know a
set of algebraic relations between $D$, $E$, $\ket{V}$ and $\bra{W}$.
In \cite{DEHP93} it was shown that sufficient conditions
for  the 
stationary solution of the master equation of the ASEP via
(\ref{int:mp}) are
\begin{eqnarray}
\label{int:DE}
DE &=& D+E \\
\label{int:DV}
D \ket{V} &=& \frac{1}{\beta} \ket{V} \\
\label{int:EW}
\bra{W} E &=& \frac{1}{\alpha} \bra{W} \;.
\end{eqnarray}

In Section~\ref{RRproof} we will prove that these relations indeed
give the desired solution of the master equation.  In the meantime, we
can at least check the necessary condition
that the steady state  particle
current $J_{i,i+1}$ across the $i\th$ bond is independent of $i$, and
further equals the currents of particles into and out of the system at
its boundaries.  To see this
note first of all that the combination $DE$ appearing in 
(\ref{int:J}) can be replaced with
$D+E=C$ using (\ref{int:DE}).  Then,
\begin{equation}
\label{int:JofZ}
J_{i,i+1} \equiv J = \frac{\bra{W} C^{N-1} \ket{V}}{Z_N} =
\frac{Z_{N-1}}{Z_N} \;.
\end{equation}
At the left boundary we have
\begin{equation}
J_L = \alpha (1-\rho_1) = \alpha \frac{ \bra{W} E C^{N-1} \ket{V}}{Z_N} =
\frac{\bra{W}C^{N-1}\ket{V}}{Z_N} = J
\end{equation}
where we have used (\ref{int:EW}). Similarly, at the right boundary one has,
using (\ref{int:DV})
\begin{equation}
J_R = \beta \rho_N = \beta \frac{ \bra{W} C^{N-1} D \ket{V} }{Z_N} =
\frac{\bra{W}C^{N-1}\ket{V}}{Z_N} = J \;.
\end{equation}

Relations \RR\ form what is referred to as a {\em quadratic algebra} which
amounts to a set of \emph{reduction relations} for matrix products.
The power of the quadratic algebra is that it allows one to consider
any arbitrary product $U$ of $D$ and $E$ matrices and determine the
element $\bra{W} U \ket{V}$ without recourse to an explicit
representation.  This is because if one has a product in which a $D$
matrix appears before an $E$ matrix, it will be possible to use
(\ref{int:DE}) to reduce to the sum of two shorter products, one of
which is missing the $D$ and the other the $E$.  If either of these
terms themselves has a $D$ before an $E$, (\ref{int:DE}) can again be
used to generate further terms in the sum.  Once all terms contain
only $D$ matrices appearing after $E$ matrices, or completely lack
either $D$s or $E$s no further reduction is possible and one has found
an expression for $U$ in the form
\begin{equation}
\label{int:reord}
U = \sum_{n,m} a_{n,m} E^n D^m
\end{equation}
where $a_{n,m}$ is a positive coefficient arising from the reduction
process.  Once this has been achieved, evaluating the element
$\bra{W}U\ket{V}$ using (\ref{int:DV}) and (\ref{int:EW}) is
simplicity itself:
\begin{equation}
\label{int:slurp}
\bra{W} U \ket{V} = \left( \sum_{n,m} a_{n,m} \frac{1}{\alpha^n}
\frac{1}{\beta^m} \right) \braket{W}{V} \;.
\end{equation}
Conventionally, the combination $\braket{W}{V}$ is taken as unity,
although its actual value never enters into any physical quantities,
since it appears as a prefactor in both the numerator and denominator
of (\ref{int:PfZ}).

One might ask whether the order with which (\ref{int:DE}) is applied
to different $DE$ pairs appearing in $U$ could affect the final
outcome (i.e., whether \RR\ describes a non-associative algebra).
That this is not the case can be seen from a  straightforward
argument.  Let $U$ comprise at least two $DE$ pairs:
$U=U_1(DE)U_2(DE)U_3$ where $U_{1,2,3}$ are products of $D$ and $E$
matrices that may, or may not, contain $DE$ pairs.  Clearly, the same
matrix product expression is obtained by applying (\ref{int:DE}) to
the two $DE$ pairs in either order.  Therefore, any non-uniqueness
would have to come at the next stage where there are four strings to reduce.
Now, each of these strings contains either no $DE$ pairs (which is
unique); one $DE$ pair (which can be reduced only in one way); or is
in the form given for $U$ above.  In the latter case, one can iterate
the argument just given to demonstrate that the  reduction is unique.
Hence, the quadratic algebra \RR\ involving only two
matrices $D$ and $E$ is associative.  However, as we shall see in
Section~\ref{2species}, with more than two matrices the requirement
for uniqueness under reduction is important and imposes restrictions
on the class of models that can be solved using the matrix product
method.

The unconvinced reader may take refuge in the fact that explicit
representations of the $D$ and $E$ matrices can be constructed---for
example, that presented in Section~\ref{diag}.  In general, these
matrices do not commute and have infinite dimension, as was shown in
\cite{DEHP93}.  An exception to this can be determined by assuming
that $D$ and $E$ do commute: then, using \RR\ one finds that
\begin{equation}
\bra{W}(DE - ED)\ket{V} = \left(\frac{1}{\alpha} + \frac{1}{\beta} -
\frac{1}{\alpha\beta}\right) \braket{W}{V} = 0
\end{equation}
which can be true only if $\alpha+\beta=1$.  Along this line, the
relations \RR\ can be satisfied by the scalar choices $D=1/\beta,
E=1/\alpha, \bra{W}=\ket{V}=1$.  This is equivalent to putting $\rho_i
= \alpha/(\alpha+\beta)$ in (\ref{int:fac}), and thus along the line
$\alpha+\beta=1$ (and only this line) the mean-field theory is exact.

\subsubsection{Nonequilibrium partition functions
and phase transitions} In Sec.~\ref{direct} we will show how the
reduction process just described can be applied systematically to
evaluate the matrix expressions (\ref{int:rho}) and (\ref{int:J}) for
the density and current in the ASEP.  Here we quote the result for the
normalisation $Z_N$
\begin{equation}
\label{open:Z}
Z_N = \sum_{p=1}^{N}
\frac{p (2N-p-1)!}{N! (N-p)!}
\frac{\big(\frac{1}{\alpha}\big)^{p+1} -
  \big(\frac{1}{\beta}\big)^{p+1}}{\frac{1}{\alpha} -
  \frac{1}{\beta}}
\end{equation}
from which the current follows via (\ref{int:JofZ}).  We now discuss
aspects of this expression which have generated recent interest.

The relationship (\ref{int:JofZ}) between the current and the
normalisation is intriguing. We may interpret $Z_N$ as the partition
function for the nonequilibrium steady state. Then assuming that $Z_N$
grows exponentially with $N$ we find that the analogue of the
thermodynamic free energy is
 \begin{equation}
-\lim_{N\to\infty} \frac{\ln Z_N}{N} 
= \lim_{N\to\infty} \ln J  \;.
\label{jfe}
\end{equation}
Thus the current plays the role of the free energy.

The phase transitions in the $\alpha$--$\beta$ plane, manifested by
different forms for the current $J(\alpha,\beta)$, arise from
different large $N$ asymptotic forms for $Z_N$ and we shall determine
these explicitly in Section~\ref{open}.  Moreover, recognising the
normalisation $Z_N$ as a nonequilibrium partition function, one can
analyse the Lee-Yang zeros of the partition function and show that
they predict precisely the location and order of the phase transitions
in the driven system\cite{BE02,BE03}. The interesting point is that
the zeros of the partition function are in the complex plane of
transition rates $\alpha,\beta$, as compared to equilibrium problems
where partition functions zeros are in the plane of some intensive
thermodynamic variable such as fugacity or temperature.

As we will discuss in Section~\ref{surf}, one can map the
normalisation (\ref{open:Z}) of the ASEP to the partition function of
a surface in thermal equilibrium with a heat bath.  In this mapping,
the transition rates $\alpha$ and $\beta$ become equilibrium
fugacities, and the phase transitions in the driven system correspond
to adsorption-desorption transitions in the equilibrium
interpretation.

The expression for the normalisation (\ref{open:Z}) involves
combinatorial coefficients known as ballot numbers
\cite{comtet74}. There has been considerable recent interest in
developing a combinatorial understanding of the origin of these
numbers and other aspects of the steady state of the ASEP \cite{DS05,Angel06,BCEPR06,CW06}.

\subsubsection{Comparison between
 matrix product states of equilibrium and nonequilibrium systems}

\label{ising}

In the latter sections of this review we shall discuss in detail other
nonequilibrium systems whose steady states are of matrix product
form. For systems with open boundaries the ansatz used is of the form
(\ref{int:mp}) where the number of matrices equals the number of
possible states for each site. For systems with periodic boundary
conditions a trace operation is used to obtain a scalar from the
matrix product rather than using boundary vectors, see
Section~\ref{ring}.  At this point it is useful to establish what the
differences are between these nonequilibrium steady states and the
equilibrium steady states which may also be written in matrix product
form.  To illustrate this we rewrite the familiar transfer matrix
solution of the the Ising model (an equilibrium system) as a matrix
product state.

Recall the one-dimensional Ising ferromagnet with $N$ spins, $s_i=\pm
1$ and Hamiltonian
\begin{equation}
{\cal H} = -J \sum_{\langle i,j \rangle} s_i s_{i+1} \;.
\end{equation}
To this chain let us add two fixed spins, $s_L$ to the left of site
$1$ and $s_R$ to the right of site $N$.  
The Boltzmann weight of a
configuration within this system when at equilibrium with a heat bath
at inverse temperature $\beta$ can be written as a matrix product
analogous to that used above for the ASEP (\ref{int:mp}).  It takes
the form
\begin{equation}
f(s_1, s_2, \ldots, s_N; s_L, s_R) = \rme^{-\beta {\cal H}} =
\bra{s_L} X_{s_1} X_{s_2} \cdots X_{s_N} \ket{s_R}
\end{equation}
where
\begin{equation}
X_{+1} = \left( \begin{array}{cc} \rme^{\beta J} & \rme^{-\beta J} \\
0 & 0 \end{array} \right) \;, \quad
X_{-1} = \left( \begin{array}{cc} 0 & 0 \\
\rme^{-\beta J} & \rme^{\beta J} \end{array} \right)
\end{equation}
and
\begin{eqnarray}
\bra{s_L=+1} &=& \left( \begin{array}{cc} 1 & 0 \end{array} \right) \;, \quad
\bra{s_L=-1} = \left( \begin{array}{cc} 0 & 1 \end{array} \right) \;, \\
\ket{s_R=+1} &=& \left( \begin{array}{c} \rme^{\beta J} \\
  \rme^{-\beta J} \end{array} \right) \;, \quad \ket{s_R=-1} = \left(
\begin{array}{c} \rme^{-\beta J} \\ 
  \rme^{\beta J} \end{array} \right) \;.
\end{eqnarray}
Note that the   algebraic relations obeyed by the matrices, such as
\begin{eqnarray}
\fl \lefteqn{W(S_{i-1}\, S_i\,S_{i+1} \to S_{i-1}\,S'_i\,S_{i+1})
X_{S_{i-1}}X_{S_i}X_{S_{i+1}}
= }\nonumber \\
W(S_{i-1}\, S'_i\,S_{i+1} \to S_{i-1}\,S_i\,S_{i+1})
X_{S_{i-1}}X_{S'_i}X_{S_{i+1}}\;,
\end{eqnarray}
where 
$W(S_{i-1}\, S_i\,S_{i+1} \to S_{i-1}\,S'_i\,S_{i+1})$ is a transition rate for flipping spin $S_i$,
do not imply any  reduction relations, rather they simply state the condition
of detailed balance in the steady state.

The partition function is  given by
\begin{equation}
Z = \bra{s_L} C^N \ket{s_R}
\end{equation}
where $C = X_{+1}+X_{-1}$. 
Calculating $Z$, for example 
by diagonalising  $C$, one finds the  free energy per spin
\begin{equation}
f = - \frac{1}{\beta} \lim_{N\to\infty} \frac{\ln Z}{N} 
= - \frac{1}{\beta} \ln (2 \cosh \rme^{\beta J}) \;.
\end{equation}
Here we notice that the boundary conditions contribute only
sub-extensively to the free energy, and that the free energy is an
analytic function of temperature---that is, there is no ordering phase
transition in the 1d equilibrium Ising model.  This contrasts strongly
with the driven nonequilibrium steady state of the ASEP, where varying
the boundary conditions induces phase transitions in bulk quantities
such as the density.

Mathematically this comes as a consequence of the $D$ and $E$ matrices
for the ASEP being infinite-dimensional whilst the 1d Ising transfer
matrices are finite-dimensional and, because their elements are
Boltzmann weights, non-negative.  For matrices of this latter class
one knows from the Perron-Frobenious theorem that the largest
eigenvalue is nondegenerate \cite{Gantmacher59,Seneta81,KT75}.  This
means that is not possible, as temperature is varied, for the largest
and second-largest eigenvalues to cross and the functional form of the
dominant contribution to the partition function to change, which would
give rise to a nonanalyticity in the free energy.  This state of
affairs applies, in fact, to all 1d spin systems with short-range
interactions \cite{Evans00}.  On the other hand, if the interactions become
long-ranged, or one goes into two or more dimensions, the transfer
matrices become infinite-dimensional and phase transitions become a
possibility.

To reiterate, the fact that one is with the matrix product
construction solving a master equation, rather than simply packaging
Boltzmann factors in an attractive way, means that one-dimensional
nonequilibrium steady states can show much richer and more interesting
behaviour than their equilibrium counterparts.  As we shall see in the
course of this article, the matrices that appear in various
nonequilibrium models often violate the conditions for the
Perron-Frobenious theorem to hold.

\subsubsection{Comparison between  matrix product states  of quantum spin
 chains and stochastic systems} Historically, matrix product states
first appeared in the context of quantum spin chains
\cite{KSZ91,FNW92}.  
There,  the   wavefunction  $|\Psi \rangle $ for the quantum
spin chain is written as a product of matrices.
Matrix product states also serve as a starting point for
approximations and  variational approaches such as Density Matrix RG
\cite{White98,CHS99}.  Recently, matrix product states have received a
lot of attention in the context of quantum information as they are
able to describe highly entangled states \cite{VC06}.

The connection between the formalism for stochastic
systems and that for quantum systems has been clearly reviewed in
\cite{RSSS98}. Here we summarise some important features.
We first note that the master equation (\ref{int:me}) can be written
in a suggestive way as
\begin{equation}
\frac{\partial}{\partial t} | P(t)\rangle  = 
-H | P(t)\rangle
\label{meqh}
\end{equation}
where $ | P(t)\rangle $ is the vector of probabilities
and $H$ is a matrix formed from the transition rates.
Clearly one can think of $H$ as a  Hamiltonian,
$| P(t) \rangle $ as a wavefunction and 
(\ref{meqh}) as a Schr\"odinger equation---this is referred to
as the `quantum
Hamiltonian
formalism' for stochastic systems (see e.g. \cite{Schutz00,Stinchcombe01}). 
The steady state of the stochastic system becomes the
ground state of the quantum system and  has eigenvalue 0.
However, it is important to  note 
some distinguishing features of (\ref{meqh}) which are consequences
of its true stochastic nature.
First, the elements of $| P(t)\rangle $ are probabilities 
therefore must all be positive (this contrasts with 
the true quantum case where the amplitude squared of the elements are
probabilities). Also, as  the master equation
conserves probability the sum of each  column of the matrix  $H$
adds to zero whereas in the true quantum case there is no such
restriction. 

For both a quantum spin chain and a nonequilibrium system the
Hamiltonian
can be written as a sum of local terms, for example representing the
exchanges between neighbouring sites,
\begin{equation}
H = \sum_i h_{i, i+1}\;.
\end{equation}
Now in the matrix product states constructed for quantum  spin chains
(see e.g. \cite{KSZ93}) one finds
\begin{equation}
h_{i, i+1}|\Psi \rangle =0
\label{hdb}
\end{equation}
for each local operator
$h_{i, i+1}$.
For a stochastic system, the simple mechanism (\ref{hdb})
to obtain a zero eigenvalue for $H$ and hence the steady
state, is only possible in the case of detailed balance, due to
conservation of
probability for each  $h_{i, i+1}$. So to 
obtain a  true  nonequilibrium steady state a more complicated
mechanism is required.
It is this more general  mechanism which is at the heart of the matrix product
formalism
for nonequilibrium systems and gives
rise to the algebraic relations, such as  (\ref{int:DE}-\ref{int:EW}),
which are a main focus of this review.

\subsection{Shocks and second-class particles}
\label{sscp}
So far we have considered the ASEP with only type of particle.
In this section  we discuss  
systems with  two types of particles: first and second-class particles.
The solution of the steady state of a system containing second-class particles
was the  second major success using the matrix product approach\cite{DJLS93}.
We shall discuss in detail the matrix product solution in Section~\ref{ring},
here we motivate the study of second-class particles  by discussing some 
physical properties of interest.

\begin{figure}
\begin{center}
\includegraphics[scale=0.4]{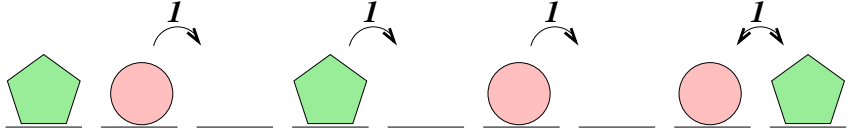}
\end{center}
\caption{\label{secondclass}
Dynamics of a system of first-class particles (circles)
and second-class  particles (pentagons).  }
\end{figure}

A system containing first-class particles (denoted 1) and second-class
particles (denoted 2) has dynamics
\begin{eqnarray}
1\,0 
\to \, 0\,1\qquad
2\,0 \to  0\,2 \qquad 1\,2 \to  2\,1 \;,
\label{scdyn}
\end{eqnarray}
all processes occurring with  rate 1 (see Figure~\ref{secondclass}). 
Thus  a  first-class particle treats the second-class particle
as  vacancy but from the point of view of a vacancy the second-class particle behaves like an ordinary (first-class)  particle.

Let us consider the velocity of a second-class  particle in an environment
of density $\rho$ of first-class particles.
The velocity of the second-class particle, denoted $v$,
is given by the probability of the second-class particle hopping forward minus the probability of it hopping backward. Ignoring correlations between  the second-class particle and local density the velocity is given by
the density of a vacancy (in front),  $1-\rho$, minus the density of  
first-class particles (behind), $\rho$
\begin{equation}
v= 1-2\rho\;.
\label{vsc}
\end{equation}
Beware that this approximation is not at all correct!---there is
structure in the local density profile around the second-class
particle, as we shall see in Section~\ref{ring}. However, the 
expression (\ref{vsc}) turns out to be  exact and was proved by Ferrari \cite{Ferrari92}. We
shall understand this result below.  For the moment, note that 
(\ref{vsc}) is
precisely the velocity of a kinematic wave of density $\rho$ discussed
in Section~\ref{sec:mfh}. Thus the second-class particle travels along
with the kinematic wave.

As described in Subsection~\ref{sec:mfh}, when two kinematic waves of
different densities meet a shock may form.  The velocity is
$1-\rho_1-\rho_2$ where $\rho_1$ is the density to the left and
$\rho_2$ is the density to the right of the shock and $\rho_2 >
\rho_1$. A second-class particle will be attracted to the position of
the shock since when it is to the left of the shock $1-2\rho_1 >
1-\rho_1-\rho_2$ and it catches up with the shock and when it is to
the right of the shock $1-\rho_1-\rho_2 > 1-2\rho_1$ so that the shock
catches up to it. Therefore we expect the second-class particle to
track the shock.  Moreover, one can use the second-class particle to
actually define the position of the shock at the microscopic
level. Then the density profile as seen from the moving frame of the
second-class particle can describe the microscopic structure of the
shock \cite{ABL88,FKS91,BCFN}.

On an infinite system exhibiting a shock, it has been shown that the
density as viewed from the second-class particle approaches a steady
state distribution with density $\rho_+$ at $+\infty$ and density
$\rho_-$ at $-\infty$ \cite{FKS91,Ferrari92}.  The exact structure of
this steady distribution has been calculated using a matrix product
approach on an infinite system which we will review in
Sections~\ref{ring} and \ref{definf}.  It demonstrates that the
density profiles decay to their limits exponentially quickly in the
distance from the second-class particle.  An interesting case is when
$\rho_+ = \rho_-$ so that there is a uniform density of particles and
no apparent shock. However, in the moving frame of a second-class
particle there is structure in the density profile: in front of the
second-class particle the density is higher and decays to the
asymptotic density according to a power law in the distance from the
second-class particle; similarly behind the second-class particle the
density is reduced and increases to the asymptotic density according
to a power law.

Another interesting aspect of the second-class particle is how its
dynamics are related to the spreading of excess mass. The central
idea, termed coupling \cite{Liggett99}, is well-known in the
mathematical community but less so within physics. Consider two
systems containing only first-class particles, identical except that
one system has $M$ particles and the other $M-1$ particles (one can
consider either a finite system or an infinite system).  The two
systems start from initial conditions differing only by the position
of the extra particle in the system with $M$ particles. In order to
implement the dynamics one can consider at each time step randomly
choosing a pair of sites $i,i+1$ to update; then if there is particle
at site $i$ and a hole at site $i+1$ the particle is moved forward.
In the dynamics let us choose the same pairs of sites in the two
systems at each update (one can think of using the same random numbers
in a Monte Carlo program).
\begin{figure}
\begin{center}
\includegraphics[scale=0.25]{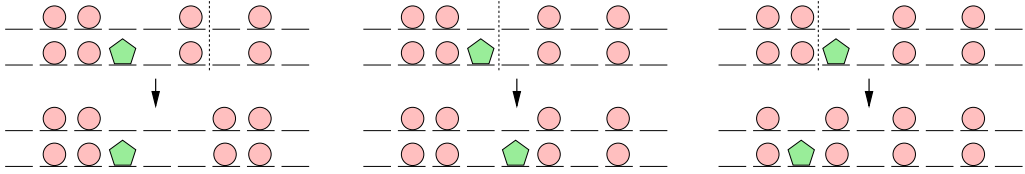}
\end{center}
\caption{\label{coupling} Illustration of coupling: the evolution of
two systems one with an excess particle. The vertical lines indicate
the bond which is updated in the two coupled systems.  The dynamics of
the excess particle (pentagon) is equivalent to that of a second-class
particle---see text for explanation.}
\end{figure}

It is easy to convince oneself that after any length of time the
configurations of the two systems will differ only by the position of
the extra particle (note that if we label the particles, the label of
the extra particle will change under the dynamics).  To see this
consider the situation where the extra particle is at site $i$ (see
also Figure~\ref{coupling}). Then, if any bond except $i-1,i$ or
$i,i+1$ is updated, the result will be the same for the two systems
and the extra particle will remain at site $i$.  If site $i+1$ is
empty and bond $i,i+1$ is updated then the extra particle at $i$ will
hop forward.  If site $i-1$ is occupied and bond $i-1,i$ is updated
then the particle at $i-1$ will hop forward in the system with $M-1$
particle but not in the system with $M$ particles.  This results in
the extra particle effectively hopping backwards.  The latter two
events result in the dynamics of the extra particle and the dynamics
are precisely those of a second-class particle.

Thus the dynamics, of a system comprising $M-1$ first-class particles
and one second-class particle describes the motion of an extra
particle added to a system of $M-1$ particles and the dynamics of
second-class particles tells us about the dynamics of mass
fluctuations.

This idea can be used to recover the velocity (\ref{vsc})
\cite{DJLS93,FKS91}. Consider adding an extra particle to a system of
density $\rho$, increasing the density by $\Delta \rho$.  The overall
current increases from $\rho(1-\rho)$ to $\rho(1-\rho) + \Delta \rho
(1-2\rho) + O(\Delta \rho^2)$.  Now, since the system can be thought of
as a system of first-class particles and an excess particle which
behaves as a second-class particle we deduce that on a large system
where $\Delta \rho \to 0$ the speed of the second-class particle must
be given by (\ref{vsc}).

An interesting result related to the spreading of mass fluctuations is
that the second-class particle exhibits {\em superdiffusive}
behaviour.  That is, with $y_t$ defined as the distance travelled by
the second-class particle, on an infinite system the variance in $y_t$
grows as \cite{Spohn91}
\begin{equation}
\langle y_t^2 \rangle - \langle y_t \rangle^2 
\sim t^{4/3}\;.
\end{equation}
On a finite periodic system of size $N$, with a density $\rho$ of first
class particles, an exact calculation has shown  \cite{DE99} 
\begin{equation}
\langle y_t^2 \rangle - \langle y_t \rangle^2 
 \simeq t N^{1/2} (\pi \rho(1-\rho))^{1/2}\;.
\label{DE99}
\end{equation}
Thus on a finite system the behaviour is diffusive but with a
diffusion constant which diverges with system size.

These results are consistent with approximate calculations such as
mode coupling \cite{vBKS85,vB91} for the spreading of excess mass
fluctuations which predict that the drift speed is $1-2\rho$ as the
spreading of density fluctuations around the drift grows as $t^{2/3}$
on an infinite system.

The dynamics of a second-class particle (\ref{scdyn})
is essentially passive in that it does not interfere with 
the dynamics of first-class particles. In this way the second-class particle can be though of as a `passive scalar' for the TASEP.
However, the density profile as seen from the second-class particle does have structure as explained earlier and gives insight into the structure of shocks. 
More recently generalisations of the second-class particle
idea have been discussed in the context of passive scalars
in  other driven diffusive systems
\cite{LMS05,CB07}.

\subsubsection{Defect particle}

The dynamics of the second-class particle can easily be generalised to hop rates
\begin{eqnarray}
1\,0 \mathop{\rightarrow}\limits^{1}
\, 0\,1\qquad
2\,0 \mathop{\rightarrow}\limits^{\alpha}  0\,2 \qquad 1\,2 
\mathop{\rightarrow}\limits^{\beta}
 2\,1 \;.
\label{defdyn}
\end{eqnarray}
In the general case $\beta \neq 1$ the dynamics of a 2 particle is no
longer passive and does affect the dynamics of the normal
particles. We will refer to the dynamics (\ref{defdyn}) as that of a
defect particle.  A related defect particle dynamics
has been considered as an exactly solvable model of
two-way traffic problem involving cars and trucks \cite{LPK97}.
\begin{figure}
\begin{center}
\includegraphics[scale=0.4]{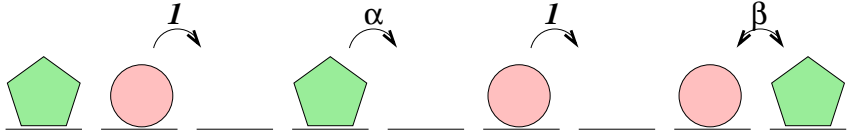}
\end{center}
\caption{\label{defect} 
Dynamics of a system of first-class particles (circles)
and defect  particles (pentagons). 
}
\end{figure}

An extreme case is $\beta = 0$ where a normal particle cannot overtake
the defect. In that case when $\alpha <1$ the defect particle is
slower than the other particles and a `traffic jam' may build up
behind the defect particle. Actually there is an interesting phase
transition which occurs on increasing the density: for low-density one
has the `traffic jam' phase where the density behind the defect
particle is large and in front of the defect particle is small.  For
sufficiently high overall density there is a transition to a uniform
density through the system.  This jamming transition is mathematically
analogous to Bose condensation, as we shall discuss in
Section~\ref{sec:multispecies}.

For general $\alpha$ and $\beta$ the steady state has been calculated
using the matrix product approach \cite{Mallick96}.  A phase diagram
emerges for a large system with density $\rho$ comprising four phases
(see Figure~\ref{fig:ringpd}) characterised by different steady-state
velocities, and density profiles surrounding the defect particle.  We
shall discuss this phase diagram in detail in Section~\ref{defpd}.

\subsection{The remainder of this review}

In this section we have introduced the asymmetric simple exclusion
process (ASEP) and some of its relatives, and through various
approximate treatments (kinematic waves, mean-field theory and the
extremal current principle) have got a feel for some of the
interesting physics, such as shocks and phase transitions, that might
emerge.  We have also presented the basic ideas underlying the exact
matrix product approach that allows the existence of these phenomena
to be confirmed.  In the remainder of this review we flesh out these
ideas considerably, presenting detailed accounts of the various
methods by which matrix product expressions can be evaluated and the
physical properties of these systems fully revealed.  As the review
progresses, we will discuss all the main models that have so far been
solved using the matrix product method.  Mostly these models are
multi-species generalisations of the ASEP in which particle number is
conserved by the dynamics (except, possibly, at the boundary sites)
whilst a small class of models further admits non-conservation in the
bulk.  Meanwhile, some of these models can be found by means of a
systematic search under certain assumptions on the existence of
reduction relations for the matrices similar to \RR.  Others involve
more complicated structures which allow, for example, treatment of
discrete-time updating schemes (such as parallel dynamics).  Although
in principle the matrix product method can be used to solve a very
wide class of nonequilibrium stochastic dynamical models in one
dimension, solutions of many interesting systems have proved elusive.
We therefore conclude our work by highlighting a few of these models
in the hope that, after reading this review, the reader will be
inspired to tackle these new and challenging problems.

For a full list of topics covered in sequence we refer the reader to
the Table of Contents presented at the start of this review.  We hope
that this will also allow the reader to locate easily any sections
that may be of particular interest.


\section{Detailed analysis of the ASEP with open boundaries}
\label{open}

Having introduced the ASEP with open boundaries (see
Figure~\ref{fig:asep}) and the matrix product expressions for the
distribution (\ref{int:mp}), density (\ref{int:rho}) and current
(\ref{int:J}) in the steady state we now explain in detail how these
are calculated using the matrix reduction relations \RR.  First,
though, we must demonstrate that the distribution implied by
(\ref{int:mp}) and \RR\ is, in fact, the stationary solution of the
master equation for the process.

\subsection{Proof of the reduction relations}
\label{RRproof}

\subsubsection{Domain-based  proof}

\begin{figure}
\begin{center}
\includegraphics[scale=0.6]{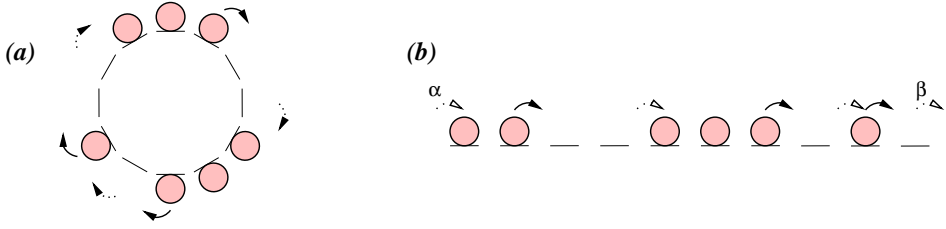}
\end{center}
\caption{\label{fig:cancel} Transitions into (dotted arrows) and out of
(solid arrows) a sample configuration for (a) the ASEP on a ring and
  (b) the ASEP with open boundaries.  Transitions occur at unit rate
  unless otherwise indicated.}
\end{figure}

As a first step towards understanding why the matrix product solution
of the ASEP with open boundaries given above is correct, we consider
the simpler case of the ASEP on a ring.  A typical configuration is
shown in Figure~\ref{fig:cancel}a.  Recall that the master equation
(\ref{int:me}) for the probability of being in configuration $\Cee$
has two contributions: a gain term from transitions into $\Cee$ and a
loss term from transitions into another state.  In the steady state,
we require these loss and gain terms to balance.

As is evident from Figure~\ref{fig:cancel}a there is for each domain of
particles on the ring one way to enter a given configuration $\Cee$
(each has a particle joining the end of a domain) and one way to exit
(when a particle leaves the front of a domain).  Therefore we have for
any given configuration $\Cee$
\begin{equation}
\sum_{\Cee'} W(\Cee' \to \Cee) = \sum_{\Cee'} W(\Cee \to \Cee') \;.
\end{equation}
The loss and gain terms in the master equation (\ref{int:me}) thus
cancel if $P(\Cee) = \mbox{const}\,\forall \Cee$.  Putting this
another way, every gain term $P(\Cee') W(\Cee' \to \Cee)$ exactly
balances one of the loss terms $P(\Cee) W(\Cee \to \Cee'')$.  Note
this is similar, but not the same, as the cancellation that occurs
when detailed balance (\ref{int:db}) is satisfied since $\Cee' \ne
\Cee''$.  This more general cancellation scheme is sometimes referred
to as {\em dynamic reversibility} \cite{Kelly79}
or  \emph{pairwise balance} \cite{SRB96}.

When one has open boundaries, as in Figure~\ref{fig:asep}, the number of
ways into a given configuration does not always equal the number of
ways out.  For example, when the left boundary site is occupied and
the right boundary empty, the number of ways in exceeds the numbers
out, and furthermore not all hops occur at the same rate (see
Figure~\ref{fig:cancel}b).  The strategy to finding the steady state
solution of the ASEP is to have a \emph{partial} cancellation between
the terms corresponding to particles attaching to the rear or
detaching from the front of domains in the bulk.  The remainder is
then cancelled with terms coming from the boundary interactions.  As
we now describe by reference to  particular examples, 
it is this cancellation scheme that is
implied by the matrix relations \RR.

\newcommand{\lat}{%
  \begin{picture}(1,1) \put(0,0){\line(1,0){1}} \end{picture}
}
\newcommand{\ent}{%
  \begin{picture}(1,1) 
    \put(0,0){\line(1,0){1}}
    \put(-0.7,2.2){\circle{1}}
    \put(0.5,1.5){\qbezier(0,0)(0,0.4)(-0.6,0.6)}
    \put(0.5,0.9){\vector(0,-1){0}}
  \end{picture}
}
\newcommand{\prt}{%
  \begin{picture}(1,1) 
    \put(0,0){\line(1,0){1}} 
    \put(0.5,0.6){\circle{1}}
  \end{picture}
}
\newcommand{\hop}{%
  \begin{picture}(1,1)
    \put(0,0){\line(1,0){1}} 
    \put(0.5,0.6){\circle{1}}
    \put(0.5,1.5){\qbezier(0,0)(0.5,0.5)(1,0)}
    \put(1.9,1.1){\vector(1,-1){0}}
  \end{picture}
}
\newcommand{\ext}{%
  \begin{picture}(1,1) 
    \put(0,0){\line(1,0){1}} 
    \put(0.5,0.6){\circle{1}}
    \put(0.5,1.5){\qbezier(0,0)(0,0.4)(0.6,0.6)}
    \put(1.5,2.5){\vector(1,1){0}}
  \end{picture}
}

\newcommand{\und}[1]{\underline{#1}}
\setlength{\unitlength}{1.2ex}

For our illustrative purposes, let us consider a particular six-site
configuration $\lat\, \prt\, \prt\, \lat\, \prt\, \lat$ for which the
master equation reads
\begin{eqnarray}
\label{hop1}
\frac{\partial}{\partial t} f( \lat\, \prt\, \prt\, \lat\, \prt\, \lat )
&=& - \alpha f(\ent\, \prt\, \prt\, \lat\, \prt\, \lat) \nonumber\\
&& {} + f( \hop\, \lat\, \prt\, \lat\, \prt\, \lat )
- f( \lat\, \prt\, \hop\, \lat\, \prt\, \lat ) \nonumber\\
&& {} + f( \lat\, \prt\, \prt\, \hop\, \lat\, \lat )
- f( \lat\, \prt\, \prt\, \lat\, \hop\, \lat ) \nonumber\\
&& {} + \beta f( \lat\, \prt\, \prt\, \lat\, \prt\, \ext ) \;.
\end{eqnarray}
In the gain terms (i.e., those with a positive sign), the arrows
indicate the hop that leads to the configuration on the left-hand
side; in the loss terms, the arrows indicate the transitions by which
this configuration can be exited. On the right-hand side we insert now
the matrix product expressions:
\begin{eqnarray}
\frac{\partial}{\partial t} f( \lat\, \prt\, \prt\, \lat\, \prt\, \lat )
&=& - \alpha  \und{\bra{W} E} DDEDE \ket{V} \nonumber\\
&& {} + \bra{W} \und{DE} DEDE \ket{V} - \bra{W} ED \und{DE} DE \ket{V}
 \nonumber\\
&& {} + \bra{W} EDD \und{DE} E \ket{V} - \bra{W} EDDE \und{DE} \ket{V}
 \nonumber\\
&& {} + \beta \bra{W} EDDED \und{D \ket{V}} \;.
\end{eqnarray}
Here we have underlined the mathematical objects (vectors or matrices)
that correspond to the sites or boundary reservoirs involved the
transitions indicated with arrows in (\ref{hop1}).  We note that all
the these underlined terms are of the form $\bra{W}E$, $DE$ or
$D\ket{V}$ and so can be reduced using \RR.
Using these reduction relations we find
\begin{eqnarray}
\fl\frac{\partial}{\partial t} f( \lat\, \prt\, \prt\, \lat\, \prt\, \lat )
&=& - \und{\bra{W}} DDEDE \ket{V} \nonumber \\
&& {} + \bra{W}\und{D}DEDE\ket{V} + \bra{W}\und{E}DEDE\ket{V} -
\bra{W}ED\und{E}DE\ket{V} - \bra{W}ED\und{D}DE\ket{V} \nonumber \\
&& {} + \bra{W}EDD\und{D}E\ket{V} + \bra{W}EDD\und{E}E\ket{V} -
\bra{W}EDDE\und{E}\ket{V} - \bra{W}EDDE\und{D}\ket{V} \nonumber \\
&& {} + \bra{W}EDDED\und{\ket{V}} \;,
\end{eqnarray}
where here the underlined vectors and matrices show what remain after
using the reduction relations.  We see that in the second and third
lines of this expressions, which originated from particles joining and
leaving domains, the middle two terms cancel; furthermore, in both
cases the first term cancels with the last term on the previous line,
and the last term with the first term on the next.  Hence the weight
$f( \lat\, \prt\, \prt\, \lat\, \prt\, \lat) = \bra{W} EDDEDE \ket{V}$
is stationary, as claimed.  If one has a configuration comprising $n$
domains of particles, none of which contact either boundary, one finds
a master equation of the form (\ref{hop1}) with $n$ pairs of terms
like those in the second and third lines of (\ref{hop1}), and the same
cancellation mechanism goes through.  For the case of the empty
lattice, $n=0$, one can verify that the term arising from the left
boundary cancels with that from the right.

It only remains to check that weights corresponding to configurations
with particles on the left, right or both boundary sites are also
stationary.  Let us consider a configuration that has a domain of
particles at the left boundary.  For such a configuration, the master
equation begins with the terms
\begin{eqnarray}
\fl \frac{\partial}{\partial t} f(\prt\, \prt\, \prt\, \lat\, \cdots) &=&
\alpha f( \ent\, \prt\, \prt\, \lat\, \cdots) - f(\prt\, \prt\, \hop\, \lat\,
\cdots) + \cdots \\
&=& \alpha \und{\bra{W}E}DDE\cdots\ket{V} -
\bra{W}DD\und{DE}\cdots\ket{V} + \cdots \\
&=& \und{\bra{W}}DDE\cdots\ket{V} - \bra{W}DD\und{E}\cdots\ket{V} - 
\bra{W}DD\und{D}\cdots\ket{V} + \cdots \;.
\end{eqnarray}
The first two terms on the last line cancel, leaving a single negative
term.  One can also verify that this term will cancel one coming from
the next domain of particles on the lattice, or from the right
boundary.  Similarly, if one has a domain of particles
in contact with the right boundary, after using the reduction
relations and performing a cancellation, a single positive term
remains that cancels with one coming from the previous domain of
particles on the lattice, or from the left boundary.  Finally, after
checking that the matrix-product expression for the statistical weight
of fully occupied lattice is also stationary, we can conclude that the
matrix product weights, defined by Equation~(\ref{int:mp}), in conjunction
with the reduction relations \RR, give the steady state solution of
the ASEP, as previously claimed.

\subsubsection{Algebraic Proof}
\label{algproof}
We now show how this pictorial representation of the cancellation
mechanism between domains can be encoded in an algebraic way.  
This will pave the way
to understanding the general results of \cite{KS97} that we 
review in Sec.~\ref{formal}.

The master
equation can be written as a sum of terms, each relating either to a pair of
neighbouring sites, or to one of the end sites, i.e., as
\begin{equation}
\label{fme}
\frac{\rmd}{\rmd t} f(\tau_1, \tau_2, \ldots, \tau_N) = 
\left[ \hat{h}_{L}+
\sum_{i=1}^{N-1}
\hat{h}_{i,i+1} +\hat{h}_{R}\right]f(\tau_1, \tau_2, \ldots, \tau_N)
\end{equation}
The operators $\hat{h}$ construct the required gain and
loss terms in the master equation via
\begin{eqnarray}
\label{int:gen}
\fl \hat{h}_{i,i+1} f(\cdots,  \tau_{i}, \tau_{i+1},
 \cdots) = W(\tau_{i+1} \tau_{i} \to \tau_{i} \tau_{i+1})
f(\cdots,  \tau_{i+1}, \tau_{i},  \cdots) \nonumber\\
 - W(\tau_{i}\tau_{i+1} \to \tau_{i+1} \tau_{i}) f(\cdots, 
\tau_{i}, \tau_{i+1},  \cdots)\\
\fl \hat{h}_{L} f(\tau_1, \cdots) = W_1(1-\tau_{1} \to \tau_{1})
f(1-\tau_1, \cdots)-W_1(\tau_{1} \to 1-\tau_{1})f(\tau_1, \cdots)
\label{int:l}\\
\fl \hat{h}_{R} f(\cdots, \tau_{N}) = W_N(1-\tau_{N}  \to \tau_{N})
f(\cdots, 1-\tau_{N})-W_N(\tau_{N}  \to 1-\tau_{N})
f(\cdots, \tau_{N}) \label{int:r}
\end{eqnarray}
in which $W(\tau\,\tau' \to \tau'\,\tau)$ gives the rate at which a
neighbouring pair of sites in configuration $(\tau,\tau')$ exchange,
$W_1(\tau_1 \to 1-\tau_1)$ gives the rate at which site 1 changes state and 
$W_N(\tau_N \to 1- \tau_N)$
gives the rate at which site $N$ changes state.
For the ASEP we have
\begin{eqnarray}
\label{int:W10}
W(10 \to 01) &=& 1\\
\label{int:W20}
W_1(0 \to 1) &=& \alpha\\
\label{int:W12}
W_N(1 \to 0) &=& \beta
\end{eqnarray}
and all other rates zero.

In terms of matrix product expressions (\ref{int:mp}) for the
weights, (\ref{int:gen}--\ref{int:r})  become
\begin{eqnarray}
\fl\hat{h}_{i,i+1} \bra{W}  \cdots X_{\tau_{i-1}} [X_{\tau_{i}} X_{\tau_{i+1}}]
 X_{\tau_{i+2}} \cdots \ket{V}
= \bra{W} \cdots X_{\tau_{i-1}} [ W(\tau_{i+1}\tau_{i} \to \tau_{i}\tau_{i+1})
X_{\tau_{i+1}} X_{\tau_i} \nonumber\\ {} - W(\tau_{i}\tau_{i+1} \to \tau_{i+1}\tau_{i})
X_{\tau_{i} }X_{\tau_{i+1}} ] X_{\tau_{i+2}} \cdots \ket{V}
\label{int:h1} \\
\fl \hat{h}_L \bra{W} X_{\tau_1} \cdots \ket{V} 
 = 
\bra{W} \left[
W_1(1-\tau_{1} \to \tau_{1})
 X_{1-\tau_1}- W_1(\tau_{1} \to 1-\tau_{1})X_{\tau_1}\right] \cdots \ket{V}
\label{int:hl1}
 \\
\fl \hat{h}_R \bra{W}\cdots\left[
W_N(1-\tau_{N} \to \tau_{N})
 X_{1-\tau_N}- W_N(\tau_{N} \to 1-\tau_{N})X_{\tau_N}\right] \ket{V} 
\label{int:hr1}
\end{eqnarray}
where we have stretched the notation so that, for example,
$\hat{h}_{i,i+1}$ now acts on the matrices $X_{\tau_i}$, $X_{\tau_{i+1}}$.

Meanwhile, we suppose there exists a set of auxiliary matrices
$\tilde{X}_i$ such that additionally
\begin{eqnarray}
\label{int:h2}
\hat{h}_{i,i+1} \bra{W} \cdots X_{\tau_{i}} X_{\tau_{i+1}} \cdots
\ket{V} = \bra{W} \cdots [ \tilde{X}_{\tau_i} X_{\tau_{i+1}} -
X_{\tau_{i}} \tilde{X}_{\tau_{i+1}} ] \cdots \ket{V} \\
\hat{h}_L  \bra{W}  X_{\tau_1} \cdots \ket{V} =  \left[ -
\bra{W}
 \tilde{X}_{\tau_1}\right] \cdots \ket{V} \label{int:hl2}\\
\hat{h}_R \bra{W} \cdots X_{\tau_N} \ket{V} = \bra{W} \cdots\left[
\tilde{X}_{\tau_N} \ket{V}\right] \;. \label{int:hr2}
\end{eqnarray}
If this is the case then the sum in (\ref{fme}) telescopes to zero
(i.e., there is a pairwise cancellation of terms).  Therefore, if the
right-hand sides of (\ref{int:h1})--(\ref{int:hr1}) and
(\ref{int:h2})--(\ref{int:hr2}) can be shown to be equal, it follows
that a stationary solution of the master equation for the process has
been determined.  Equating the terms in square brackets in
(\ref{int:h1})--(\ref{int:hr1}) and (\ref{int:h2})--(\ref{int:hr2}) is
sufficient to ensure overall equality of the two expressions.  That
is, if
\begin{eqnarray}
\fl W(\tau_{i+1}\tau_{i} \to \tau_{i}\tau_{i+1}) X_{\tau_{i+1}} X_{\tau_i}  -
W(\tau_{i}\tau_{i+1} \to \tau_{i+1}\tau_{i}) X_{\tau_{i}} X_{\tau_{i+1}} = \tilde{X}_{\tau_i}
X_{\tau_{i+1}} - X_{\tau_i} \tilde{X}_{\tau_{i+1}}\label{int:h3}\\
\fl W_1(1-\tau_{1} \to \tau_{1})
\bra{W} X_{1-\tau_1}- W_1(\tau_{1} \to 1-\tau_{1})\bra{W} X_{\tau_1}
= -\bra{W}
 \tilde{X}_{\tau_1}\label{int:hl3}\\
\fl W_N(1-\tau_{N} \to \tau_{N})
 X_{1-\tau_N}\ket{V}- W_N(\tau_{N} \to 1-\tau_{N})X_{\tau_N}\ket{V}
= \tilde{X}_{\tau_N} \ket{V}\label{int:hr3} \;.
\end{eqnarray}

Note that (\ref{int:h3}--\ref{int:hr3}) amount to a total of eight conditions
which become with the  set of rates
(\ref{int:W10})--(\ref{int:W12})
\begin{eqnarray}
0 &=&\tilde{D}D - D \tilde{D} =\tilde{E}E - E \tilde{E} \\
DE &=& \tilde{E}D - E \tilde{D}=-  \tilde{D}E +  D\tilde{E}  \\ 
\alpha \bra{W}E &=& -\bra{W}\tilde{D} = \bra{W}\tilde{E} \\
\beta D\ket{V}  &=& \tilde{E}\ket{V} = -\tilde{D}\ket{V} \;.
\end{eqnarray}
With a \emph{scalar} choices for the auxiliaries, viz $\tilde{D}=-1,
\tilde{E}=1$ and $\tilde{A} = 0$, these reduce
to the three equations  (\ref{int:DE}--\ref{int:EW}) as required.

The algebraic proof can be easily generalised to other models,
for example those with more general bulk dynamics than exchanges, as shall be carried out in Section~\ref{formal}.

\subsection{Calculation of the nonequilibrium partition function $Z_N$}
\label{calcZ}

Our first calculational task is to compute $Z_N$ given by
(\ref{int:Z}), 
from which the current follows via (\ref{int:JofZ}).  There are several approaches to the task 
and these are representative of the three main approaches to calculation
which will be employed
at different points in this review: direct matrix reordering,
diagonalisation of an explicit representation of the matrices and
generating function techniques.

\subsubsection{Direct calculation by matrix reordering}
In Section~\ref{mpa} we explained how the reduction relations can, in
principle, be used repeatedly to bring any arbitrary matrix product
$U$ into a standard form (\ref{int:reord}) which can then be
straightforwardly evaluated (\ref{int:slurp}).  We now put this
principle into practice, and 
begin with the product $ C^N = (D+E)^N$ required to calculate
the normalisation (\ref{int:Z}) and thence the current (\ref{int:J}).
By performing the reduction longhand for the first few $N$, one finds
\begin{eqnarray}
C &=& D + E \;, \\
C^2 &=& DD + DE + ED + EE = DD + ED + EE + D + E \;, \\
C^3 &=& DDD + DDE + DED + DEE + EDD + EDE + EED + EEE \nonumber\\
    &=& DDD + EDD + EED + EEE + \nonumber\\
&& \qquad DD + DE + DD + ED + DE + EE + ED + EE
\nonumber\\
    &=& DDD + EDD + EED + EEE + 2 (DD +  ED +  EE ) + 2 (D + E) \;.\nonumber\\
\end{eqnarray}
Eventually one is drawn to the general formula
\begin{equation}
\label{open:CNsum}
C^N = \sum_{p=0}^{N} \frac{p (2N-p-1)!}{N! (N-p)!} \sum_{q=0}^{p} E^q
D^{p-q} \;.
\end{equation}
which can be proved by induction \cite{DEHP93}.  For this two
preliminary identities are required.  First,
\begin{equation}
\label{open:DNCsum}
D^N C = D + D^2 + \cdots + D^{N+1} + E \;,
\end{equation}
which can be proved inductively by multiplying both sides from the
left by $D$ and using (\ref{int:DE}).  The second identity involves
the combinatorial factors
\begin{equation}
\label{open:BNp}
B_{N,p} = \left\{ \begin{array}{ll}
\frac{p (2N-p-1)!}{N! (N-p)!} & 0 < p \le N \\
0 & \mbox{otherwise}
\end{array}
\right. \;.
\end{equation}
It reads
\begin{equation}
\label{open:Brec}
B_{N,p} - B_{N,p+1} = B_{N-1,p-1} \quad\forall N>1, p>0 
\end{equation}
which can be verified by substituting in the explicit expressions from
(\ref{open:BNp}). 

We now assume that the identity (\ref{open:CNsum}) for $C^N$ has been shown
to hold for some $N$, and derive an expression for $C^{N+1}$ by
multiplying from the right with $C$.  Then, (\ref{open:DNCsum}) implies
\begin{equation}
C^{N+1} = \sum_{p=0}^{N} B_{N,p} \sum_{q=0}^{p} \left[ E^{q+1} + E^q (D +
    D^2 + \cdots + D^{p-q+1}) \right] \;.
\end{equation}
Now, using (\ref{open:Brec}) we find
\begin{eqnarray}
C^{N+1} &=& \sum_{p=1}^{N+1} B_{N+1,p} \sum_{q=0}^{p-1} \left[ E^{q+1}
  + E^{q} (D + D^2 + \cdots + D^{p-q})\right] - {}\nonumber\\
&& \quad \sum_{p=2}^{N+1} B_{N+1,p} \sum_{q=0}^{p-2} \left[ E^{q+1}
  + E^{q} (D + D^2 + \cdots + D^{p-q-1})\right]\\
&=& \sum_{p=1}^{n+1} B_{N+1,p} \left[ E^p + E^{p-1} D + E^{p-2} D^2 +
  \cdots + D^p \right]
\end{eqnarray}
which is equal to the right-hand side of (\ref{open:CNsum}) with $N \to
N+1$.  Since (\ref{open:CNsum}) is obviously correct for $N=1$ ($C=D+E$),
it follows by induction that it is true for all $N$.  Using
(\ref{int:DV}), (\ref{int:EW}) and the convention $\braket{W}{V}=1$ we
immediately find the exact expression
\begin{equation}
\label{open:Z2}
Z_N = \sum_{p=0}^{N} B_{N,p} \sum_{q=0}^{p}
\left(\frac{1}{\alpha}\right)^q \left(\frac{1}{\beta}\right)^{p-q} =
\sum_{p=0}^{N} B_{N,p} \frac{\big(\frac{1}{\alpha}\big)^{p+1} -
  \big(\frac{1}{\beta}\big)^{p+1}}{\frac{1}{\alpha} -
  \frac{1}{\beta}}\;.
\end{equation}

\subsubsection{Diagonalisation of an explicit representation}\label{diag}
So far we have not given any explicit representations
of the  matrices
 $D,E$ and vectors
 $\bra{W}, \ket{V}$ that satisfy (\ref{int:DE}--\ref{int:EW}).
Several possible representations were proposed in \cite{DEHP93}.
For example, one can take:
\begin{eqnarray}
{D} = \left( \begin{array}{ccccc}
		  1&1&0&0& \cdots \\
		  0&1&1&0& \\
		  0&0&1&1&\\
		  0&0&0&1&  \\
		  \vdots &&&&\ddots  \\
		   \end{array}
		   \hspace{0.1in} \right)
\hspace{0.2in}
{E} = \left( \begin{array}{ccccc}
		  1&0&0&0&\cdots \\
		  1&1&0&0& \\
		  0 &1&1&0&\\
		  0 &0&1&1&\\
		  \vdots &&&&\ddots \\

		   \end{array}
		   \hspace{0.1in} \right)
\label{D1E1}\\
\nonumber \\
\bra{W}=  \kappa \left( \begin{array}{ccccc}
		 1, &a,&
		a^2&
		 .&.
		 \end{array} \right)
\hspace{0.2in}
\ket{V}= \kappa \left( \begin{array}{c}
		 1 \\
		 b \\
		 b^2 \\
		 .\\.
		 \end{array} \right)\;,
\label{V1W1}
\end{eqnarray}
where 
\begin{equation}
a= \frac{1-\alpha}{\alpha}\qquad
b= \frac{1-\beta}{\beta}\;.
\end{equation}
$\kappa^2 = ( \alpha + \beta -1)/ \alpha \beta$ 
is chosen to ensure that
$\langle W | V \rangle = 1 $. Note that for this  we require
$\alpha + \beta >1$, but as we shall discuss later (in Section~\ref{pasymp}) one can 
analytically continue results to $\alpha + \beta \leq 1$ in a 
straightforward manner. Some other representations
are presented in Appendix~\ref{appreps}.

Let us consider the eigenvectors of $C$, given
in this representation by
\begin{eqnarray}
{C} = \left( \begin{array}{cccccc}
		  2&1&0&0&. &.\\
		  1&2&1&0&& \\
		  0&1&2&1&&\\
		  0&0&1&2&.  &\\
		  . &&&&. &. \\
		  . &&&&&.
		   \end{array}
		   \hspace{0.2in} \right)\;.
\end{eqnarray}
We parametrise
the eigenvalues  by   $2(1+\cos \theta)$ where $0< \theta <2\pi$
and denote the 
associated  eigenvector by $\ket{\cos \theta}$ so that
\begin{equation}
C \ket{\cos \theta}= 2(1+\cos\theta) \ket{\cos \theta}\label{Cevec}
\end{equation}
where
\begin{eqnarray}
\ket{\cos \theta}= \frac{1}{\sin \theta}  \left( \begin{array}{c}
		 \sin \theta \\
		 \sin 2\theta \\
		 \sin 3\theta \\
		 .\\.
		 \end{array} \right)\;.
\end{eqnarray}
Note that the elements of $\ket{\cos \theta}$ are
precisely the Chebyshev polynomials of the second kind \cite{AAR00}.
It is easy to show (by summing geometric series) that
\begin{equation}
\braket{W}{\cos \theta}
= \frac{\kappa}{\sin \theta} \sum_{n=0}^{\infty}
a^n \sin (n+1)\theta 
= \frac{\kappa}{(1-a \rme^{i\theta})(1-a \rme^{-i\theta})}
\end{equation}
and $\braket{\cos \theta}{V}$ is given by a similar expression with
$a$ replaced by $b$.
Also it is easy to show  (using orthogonality of $\sin n \theta$ for different $n$) that
\begin{equation}
\label{pasep:HP1}
\One = \frac{1}{\pi} \int_0^{2\pi} \rmd\theta
\sin^2\theta
\ket{\cos\theta}\bra{\cos\theta}\;.
\end{equation}
Therefore $Z_N$ can be evaluated as follows
\begin{eqnarray}
Z_N &=& \bra{W} C^N \ket{V}\\
     &=& \frac{1}{\pi} \int_0^{2\pi} \rmd\theta
\sin^2 \theta
\bra{W} C^N \ket{\cos\theta}\braket{\cos\theta}{V} \\
&=&
\frac{\kappa^2}{\pi} \int_0^{2\pi} \rmd\theta \frac{ \sin^2\theta
\left[2(1+\cos \theta)\right]^N}
{(1-a \rme^{i\theta})(1-a \rme^{-i\theta})
(1-b \rme^{i\theta})(1-b \rme^{-i\theta})}\;. \label{asep:Z0}
\end{eqnarray}
This integral representation of the normalisation can be shown to
be equivalent to (\ref{open:Z2}) by using the calculus of residues
(see Appendix~\ref{intsum}). 
\subsubsection{Generating function approach}\label{gfa}
Let us define the generating function of the normalisation as
\begin{equation}
\Zed(z) = \sum_{N=0}^\infty z^N Z_N\;.
\label{Zeddef}
\end{equation}
In order to determine this quantity in a fast way
the idea is to work with formal power
series involving matrices, i.e., expressions of the form
\begin{equation}
f(U) = \sum_{n=0}^{\infty} a_n U^n
\end{equation}
where $U$ is a matrix.
The ``formal'' component of this approach lies in the fact that we do
not worry about convergence properties of these series.  Rather, we
simply use such expressions as a device for manipulating matrix
product expressions.  For example, one might multiply both sides by
some other formal series in a matrix $V$, bearing in mind that if $U$
and $V$ are noncommuting, one must multiply both sides from the same
direction.

Consider the formal series
\begin{equation}
\label{open:invC}
\frac{1}{1-z C} = \sum_{n=0}^{\infty} z^n C^n
\end{equation}
so that the generating function of the normalisation (\ref{Zeddef})
can be written as
\begin{equation}
\Zed(z) = \bra{W}\frac{1}{1-zC}\ket{V} \;.
\end{equation}
As was observed by Depken \cite{DepkenPhD}, Equation~(\ref{int:DE}) has the
simple consequence that
\begin{equation}
(1-\eta D) (1-\eta E) = 1 - \eta (D+E) + \eta^2 DE = 1 - \eta(1-\eta) C \;.
\end{equation}
Putting $z = \eta(1-\eta)$, and multiplying both sides from the left
by $\frac{1}{1-\eta E} \frac{1}{1-\eta D}$ and from the right by
$\frac{1}{1-zC}$ one learns that
\begin{equation}
\frac{1}{1-zC} = \frac{1}{1-\eta E} \frac{1}{1-\eta D} \;.
\end{equation}
Using (\ref{int:DV}) and (\ref{int:EW}) one then finds that
\begin{equation}
\Zed(z) = \left(1 - \frac{\eta(z)}{\alpha}\right)^{-1} \left(1 -
\frac{\eta(z)}{\beta}\right)^{-1}\;.
\label{open:Zed}
\end{equation}
Note that the condition $\Zed(0)=1$ implies $\eta(0)=0$
which means  that we must choose the root
\begin{equation}
\label{open:xi}
\eta(z) = \frac{1}{2}\left(1 - \sqrt{1-4z}\right)\;.
\end{equation}

The generating function approach provides a
very fast route to the phase diagram and current via the asymptotic
analysis of (\ref{open:Zed}) that we perform Sec.~\ref{asymp}.
Furthermore, the exact expression (\ref{open:Z2}) can also be recovered
with very little work through an application of the Lagrange inversion
formula for generating functions which we recap in Appendix~\ref{gf}.
As indicated in Appendix~\ref{gf}, this formula can be
applied because the combination $\eta(z)/z$ can be expressed in terms
of $\eta(z)$ alone, i.e., its dependence on $z$ enters only implicitly
through the argument of $\eta(z)$.  Then, the Lagrange inversion
formula implies that the normalisation $Z_N$ is given
by
\begin{eqnarray}
\fl\coeff{ z^N } \Zed[\eta(z)] &=& \frac{1}{N} \coeff{\eta^{N-1}} \left(
\frac{\rmd}{\rmd \eta} \frac{1}{1-\frac{\eta}{\alpha}}
\frac{1}{1-\frac{\eta}{\beta}} \right) \left( \frac{1}{1-\eta} \right)^N \\
\fl
&=& - \frac{1}{N} \left( \alpha \frac{\partial}{\partial \alpha} +
\beta \frac{\partial}{\partial \beta} \right) \coeff{ \eta^N } 
\frac{1}{1-\frac{\eta}{\alpha}} \frac{1}{1-\frac{\eta}{\beta}} 
\left( \frac{1}{1-\eta} \right)^N\\
&=& \sum_{p=0}^{N} \frac{p (2N - p -1)!}{N!(N-p)!} \sum_{q=0}^{p}
\left(\frac{1}{\alpha}\right)^p \left(\frac{1}{\beta}\right)^{q-p} 
\label{Zagain}
\end{eqnarray}
where we introduce an important notation:
\[\fl\hspace{7ex}\mbox{
\fbox{$\coeff{x^n} f(x)$ denotes  the coefficient
of $x^n$ in the formal power series $f(x)$.}}
\]
Expression (\ref{Zagain}) is in perfect
agreement with (\ref{open:Z2}), as required.

\subsection{Exact density profile from direct matrix reordering}
\label{direct}
In order to calculate the density profile we use the approach of
matrix reordering. This has the advantage of giving an exact
finite sum expression
(\ref{open:rho})
for finite systems.
We start from
(\ref{int:rho}), which contains the matrix product $D C^j$.  Again,
one can perform the reduction manually for small $j$ (we shall omit
this here) and deduce the identity
\begin{equation}
\label{open:DCsum}
DC^j = \sum_{k=0}^{j-1} B_{k+1,1} C^{j-k} + \sum_{k=2}^{j+1} B_{j,k-1}
D^k \;.
\end{equation}
This was proved in \cite{DEHP93}, again using an inductive
argument.  Assuming (\ref{open:DCsum}) holds for some $j \ge 1$, we
find this implies for $j+1$ that
\begin{eqnarray}
\fl(DC^{j})C &=& \sum_{k=0}^{j-1} B_{k+1,1} C^{j+1-k} + \sum_{k=2}^{j+1}
  B_{j,k-1} D^k C\\
\fl&=& \sum_{k=0}^{j-1} B_{k+1,1} C^{j+1-k} + \sum_{k=2}^{j+1} (B_{j+1,k}
  - B_{j+1,k+1}) (C + D^2 + D^3 + \cdots + D^{k+1})
\end{eqnarray}
where both (\ref{open:DNCsum}) and (\ref{open:Brec}) have been used.  Since,
for $N\ge1$, $B_{N+1,1} = B_{N+1,2}$, a little more algebra reveals that
\begin{equation}
\fl
DC^{j+1} = \sum_{k=0}^{(j+1)-1} B_{k+1,1} C^{(j+1)-k} + B_{j+1,1} D^2
+ B_{j+1,2} D^3 + \cdots + B_{j+1,j+1} D^{j+2}
\end{equation}
which is (\ref{open:DCsum}) with $j\to j+1$ as we desired.  The case
$j=1$ is easily verified using the reduction relation (\ref{int:DE}),
which completes the proof of (\ref{open:DCsum}) for all $j$.  An
expression for the density then follows by substituting
(\ref{open:DCsum}) into (\ref{int:rho}).  We find
\begin{equation}
\label{open:rho}
\rho_i = 
\sum_{n=1}^{N-i} B_{n,1} \frac{Z_{N-n}}{Z_N} + \frac{Z_{i-1}}{Z_N}
\sum_{p=1}^{N-i} B_{N-i,p} \frac{1}{\beta^{p+1}}
\end{equation}
for $1 \le i < N$.  Note that the case $i=N$ is special: $\rho_N =
Z_{N-1}/(\beta Z_N)$.

\begin{figure}
\begin{center}
\includegraphics[scale=0.33]{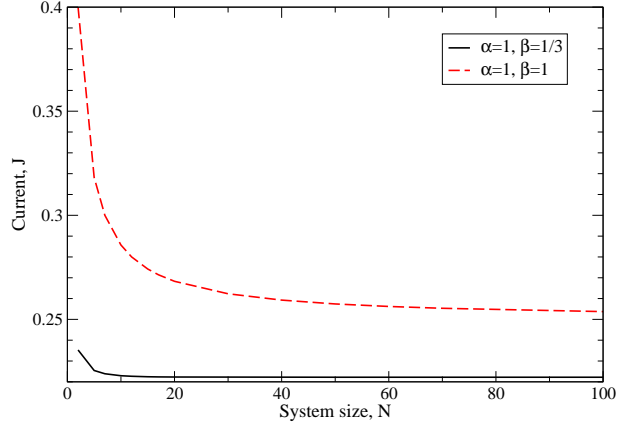}
\end{center}
\caption{\label{CurrentOfN} Particle current in the ASEP at finite
  system size for two different combinations of $\alpha$ and $\beta$.}
\end{figure}

In the next subsection, we shall analyse the large-$N$ behaviour of
the current and density profiles.  First, though, we remark that the
formul\ae\ (\ref{open:Z2}) and (\ref{open:rho}) are of utility if one
wishes to study \emph{finite} systems, e.g., in the context of the
biophysical applications of the ASEP where the thermodynamic limit is
not appropriate.  For example, the particle current as a function of
system size is plotted in Figure~\ref{CurrentOfN} for the combinations
$\alpha=\beta=1$ and $\alpha=1, \beta=\frac{1}{3}$.  The figure is
suggestive that the decay to the asymptotic current with increasing
system size is slower for the case $\alpha=\beta=1$ compared to
$\alpha=1, \beta=\frac{1}{3}$.  As one might guess from the mean-field
treatment of the ASEP in Section~\ref{asep}, these different decay
forms arise from being in different phases of the model.

\subsection{Asymptotic analysis of the current and density profiles}
\label{asymp}

In order to obtain the exact phase diagram for the ASEP, it is
necessary to determine the asymptotics (large-$N$ forms) of the
current and density profiles.  There are a number of ways to achieve
this.  A method that is systematic and easy to apply in a wide range
of cases is to construct the generating functions functions of the
desired quantities.  Then one can apply standard techniques for
extracting the asymptotics of the coefficients from a generating
function (see, e.g., \cite{Wilf94}).  For convenience, we have
recapitulated these techniques in Appendix~\ref{gf}.

It will also be useful to have at our disposal some generating functions
for common combinatorial quantities.
We first note  that
$\eta(z)$
(\ref{open:xi})
that appears in (\ref{open:Zed}) is  actually
(up to a factor of $z$)
the generating function for Catalan numbers:
\begin{equation}
\label{open:G1}
\eta(z)  = \sum_{N=1}^{\infty} z^N \frac{1}{N} {2(N-1) \choose N-1}
\;.
\end{equation}
The  Catalan numbers
$C_n = \frac{1}{n+1} {2n \choose n}$
appear  in a remarkably large number of combinatorial contexts
\cite{Stanley01} and will also be used many times in the course of this
article.  
Equation (\ref{open:G1})  can be seen by expanding the square root in 
(\ref{open:xi}) in powers of $z$. 
It is also  instructive to consider the generating function for ballot numbers
\begin{equation}
G_p(z) = \sum_{N=p}^{\infty} z^N B_{N,p}
\end{equation}
where the coefficients $B_{N,p}$, are defined in
(\ref{open:BNp}),  and we have that $G_0(z) = 1$
and $G_1(z) = \eta(z)$.
It turns out that
$G_p(z)$  is simply the $p\th$ power of $\eta(z)$.  To see
this, note that (\ref{open:Brec}) implies
\begin{equation}
G_p(z) = G_{p+1}(z) + z G_{p-1}(z)
\end{equation}
and so that if $G_p(z)=[\eta(z)]^p$ one must have $\eta(z) [ 1 - \eta(z)
] = z$ which is indeed satisfied by (\ref{open:xi}).

\begin{figure}
\begin{center}
\includegraphics[scale=0.6]{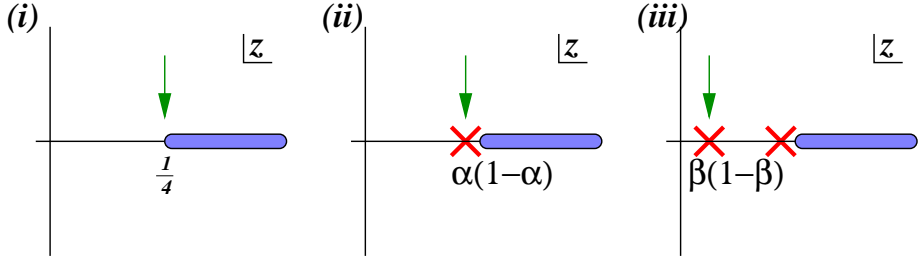}
\end{center}
\caption{\label{fig:gfsing} Poles (crosses) and branch cuts (solid
  lines) in the complex-$z$ plane of the generating function
  $\Zed(z;\alpha,\beta)$ at three different values of $\alpha$ and
  $\beta$. (i) $\alpha>{1\over2}$, $\beta>{1\over2}$; (ii)
  $\alpha<{1\over2}$, $\beta>{1\over2}$; and (iii) $\alpha<{1\over2}$,
  $\beta<{1\over2}$ and $\beta<\alpha$.  In each case the singularity
  dominating the asymptotics of the normalisation is the one closest
  to the origin and indicated with an arrow.}
\end{figure}

As noted in Appendix~\ref{gf}, the asymptotics of the coefficients of
a generating function (here, the normalisation $Z_N$) are dominated by
the singularity nearest the origin.  Now, the function $\eta(z)$ has a
square-root singularity at $z={1\over4}$ which is always present.
Additionally there is a pole at $\eta(z)=\alpha$; however, this is
present on the positive  branch of the square-root that appears in
$\eta(z)$ \emph{only} if $\alpha<{1\over2}$.  Similarly, there is a
pole at $\eta(z)=\beta$ if $\beta<{1\over2}$.  As $\eta(z)$ increases
monotonically from zero in the range $0 \le z \le {1\over4}$, the
dominant singularity is the square-root singularity at $z={1\over4}$
if both $\alpha$ and $\beta$ exceed $1\over2$.  Otherwise, whichever
of $\alpha$ or $\beta$ is smaller supplies the dominant
singularity---see Figure~\ref{fig:gfsing}.  Note that the regions
within which particular singularities dominate coincide with the
phases previously established using a mean-field approach, shown in
Figure~\ref{fig:asepmfpd}.  In other words, this analysis of the
generating function reveals the phase diagram to be exact.

In the region identified as the low-density phase from the mean-field
calculation, $\alpha<\beta, \alpha<{1\over2}$, the pole in
(\ref{open:Zed}) at $z_0=\alpha(1-\alpha)$ implies---via the results
presented in Appendix~\ref{gf}---that
\begin{eqnarray}
Z_N(\alpha, \beta) &\sim& \left[ \lim_{z\to z_0} \left( 1 -
  \frac{z}{z_0} \right) \Zed(z; \alpha, \beta) \right] z_0^{-N} \\
&\sim& \frac{\alpha\beta}{z_0 \eta^\prime(z_0) \left[ \beta - \eta(z_0) \right]}
  z_0^{-N}\\
\label{open:ZLD}
&\sim& \frac{\alpha\beta(1-2\alpha)}{(\beta-\alpha)}
  \left[\alpha(1-\alpha)\right]^{-N-1} \;.
\end{eqnarray}
Using (\ref{int:JofZ}), one finds the current is $J = z_0 =
\alpha(1-\alpha)$.  In the High-density phase, $\beta<\alpha,
\beta<{1\over2}$ the pole at $z_0=\beta(1-\beta)$ dominates and one
obtains the same expression for the normalisation but with $\alpha$
and $\beta$ exchanged.  This is in accordance with the particle-hole
symmetry of the model.  In the maximal-current phase, where both
$\alpha$ and $\beta$ are greater than ${1\over2}$, we expand up the
square-root in (\ref{open:Zed}) to find
\begin{equation}
\fl
\Zed(z; \alpha, \beta) = \frac{4 \alpha\beta}{(2\alpha-1)(2\beta-1)} \left( 1 -
\left[ \frac{1}{2\alpha-1} + \frac{1}{2\beta-1}\right] \sqrt{1-4z} +
     \Order(1-4z) \right) \;.
\end{equation}
It then follows from the results of Appendix~\ref{gf} (Equation~\ref{gf:sing})
that
\begin{equation}
\label{open:ZMC}
Z_N(\alpha, \beta) \sim \frac{4 \alpha \beta
  (\alpha+\beta-1)}{\sqrt{\pi} (2\alpha-1)^2 (2\beta-1)^2}
  \frac{4^N}{N^{3/2}}
\end{equation}
which can be shown to agree with the expression given in \cite{DEHP93}
after a little algebra.  In this phase, the current $J = z_0 =
{1\over4}$, thus showing that the mean-field predictions for the
asymptotic current in all three phases, as presented in Table~\ref{tab:asepmf},
are exact.

We now turn to the exact formula for the density profile
(\ref{open:rho}) which for convenience we give again here:
\begin{equation}
\label{open:rho2}
\rho_i = 
\sum_{n=1}^{N-i} B_{n,1} \frac{Z_{N-n}}{Z_N} + \frac{Z_{i-1}}{Z_N}
\sum_{p=1}^{N-i} B_{N-i,p} \frac{1}{\beta^{p+1}} \;.
\end{equation}
Although formidable at a first glance, its analysis is in fact
reasonably straightforward when both $N$ and $i$ are large.  More
precisely, we shall take $N\to\infty$ and $i\to\infty$ with the
distance $j=N-i$ from the right boundary fixed.  It is easily verified
that in this limit one has in all phases
\begin{equation}
\lim_{N\to\infty} \frac{Z_{N-j}}{Z_N} = J^j
\end{equation}
for any fixed finite $j$.  Recall that $J$ is the current the
thermodynamic limit, as determined above.  Thus
\begin{equation}
\label{rhosimple}
\lim_{N\to\infty} \rho_{N-j} = \sum_{n=1}^{j} B_{n,1} J^n +
\frac{J^{j+1}}{\beta} \sum_{p=1}^{j} \frac{B_{j,p}}{\beta^p} \;.
\end{equation}
The second sum in this expression we have met before: it is simply the
normalisation for the ASEP (\ref{open:Z}) in the limit
$\alpha\to\infty$.  Hence, for $j \gg 1$ we can use the asymptotic
analyses from above to determine that
\begin{equation}
\label{open:Ztilde}
\sum_{p=1}^{j} \frac{B_{j,p}}{\beta^p} \sim \left\{ \begin{array}{ll}
\displaystyle \frac{\beta (1-2\beta)}{[\beta(1-\beta)]^{j+1}} & \beta < \frac{1}{2} \\[2ex]
\displaystyle \frac{4^j}{\sqrt{\pi j}} & \beta = \frac{1}{2} \\
\displaystyle \frac{\beta 4^j}{\sqrt{\pi} (2\beta-1)^2 j^{3/2}} & \beta >
\frac{1}{2}
\end{array}\right. \;.
\end{equation}
The case $\beta=\frac{1}{2}$ is special because the pole and
square-root singularities in the generating function coincide, thus
modifying the nature of the dominant singularity.

The first sum can be analysed through its generating function in a
manner similar to that described for the normalisation above.  One
ultimately finds
\begin{equation}
\sum_{n=1}^{j} B_{n,1} J^{j} =
\coeff{y^j} \frac{\eta(yJ)}{1-y}
\end{equation}
where we recall that $\coeff{y^j}$ picks out the coefficient of $y^j$
from the generating function.  In the high- and low-density phases,
where $J<{1\over4}$, the pole provides the dominant contribution to
the sum for $j\to\infty$, with finite-$j$ corrections supplied by the
square-root singularity in $\eta(yJ)$ at $y={1\over4J}$.  The result
is
\begin{equation}
\sum_{n=1}^{j} B_{n,1} J^{j} \sim \eta(J) - \frac{J}{1-4J}
\frac{(4J)^j}{\sqrt{\pi} j^{3/2}} + \cdots \;.
\end{equation}
Note that $\eta(J) = \min\{\alpha,\beta\}$ in these phases.  Meanwhile,
in the maximal-current phase, the pole and square-root singularity
coincide and then
\begin{equation}
\sum_{n=1}^{j} B_{n,1} J^{j} \sim \frac{1}{2} \left( 1 -
\frac{1}{\sqrt{\pi j}} + {\rm O}(j^{-3/2}) \right) \;.
\end{equation}

\subsection{Exact phase diagram}

We can now put these results together to determine the density profile
at a fixed (but large) distance $j$ from the right boundary in the
limit $N\to\infty$ using (\ref{open:rho}).  We do not need to consider
the left boundary explicitly because of a particle-hole symmetry in
the model that implies
\begin{equation}
\rho_{N+1-i}(\alpha,\beta) = 1-\rho_i(\beta,\alpha)
\label{open:phs}
\end{equation}
since holes enter the system at the right boundary at a rate $\beta$,
hop with unit rate to the left and exit at rate $\alpha$.  There are a
number of different cases to consider in obtaining the density
profile, since not only does the first term in (\ref{open:rho2}), and
prefactor of the second, depend on which phase the system is in, but
the second sum additionally takes two different functional forms at
the right boundary in the low-density phase.  The various distinct 
cases are as follows.

\begin{itemize}
\item \textbf{Maximal-current phase} $\beta>{1\over2},
  \alpha>{1\over2}$.  Here, the generating-function analysis of
  (\ref{open:rho}) involves only algebraic singularities, rather than
  poles.  Hence one finds a power-law decay of the the density from
  the bulk value of ${1\over2}$, viz,
\begin{equation}
\rho_{N-j} \sim \frac{1}{2} \left[ 1 - \frac{1}{\sqrt{\pi j}} +
\Order(j^{-3/2}) \right] \;.
\end{equation}
  Although the mean-field theory predicts the bulk density correctly,
  the decay exponent is incorrectly predicted as $1$ rather than
  ${1\over2}$.  Note also the universal (i.e., independent of $\alpha$
  and $\beta$) prefactor of this decay which is also found at the left
  boundary as a consequence of the particle-hole symmetry.
\item \textbf{Low-density phases}  $\alpha<{1\over2}, \alpha <
  \beta$.  As noted above, the second term in
  (\ref{open:rho}) takes different forms depending on whether
  $\beta$ is less than or greater than $1\over2$.  This gives rise to
  two sub-phases.
\begin{itemize}
\item \textbf{LD-I} $\beta<{1\over2}$.  Here, the generating function
  for the density contains only poles, and so a purely exponential
  decay is obtained at the right boundary:
\begin{equation}
\rho_{N-j} \sim \alpha + (1-2\beta) \left(
\frac{\alpha(1-\alpha)}{\beta(1-\beta)} \right)^{j+1} \;.
\end{equation}
\item \textbf{LD-II} $\beta>{1\over2}$. In this instance a pole and
  algebraic singularity combine to give an exponential decay modulated
  by a power law:
\begin{equation}
\rho_{N-j} \sim \alpha + \left[ \frac{1}{(2\beta-1)^2} - \frac{1}{(2\alpha-1)^2} \right]
\frac{1}{\sqrt{\pi} j^{3/2}} 4^j \left[ \alpha(1-\alpha) \right]^{j+1} \;.
\end{equation}
\end{itemize}
\item \textbf{High-density phases} $\beta<{1\over2}, \beta<\alpha$.
In the high-density phases, all variation in the density is at the
left boundary (which is out at infinity in the foregoing analysis).
At the right boundary the density takes the constant value
\begin{equation}
\rho_{N-j} \sim 1 - \beta \;,
\end{equation}
as was seen in the mean-field analysis.
\item \textbf{High/low-density to maximal-current transition lines}
  Along the boundaries between the high- or low-density and the
  maximal-current phases, the bulk density is $1\over2$ with
  deviations of order $j^{-3/2}$ (the prefactor can be calculated by
  developing the asymptotic expansions given above: this is left as an
  exercise for the reader).
\item \textbf{High- to low-density transition line} Here, the
  situation is more interesting.  Since $\alpha=\beta<{1\over2}$, the
  generating function for the normalisation has a double pole, and so
  $Z_N \sim (\mbox{const}) N J^{-N}$.  This factor of $N$ in the
  normalisation gives rise to a linear density profile from the
  prefactor $Z_{i-1}/Z_N$ of the second sum in (\ref{open:rho}).
  That is,
  \begin{equation}
    \rho_i = \alpha + (1 - 2\alpha) \frac{i}{N} \;.
  \end{equation}
  This density profile can be understood from the study of shock
  fronts discussed in Section~\ref{sscp}.  A shock, at $x=0$, is
  stationary if the density approaches $\rho_{L}<{1\over2}$ as the
  spatial coordinate $x\to-\infty$ and a value $\rho_{R}=1-\rho_{L} >
  \rho_{L}$ as $x\to\infty$.  One notes that this is the situation
  imposed by the boundary conditions on the open system in the $N\to
  \infty$ limit along this transition line: $\rho_{L}=\alpha$,
  $\rho_{R}=1-\alpha$.  One has a \emph{diffusing} shock front
  separating a low-density region to the left and a high-density
  region to the right.  Given that this shock front performs a random
  walk, its position on the lattice is given by a flat distribution
  which in turn implies a linear \emph{average} density profile of the
  form given above.
\end{itemize}

\begin{figure}
\begin{center}
\includegraphics[scale=0.85]{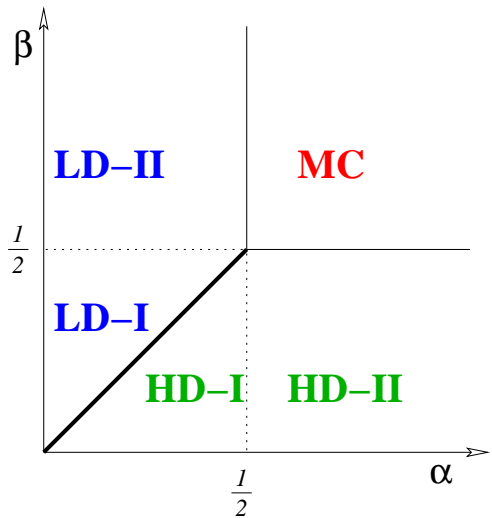}
\end{center}
\caption{\label{fig:aseppd} Exact phase diagram for the ASEP.  The two
high- and low-density sub-phases differ in the decay of the density
profile to its bulk value from the boundaries.}
\end{figure}

With these results we are now able to present the complete phase
diagram for the ASEP.  It is shown in Figure~\ref{fig:aseppd} and
differs from the mean-field version (Figure~\ref{fig:asepmfpd}) in
that the high- and low-density phases have acquired two sub-phases in
which the density decay near the boundary takes a different form. 
The lengthscale characterising this density decay diverges
as the maximal-current phase is approached from either the high- or
low-density phases.  This type of physics is suggestive of a
continuous phase transition \cite{SD93}.  
Meanwhile, the phase coexistence at the
high- and low-density phase boundary is reminiscent of phase
coexistence at a first-order
phase transition in an equilibrium system.  Motivated by these
observations, we now explore more deeply the relationship with
equilibrium phase transitions.

\subsection{Connection to equilibrium phase transitions}
\label{surf}
In Section~\ref{asymp} we saw that the generating function of Catalan
numbers played a key part in the analysis of the ASEP's phase
behaviour.  This fact can be used to make contact with
\emph{equilibrium} phase transitions in two-dimensional surface models
\cite{BE04,BdGR04,BJJK04a} since one of the many combinatorial objects
counted by Catalan numbers are excursions on the rotated square
lattice, known also as \emph{Dyck paths} (see e.g.,
\cite{vanRensburg00} for further applications and references).

\begin{figure}
\begin{center}
\includegraphics[scale=0.66]{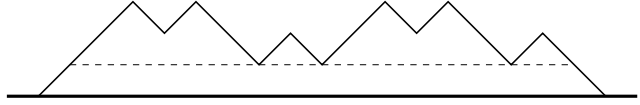}
\end{center}
\caption{\label{fig:dyck} A Dyck path is a random walk on the rotated
  square lattice that returns to the origin for the first time at a
  fixed end-point.  Between the initial upward and final downward
  step, the walk comprises a some number (possibly zero) of Dyck paths
  placed end-to-end.}
\end{figure}

Specifically, a Dyck path of length $2N$ comprises $N$ up-steps and
$N$ down-steps, so that it contacts the origin both at the start and
the end.  It is further constrained never to contact the origin in
the intermediate steps---see Figure~\ref{fig:dyck}.  
The coefficient of
$z^N$ in the generating function $G(z)$ counts the number of paths of
length $2N$. Referring to Figure~\ref{fig:dyck} we see that a 
Dyck path of a given length $2N$ begins with an up-step, ends on a
down-step and comprises any number of excursions that together have
length $2(N-1)$ as can be seen from Figure~\ref{fig:dyck}.
Therefore the generating function for Dyck path obeys
\begin{eqnarray}
G(z) &=& z \left[ 1 + G(z) + G^2(z) + G^3(z) + \cdots \right] \label{G(z)} \\
&=& \frac{z}{1-G(z)}\label{G(z)2}\;,
\end{eqnarray}
where on the rhs of (\ref{G(z)}) 
the factor $z$ comes from the pair of initial and
final up and down steps and the sum of terms $G^n(z)$ come from
excursions which return exactly $n$ times to site 1 in between.
We see then from (\ref{G(z)2}) that $G(z) = \eta(z)$, the function defined in
(\ref{open:xi}).

\begin{figure}
\begin{center}
\includegraphics[scale=0.33]{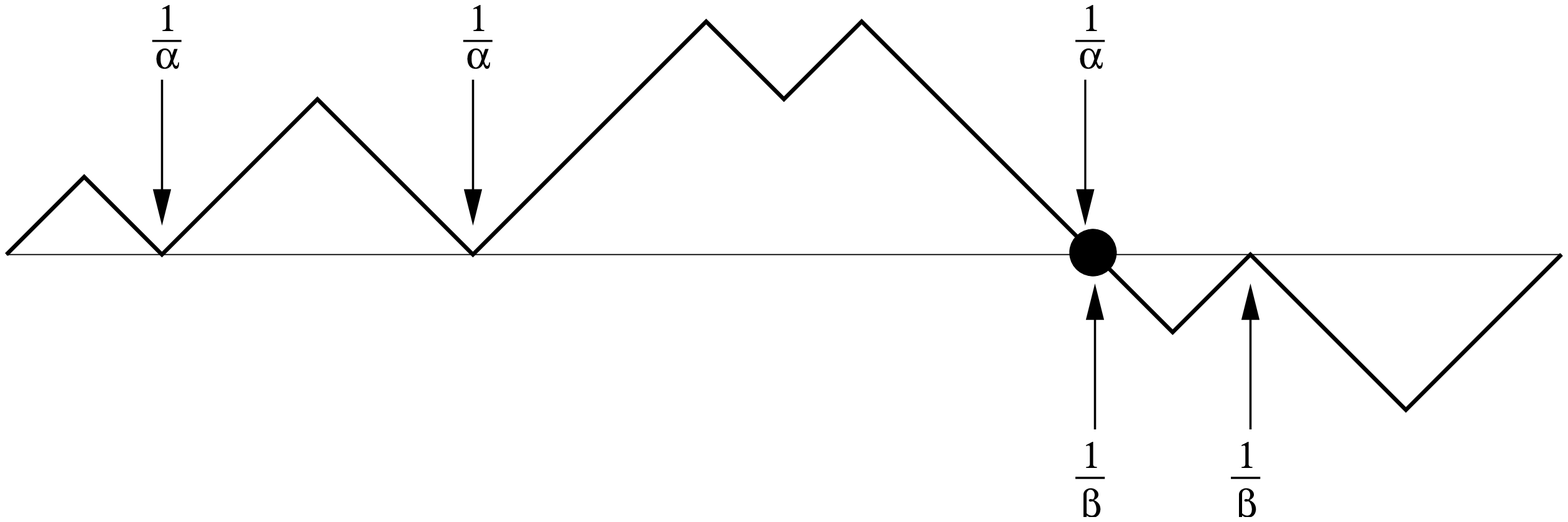}
\end{center}
\caption{\label{fig:OTW} A one-transit walk whose statistical weight in
  the fixed-length ensemble depends on the density of contacts from
  above and below.}
\end{figure}

Similar reasoning implies that the
the coefficient of $z^N \alpha^{-n} \beta^{-m}$ in the generating function
(\ref{open:Zed})
\begin{equation}
\Zed(z) = \left( 1-\frac{\eta(z)}{\alpha} \right)^{-1} \left(
  1-\frac{\eta(z)}{\beta} \right)^{-1}
\end{equation}
counts the number of walks of length $2N$ that return to the
origin a total of $n+m$ times (the last of those returns being the end
of the path).  It is convenient to have the first $n$ of these
excursions take place above the origin, and the remaining $m$ below.
Then, the coefficient of $z^N$, which is the normalisation of the ASEP
$Z_N$ for an $N$-site system, is also the partition function for the
equilibrium ensemble of surfaces of length $2N$ that is constrained to
start and end at the same height, crosses the origin at most once (and
then only from above), and has a fugacity $1/\alpha$ associated with
each contact with the origin from above, and $1/\beta$ with those from
below---see Figure~\ref{fig:OTW}.  This construction has been dubbed a
\emph{one-transit walk} \cite{BdGR04}.

It is an elementary exercise in statistical mechanics
to derive the
free energy per site in the thermodynamic limit as a function of the
density of contacts $\rho_a$ from above and $\rho_b$ from below, these
being the order parameters of the surface.  This free energy is \cite{BdGR04,BJJK04b}
\begin{eqnarray}
f(\rho_a, \rho_b) = \left\{ \rho_a \ln\alpha + \rho_b \ln\beta
\right\} \nonumber\\
  \qquad {} - \left\{ (2-\rho_a-\rho_b) \ln (2-\rho_a-\rho_b) -
  (1-\rho_a-\rho_b)\ln(1-\rho_a-\rho_b) \right\}
\end{eqnarray}
in which the first term in curly brackets is an energy-like quantity
that is lowered by the surface contacting the origin and the second an
entropy-like quantity that is increased by the surface making large
excursions away from the origin.  The equilibrium state arising from
this energy-entropy competition is found in the conventional manner,
i.e., by minimising the free energy with respect to the order
parameters $\rho_a$ and $\rho_b$.

\begin{figure}
\begin{center}
\includegraphics[scale=0.5]{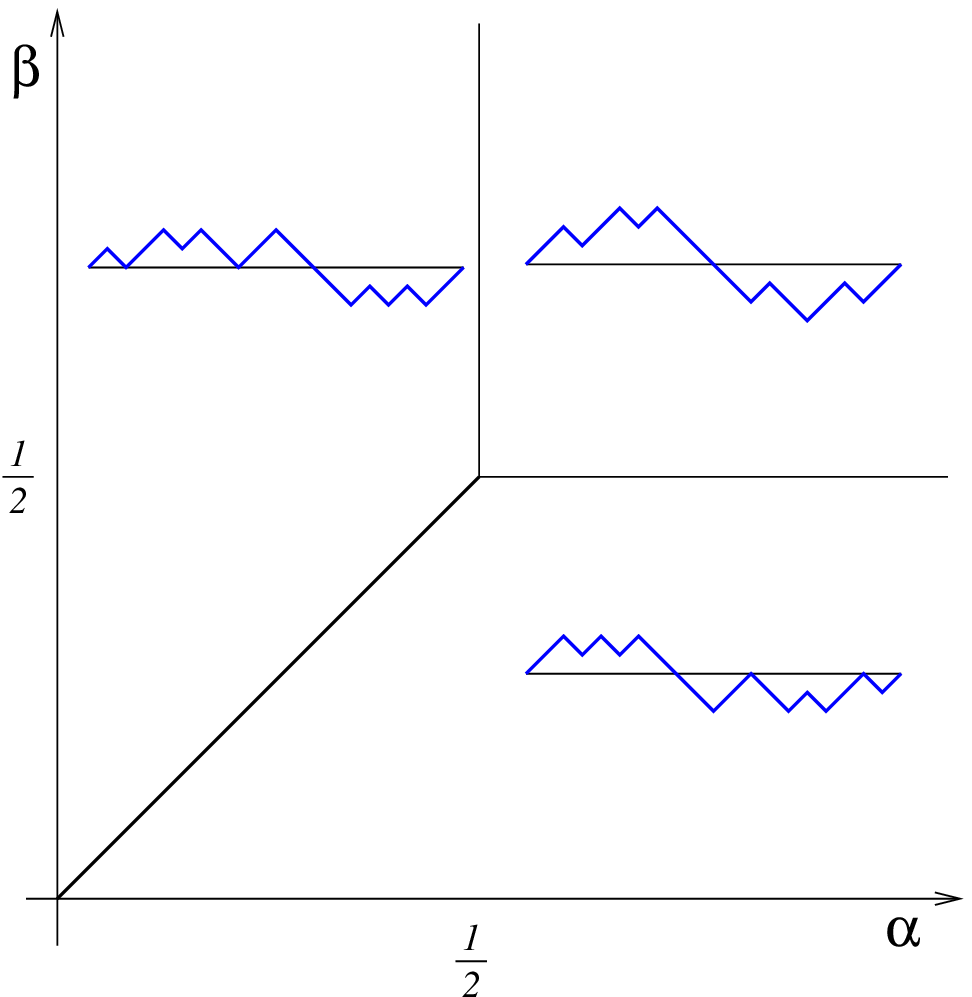}
\end{center}
\caption{\label{fig:OTWphase} Phase diagram of the one-transit walk in
the plane of equilibrium fugacities $\alpha$ and $\beta$.}
\end{figure}

The free energy is minimised by choosing $\rho_a,
\rho_b$ to lie somewhere on the boundary of the physical region $0 \le
\rho_a+\rho_b \le 1$.  When $\alpha$ and $\beta$ are large, the
entropy of walks that rarely contact the origin dominates and the
surface is desorbed: $\rho_a=\rho_b=0$.  On the other hand, if one of
$\alpha$ or $\beta$ is smaller than the critical value of
$\frac{1}{2}$, contacting the origin is energetically favourable.  If
$\alpha<\beta$, it is most favourable for all contacts to come from
above, otherwise from below.  Thus the low-density, high-density and
maximal-current phases of the ASEP correspond to two adsorbed phases,
one from above and one from below, and a desorbed phase
respectively---see Figure~\ref{fig:OTWphase}.
We remark that a similar unbinding transition was explored using
the mapping from ASEP to the disordered
equilibrium problem of directed polymers in a random
medium \cite{KT94}.

An interesting feature
of this association of nonequilibrium and equilibrium
phase behaviour is that the orders of the phase transitions must agree in
both cases.  In the equilibrium case, these are defined in the
conventional way, i.e., through discontinuities in derivatives of the
equilibrium free energy with respect to the contact fugacities
$1/\alpha, 1/\beta$.  In the nonequilibrium case,  on the other hand,
derivatives of the
``free energy'' (the logarithm of the normalisation---see Equation \ref{jfe}) with respect to
the transition rates $\alpha$ and $\beta$ have no obvious connection
to moments of physical observables as they do for an equilibrium
system. 
We remark that this phenomenon had previously
been observed in a study of the Lee-Yang zeros of the ASEP
normalisation \cite{BE02} and direct the interested reader to two
recent thorough reviews of this subject \cite{BE03,BDL05} for details.

\subsection{Joint current-density Distribution}
\label{series}

Finally, in this section on the ASEP with open boundaries, we 
develop the generating function approach further
to compute certain  more complicated quantities.
It is useful to 
introduce a pair
of matrices
\begin{eqnarray}
\label{open:DD}
\DD &=& (1-x-y) \frac{D}{1-xD} \\
\label{open:EE}
\EE &=& (1-x-y) \frac{E}{1-yE} \;.
\end{eqnarray}
Then, for example, the combination $\frac{x}{1-x-y} \DD$ in a
generating function constructs a domain of particles, the size of
which is conjugate to the parameter (fugacity) $x$.  These matrices
have the special property that they satisfy the familiar relation
\begin{equation}
\label{open:DDEE}
\DD \EE = \DD + \EE
\end{equation}
as a consequence of (\ref{int:DE}).  This means that results already
obtained by, for example, direct ordering of $D$ and $E$ matrices can
be carried across to expressions involving $\DD$ and $\EE$.  One
quantity calculated in this way is a joint current-density
distribution was calculated in \cite{DS04}, the derivation of which we
now briefly review.

The interest here is in the quantity
\begin{equation}
\label{open:rhoj}
\fl P_N(M,K) = \frac{1}{Z_N} \coeff{ x^M y^{N-M} } \bra{W} \frac{1}{1-yE}
\left( \frac{xD}{1-xD} \frac{yE}{1-yE} \right)^K \frac{1}{1-xD}
\ket{V} \;.
\end{equation}
This gives the probability that a configuration of $N$ sites
containing $M$ particles and $K$ particle-hole domain walls is
realised in the steady state of the ASEP.  Such configurations support
$K$ units of flux in the bulk, and so $\lim_{N\to\infty}
P_N(N\rho,NJ)$ gives the joint probability of observing a density
$\rho$ and current $J$ in the thermodynamic limit.

To make use of earlier results for the ASEP, we note that
\begin{equation}
\fl
\left( \frac{xD}{1-xD} \frac{yE}{1-yE} \right)^K = \left[
\frac{xy}{(1-x-y)^2} \DD \EE \right]^K =
\frac{(xy)^K}{(1-x-y)^{2K}} ( \DD + \EE )^K \;.
\end{equation}
Then, from (\ref{open:DDEE}) and the relations \RR\ we have
\begin{equation}
\label{open:rhojgf}
\fl
P_N(M,K) = \frac{1}{Z_N(\alpha,\beta)} \coeff{ x^{M-K} y^{N-M-K} } 
\frac{1}{(1-\frac{y}{\alpha})(1-\frac{x}{\beta})}
\frac{Z'_K}{(1-x-y)^{2K}}
\end{equation}
in which $Z'_N$ is the normalisation for the ASEP as given in
(\ref{open:Z2}) but with the boundary rates $\alpha$ and $\beta$
replaced by their primed versions
\begin{eqnarray}
\alpha' &=& \frac{\alpha-y}{1-x-y} \\
\beta' &=& \frac{\beta-x}{1-x-y}
\end{eqnarray}
which arise from the requirement that $\DD \ket{V} = (1/\beta')
\ket{V}$ and $\bra{W} \EE = (1/\alpha') \bra{W}$ in order for
(\ref{open:rhojgf}) to be correct.  This expression is in agreement
with Equation (4) in \cite{DS04} which was obtained using almost the
same approach.  

To analyse the joint current-density distribution $P_N(M,K)$ in the
thermodynamic limit, requires, once again, a careful analysis of the
singularities in the generating function \cite{DS04}.  One anticipates
expressions of the form $P_N(M,K) \approx \exp[-N r(\rho,J)]$ for
large $N$, with the function $r(\rho,J)=-\lim_{N\to\infty} \ln
P_N(\rho N,JN)/N$ taking its minimum value of zero at the
thermodynamic values of the current and density given in
Table~\ref{tab:asepmf}.  In the maximal-current phase one finds (up to
additive constants relating to the normalisation of $P$)
\begin{equation}
r(\rho,J) = 2J\ln J + (\rho-J)\ln(\rho-J) +
  (1-\rho-J)\ln(1-\rho-J)
\end{equation}
and in the low-density phase
\begin{eqnarray}
r(\rho,J) = 2J\ln J + (\rho-J)\ln(\rho-J) +
  (1-\rho-J)\ln(1-\rho-J) \nonumber\\ 
\qquad{}+ (1-\rho)\ln \alpha + \rho \ln
  (1-\alpha) - \rho\ln\rho - (1-\rho)\ln(1-\rho) \;,
\end{eqnarray}
from which the result for the high-density phase follows from the
particle-hole symmetry.  In each case, the thermodynamic values give
the desired minimum, and one notes the presence of non-Gaussian
fluctuations.  In \cite{DS04}, such
fluctuations are suggested to be a signature of the long-range
correlations that are present in the model.

\section{ASEP on a ring with two particle species}
\label{ring}
\begin{figure}
\begin{center}
\includegraphics[scale=0.33]{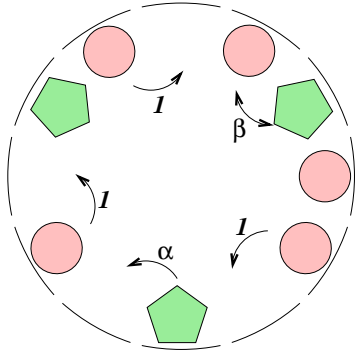}
\end{center}
\caption{\label{fig:ring} Asymmetric exclusion process on a ring with
  a second species of  particle (shown as pentagons).  The labels indicate the
  rates at which the various particle moves can occur.}
\end{figure}

In Section~\ref{sscp} we reviewed  second-class particles and their various
roles, and introduced a generalisation to two species of particles
(see Figure~\ref{fig:ring}):
\begin{eqnarray}
1\,0 \mathop{\rightarrow}\limits^{1}
\, 0\,1\qquad
2\,0 \mathop{\rightarrow}\limits^{\alpha}  0\,2 \qquad 1\,2 
\mathop{\rightarrow}\limits^{\beta}
 2\,1 \;.
\label{defdyn2}
\end{eqnarray}
 The case $\alpha= \beta =1$ corresponds  to
species 1 being first-class particles and
species 2 being   second-class particles.
Let us stress here that we denote by a two-species exclusion process a
system with two-species of particles and in addition vacancies.  In
some works it has been convenient to consider the vacancies as a
species of particles and thus sometimes the two species system we
study here has been referred to as a three species system in the literature.

\subsection{Matrix product expressions}

A microscopic configuration is completely specified by the set of
occupation numbers $\tau_i$ where $\tau_i=0$ indicates that site $i$
of the ring (relative to an arbitrary origin) is vacant and
$\tau_i=1,2$ indicate occupancy by a particle of species 1 or species 2
respectively.  We shall again deal with statistical weights
(unnormalised probabilities) as in (\ref{int:PfZ}).  On this occasion,
\begin{equation}
\label{ring:PfZ}
P(\tau_1, \tau_2, \ldots, \tau_N) = \frac{f(\tau_1, \tau_2, \ldots,
  \tau_N)}{Z_{N,M,P}}
\end{equation}
where here the normalisation is the sum of the weights over all
configurations with $M$ species 1  and $P$ species 2 particles on
the $N$-site ring, since these quantities are all conserved by the
dynamics.  We extract a scalar from  the matrix product appearing in
(\ref{int:mp})  by the trace operation
\begin{equation}
\label{ring:mp}
f(\tau_1, \tau_2, \ldots, \tau_N) = \tr( X_{\tau_1} X_{\tau_2} \cdots
X_{\tau_N} )
\end{equation}
since this reflects the rotational invariance of the system.  As
before we use the notation $D$ and $E$ for the matrices relating to
species 1 particles and vacant sites, and introduce $A$ for
species 2  particles.

The reduction relations for these matrices are
\begin{eqnarray}
\label{ring:DE}
DE &=& D + E \\
\label{ring:AE}
AE &=& \frac{1}{\alpha} A \\
\label{ring:DA}
DA &=& \frac{1}{\beta} A \;.
\end{eqnarray}
One notes the similarity with the relations \RR\ for the ASEP with
open boundaries.  Indeed, two sets of relations are equivalent if one
chooses for $A$ the outer product, or projector,
\begin{equation}
A = \ket{V} \bra{W} \;.
\label{ring:A}
\end{equation}
In particular, if one has a single species 2  particle ($P=1$), the
weights become
\begin{equation}
\label{ring:f1}
f(\tau_1, \tau_2, \ldots, \tau_{N-1}) = \bra{W} X_{\tau_1} X_{\tau_2}
\cdots X_{\tau_{N-1}} \ket{V}
\end{equation}
if one chooses the location of the origin as being that of the
second-class particle.  Although the weight is   identical to the
matrix-product expression (\ref{int:mp}) for the case of
open boundaries, the distributions (\ref{int:PfZ}) and
(\ref{ring:PfZ}) are, in principle, different because the ensembles, and hence
normalisations, are not the same.  We shall return to this important
point again below in Subsections~\ref{ss2} and \ref{gce}.  First, we show that this set of matrix product relations satisfies
the stationary conditions  for
the system (\ref{defdyn2}) (i.e., that illustrated in Figure~\ref{fig:ring}).

\subsection{Proof of the reduction relations}
\label{RRRproof}

We use a simple generalisation of the algebraic approach of Sec.~\ref{algproof}
to prove the reduction relations. The master
equation can be written as a sum of terms, each relating to a pair of
sites,  as
\begin{equation}
\label{ring:me}
\frac{\rmd}{\rmd t} f(\tau_1, \tau_2, \ldots, \tau_N) = \sum_{i=1}^{N}
\hat{h}_{i,i+1} f(\tau_1, \tau_2, \ldots, \tau_N)
\end{equation}
where in terms of matrix product expressions (\ref{ring:mp}) for the
weights
\begin{eqnarray}
\label{ring:h1}
\fl\hat{h}_{i,i+1} \tr( \cdots X_{\tau_{i-1}} [X_{\tau_i} X_{\tau_{i+1}}] X_{\tau_{i+2}} \cdots )
= \tr( \cdots X_{\tau_{i-1}} [ W(\tau_{i+1}\tau_{i} \to \tau_{i}\tau_{i+1})
X_{i+1} X_i \nonumber\\ {} - W(\tau_{i}\tau_{i+1} \to \tau_{i+1}\tau_{i})
X_{\tau_{i}} X_{\tau_{i+1}} ] X_{\tau_{i+2}} \cdots ) \;,
\end{eqnarray}
in which $W(\tau,\tau' \to \tau',\tau)$ gives the rate at which a
neighbouring pair of sites in configuration $(\tau,\tau')$ exchange.
For  two species of particles on a ring  we have
\begin{eqnarray}
\label{ring:W10}
W(10 \to 01) &=& 1\\
\label{ring:W20}
W(20 \to 02) &=& \alpha\\
\label{ring:W12}
W(12 \to 21) &=& \beta
\end{eqnarray}
and all other rates zero.

Introducing auxiliary matrices, $\tilde{X}_{\tau}$, $\tau=0,1,2$, to generate a pairwise cancellation,
we arrive at a sufficient condition
\begin{equation}
\fl W(\tau_{i+1}\tau_{i} \to \tau_{i}\tau_{i+1}) X_{\tau_{i+1}} X_{\tau_i} \nonumber -
W(\tau_{i}\tau_{i+1} \to \tau_{i+1}\tau_{i}) X_{\tau_{i}} X_{\tau_{i+1}} = \tilde{X}_{\tau_i}
X_{\tau_{i+1}} - X_{\tau_i} \tilde{X}_{\tau_{i+1}}\;.
\label{ring:gen2}
\end{equation}

With a {\em scalar} choice  for the auxiliaries
and the set of rates
(\ref{ring:W10})--(\ref{ring:W12})  the 9 equations in 
(\ref{ring:gen2}) reduce to three equations
\begin{eqnarray}
DE &=& \tilde{E}D -  \tilde{D} E \\
\alpha AE &=& \tilde{E}A -  \tilde{A} E\\
\beta DA &=& \tilde{A}D -  \tilde{D} A \;.
\end{eqnarray}
In this case, 
the  choice  $\tilde{D}=-1,
\tilde{E}=1$ and $\tilde{A} = 0$ yields \RRR\, as required.

\subsection{Normalisation for a single defect particle}
\label{ss2}
On a periodic system
the numbers of each species of  particle are fixed, which puts 
constraints on the set of allowed configurations.  For
example, when $P=1$, so that there is a single species 2 particle (which we refer to as a defect), the
normalisation for an $N$-site ring populated by $M$ particles is given by
\begin{equation}
Z_{N,M} = \sum_{\tau_1=0}^{1} \sum_{\tau_2=0}^{1} \cdots
\sum_{\tau_{N-1}=0}^{1} \bra{W} \prod_{j=1}^{N-1}[\tau_j E +
  (1-\tau_j) D]^{\tau_1} \ket{V} \delta_{\sum_{j} \tau_j, M} 
\end{equation}
where $\delta_{i,j}$ is the usual Kronecker $\delta$-symbol and
enforces the fixed particle number constraint.  Although this
expression can be evaluated directly, the
general expression is rather cumbersome and the derivation relies on
choosing a particular representation of the matrices  \cite{DEM93,Mallick96}. 
We consider instead the generating function
\begin{equation}
\label{ring:FN}
F_N(u) = \sum_{M=0}^{N-1} Z_{N,M} u^{M} = \bra{W} (u D + E)^{N-1}
\ket{V}
\end{equation}
which, by virtue of being now rather similar to (\ref{int:Z}) can be
readily evaluated.  To do this, one transforms to a new pair of
matrices $\DD$ and $\EE$ via
\begin{equation}
\label{ring:DDEE}
D = \frac{1}{\sqrt{u}} \left( \DD - \One \right) + \One \quad\mbox{and}\quad
E = \sqrt{u} \left( \EE - \One \right) + \One
\end{equation}
which, by (\ref{ring:DE}), obey the relations $\DD\EE = \DD+\EE$,
$\DD\ket{V} = 1/\beta' \ket{V}$,
$\bra{W}\EE = 1/\alpha' \bra{W}$
with the boundary
parameters
\begin{eqnarray}
\frac{1}{\alpha'} =  \left[ 1 + \frac{1}{\sqrt{u}} \left(
\frac{1}{\alpha} - 1 \right) \right] \quad\mbox{and}\quad
\frac{1}{\beta'} = \left[ 1 + \sqrt{u} \left(
\frac{1}{\beta} - 1 \right) \right] \;.
\end{eqnarray}
This transformation is very similar to that described in
Sec.~\ref{series} to calculate the joint current-density
distribution for the ASEP with open boundaries.
Then, (\ref{ring:FN}) may be written
\begin{eqnarray}
F_N(u) &=& \bra{W} \left( \sqrt{u} \left[ \DD + \EE
\right] + \left[ \sqrt{u} - 1 \right]^2 \right)^{N-1} \ket{V}\;.
\end{eqnarray}

We can now use  the formal power series approach of Section~\ref{gfa} 
to evaluate the generating function of $F_N(u)$.
Consider the formal series
\begin{eqnarray}
\Eff(u,z) = \sum_{N=1}^{\infty} z^{N-1} F_N(u) = \sum_{N=0}^\infty
\bra{W}z^N (\sqrt{u}C' + (\sqrt{u}-1)^2 )^N \ket{V}\nonumber \\
= \frac{1}{1-z(\sqrt{u} -1)^2} \bra{W} \left[1- \frac{zu^{1/2}}{1-z(\sqrt{u} -1)^2}C'\right]^{-1}\ket{V} \;,  \label{ring:Eff}
\end{eqnarray}
where $C' = D' + E'$. Now, we may write
\begin{equation}
1- \frac{zu^{1/2}}{1-z(\sqrt{u} -1)^2}C'
= \left[1- \omega(u,z) D'\right]\left[1- \omega(u,z) E'\right]\;,
\end{equation}
where
\begin{equation}
\omega(1-\omega) = \frac{z\sqrt{u}}{1-z(\sqrt{u}-1)^2}\;.
\end{equation}
Therefore,
\begin{equation}
\left[1- \frac{zu^{1/2}}{1-z(\sqrt{u} -1)^2}C'\right]^{-1}
= \left[1- \omega(u,z) E'\right]^{-1}\left[1- \omega(u,z) D'\right]^{-1}
\end{equation}
and we deduce
\begin{equation}
\Eff(u,z) = \left[\left(1-z(\sqrt{u} -1)^2\right)\left(1- \frac{\omega}{\alpha'}\right)\left(1- \frac{\omega}{\beta'}\right)\right]^{-1}\;.
\label{eff}
\end{equation}

\subsubsection{The grand canonical ensemble}
\label{gce}
An explicit expression for 
$Z_{N,M}$ could be extracted from $F_N(u)$ (\ref{ring:FN})
via, for example, the residue theorem
\begin{eqnarray}
Z_{N,M} &=& \coeff{ u^{M}}  F_N(u)     = \oint \frac{\rmd u}{2\pi i} \frac{1}{u^{M+1}}
F_N(u)\;.
\end{eqnarray}
Equivalently  we can work in the grand canonical ensemble.
To proceed we view (\ref{ring:FN}) as a partition function for a
system where the number of
first-class particles can fluctuate.  Then, $u$ is a fugacity that
controls the density of occupied sites.  We remark that it is, in
fact, possible to generate this grand canonical ensemble dynamically
by introducing certain specific dynamical rules which do not conserve
the particle number as will be described in Section~\ref{gcmat}.

In the grand canonical ensemble the mean particle number $\rho$ 
 is given by
\begin{equation}
\label{ring:barrho}
\rho = \lim_{N\to\infty} \frac{\avg{M}}{N} = \lim_{N \to
  \infty} \frac{1}{N} u \frac{\partial}{\partial u} \ln F_N(u)\;.
\end{equation}
From our understanding of the grand canonical ensemble in equilibrium
statistical physics, we anticipate that the both the mean number of
particles $\avg{M}$ and its variance $\avg{M^2}-\avg{M}^2$ in the
ensemble will be proportional to the system size $N$ in the
thermodynamic limit.  Thus, the particle number $\avg{M}$ will have
fluctuations $O(N^{1/2})$ about the mean and the density will be
sharply peaked about $\rho$.  The grand canonical and canonical
ensembles should yield equivalent macroscopic properties.  Since
working at a fixed fugacity $u$ is, in the thermodynamic limit,
equivalent to working at fixed particle density ${\rho}$  there is
no need to invert the generating function (\ref{ring:FN}).  Instead
(for large $N$) one can substitute $u$ for $\rho$ via
(\ref{ring:barrho}) in the expression for some physical quantity.

\subsection{Phase diagram for a single defect particle}\label{defpd}
The phase diagram in the $\alpha$--$\beta$ plane for the model with a
single species 2 particle was determined in \cite{Mallick96} by
considering the exact expressions for the $Z_{N,M}$ for $N$ large. An
alternative approach was presented in \cite{DE99} where the Bethe
ansatz was used to develop contour integral representations of
$Z_{N,M}$ which were then evaluated in the large $N$ limit.  Here we
will derive the phase digram from the asymptotic forms of the grand
canonical partition function (\ref{ring:FN}).  We therefore proceed to
evaluate the asymptotics of (\ref{ring:FN}) from (\ref{eff}).

We first note that the possible singularities of (\ref{eff}) as a
function of $z$ are (since $u$ is positive) a square root at
\begin{equation}
z_0 = (
1+\sqrt{u})^{-2}\;,
\end{equation}
if $\alpha'<1/2$ a  pole at $\omega(z_\alpha) = \alpha'$ and 
if $\beta'<1/2$ a pole at  $\omega(z_\beta) = \beta'$.
The locations of the two poles  reduce to
\begin{eqnarray}
z_\alpha &=& \frac{\alpha(1-\alpha)}{\alpha(u-1) +1}\\
z_\beta &=& \frac{\beta(1-\beta)}{u(1-\beta)+\beta}\;.
\end{eqnarray}
There is also an (irrelevant) pole at $z_1 = (1-\sqrt{u})^{-2}$.

Using the results of  Appendix~B one can compute   the following
asymptotic contributions
to $F_N(u)$ :
\begin{eqnarray}
\mbox{from}\quad z_0 \qquad
\frac{1}{z_0^N} \frac{1}{2\pi (1-z_0/z_1)^{3/2}} 
\frac{2\alpha'}{\alpha'-1} \frac{2\beta'}{\beta'-1}
\left[ \frac{2\alpha'}{\alpha'-1} +\frac{2\beta'}{\beta'-1}\right]
\label{F0}
\\
\mbox{from}\quad z_{\alpha} \qquad
\frac{1}{z_\alpha^N}
\frac{\left[ (1-\alpha)/\alpha - u\alpha/(1-\alpha)\right]}
{\left[ (1-\alpha)/\alpha - u(1-\beta)/\beta\right]}
\label{Falpha}
\\
\mbox{from}\quad z_{\beta} \qquad
\frac{1}{z_\beta^N}
\frac{\left[ \beta/(1-\beta) - u(1-\beta)/\beta\right]}
{\left[ (1-\alpha)/\alpha - u(1-\beta)/\beta\right]}\;.
\label{Fbeta}
\end{eqnarray}

Since the asymptotic form of $F_N$ is always $\sim z^{-N}$, 
with $z$ the dominant singularity, one can associate
via
(\ref{ring:barrho}) a density with each of the singularities
\begin{eqnarray}
\rho_0 &=&  -u \frac{\partial \ln z_0}{\partial u} =
\frac{u^{1/2}}{1+u^{1/2}}\label{rho0}\\
\rho_{\alpha} &=& -u \frac{\partial \ln z_\alpha}{\partial u} =
 \frac{u\alpha}{\alpha u + 1-\alpha}\label{rhoa}\\
\rho_{\beta} &=& 
-u \frac{\partial \ln z_\beta }{\partial u} =\frac{u(1-\beta)}{ u(1-\beta) + \beta}\;.
\label{rhob}
\end{eqnarray}
Note that since $z_\alpha < z_0$ and $z_\beta <z_0$ a contribution
to $F_N(u)$
from $z_{\alpha}$ or $z_\beta$, if it exists, will dominate the
contribution from $z_0$.  For example, the condition for the
contribution from $z_{\alpha}$ to exist is $u^{1/2} <
(1-\alpha)/\alpha$.  This becomes, using $\rho=\rho_\alpha$,
(\ref{rhoa}) to give $u$ as a function of $\rho$, $1-\alpha >\rho$.
Similarly, the condition for the contribution from $z_{\beta}$ to
exist is $u^{1/2} > \beta/(1-\beta)$ and this becomes using
(\ref{rhob}) $\beta >\rho$.

Finally, we note that in both the $z_\alpha$ and $z_\beta$
contributions to $F_N(u)$, Eqs.~(\ref{Falpha}) and (\ref{Fbeta}),
there is a pole at
\begin{equation}
u_p = \frac{\beta(1-\alpha)}{(1-\beta)\alpha}\;.
\end{equation}The presence of this pole means that as the 
parameters $\alpha$ and $\beta$ change, $u$ cannot cross $u_p$.
For the contribution coming from $z_\alpha$, one finds that
as $\beta$ decreases to $\rho$, $ u$ increases to $u_p$. 
Similarly, for  the contribution coming from $z_\beta$,
as $\alpha$ decreases to $1-\rho$, $ u$ decreases to $u_p$. 
Therefore in the region $\alpha <1-\rho$, $\beta<\rho$, $u$ is fixed  at
$u=u_p$. Actually, this holds  strictly  in the thermodynamic
limit;  to correctly achieve the density $\avg{\rho}=\rho$
one should choose $u = u_p + \gamma(\alpha,\beta,\rho)/N$,
where the function $\gamma$ is some function to be determined,
then let $N \to \infty$.

In the thermodynamic limit, we can   can identify four phases as follows
\begin{enumerate}
\item $1-\alpha < \rho < \beta$\qquad $z=z_0= (1-\rho)^2$ and 
$\ds u=\frac{\rho}{(1-\rho)^2}$ 
\item $1-\alpha > \rho$ and $\rho<\beta$\qquad $z= z_\alpha= \alpha(1-\rho)$
and $\displaystyle u= \frac{\rho (1-\alpha)}{(1-\rho)\beta}$
\item $1-\alpha < \rho$ and $\rho > \beta$\qquad $z= z_\beta= (1-\beta)(1-\rho)$ and $\displaystyle u= \frac{\rho \beta}{(1-\rho)(1-\beta)}$
\item $1-\alpha > \rho$ and $\rho > \beta$\qquad
$z= z_\alpha= z_\beta = \alpha(1-\beta)$ 
and $\displaystyle u= \frac{(1-\alpha)\beta }{\alpha(1-\beta)}$
\end{enumerate}

We now calculate the velocity of the defect particle.  In the
canonical ensemble this is given by $\alpha(1- \rho_+)$, where
$\rho_+$ is the probability of a hole in front of the defect minus
$\beta \rho_-$, where $\rho_-$ is the probability of a particle behind
the defect particle: using (\ref{ring:AE}) and (\ref{ring:DE})
\begin{equation}
v = \frac{Z_{N-1,M}- Z_{N-1,M-1}}{Z_{N,M}}\;.
\end{equation}
In the grand canonical ensemble the corresponding expression is
\begin{eqnarray}
v &=& \frac{\alpha \bra{\alpha}E (uD +E)^{N-1}\ket{\beta}-
\beta \bra{\alpha} (uD +E)^{N-1}u D \ket{\beta}}{F_{N}(u)}\\
&=& \frac{F_{N-1}(u) - u F_{N-1}(u)}{F_{N}(u)}\;.
\end{eqnarray}
We find using the form of the asymptotic behaviour, $F_N(u)\sim z^{-N}$,
that \begin{equation}
v \to z(1-u)\quad \mbox{for $N$ large}\;.
\end{equation}

\begin{figure}
\begin{center}
\includegraphics[scale=0.6]{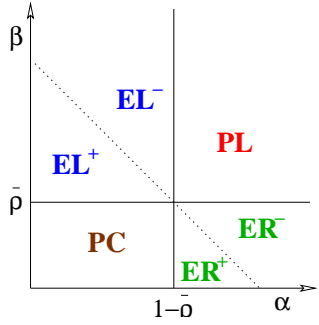}
\end{center}
\caption{\label{fig:ringpd} Phase diagram for the ASEP with a single
  second-class particle on a ring.}
\end{figure}

\begin{figure}
\begin{center}
\includegraphics[scale=0.5]{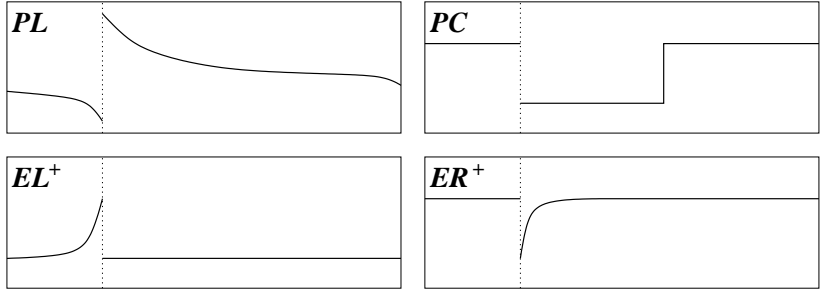}
\end{center}
\caption{\label{fig:ring2profiles} Density profiles as seen from the
defect particle (shown as a vertical dotted line) in the four phases
of the ASEP on a ring with a defect particle: power law (PL), phase
coexistence (PC), exponential decay to the right (ER) and left
(EL) of the defect particle.}
\end{figure}

The different phases are characterised by the values in
Table~\ref{tab:sd}.
\begin{table}
\begin{center}
\begin{tabular}{c|c|c|c|c}
Region & Phase & $v$ & $\rho_+$ & $\rho_-$\\\hline
(i)& PL & $1-2\rho$ & $1-(1-\rho)^2/\alpha$  & $\rho^2/\beta$\\ 
(ii)& ER &$\alpha-\rho$& $\rho $  & $\rho(1-\alpha)/\beta$\\ 
(iii)& EL &$1-\beta-\rho$& $1-(1-\rho)(1-\beta)/\alpha$  & $\rho$\\ 
(iv)& PC  &$\alpha -\beta$& $\beta$  & $1-\alpha$
\end{tabular}
\end{center}
\caption{\label{tab:sd} Velocity, density in front of the defect
and density behind the defect in the various phases of  the single defect system}
\end{table}
The phase diagram which results is shown in Figure~\ref{fig:ringpd},
and the corresponding density profiles in Figure~\ref{fig:ring2profiles}.
Most of the phases are recognisable from the ASEP with open
boundaries.  First, the 
region  $\beta > \rho > 1-\alpha$, which includes the case
of a second-class particle ($\alpha= \beta=1$), has $v=1-2\rho$
and is  marked PL (power law) in Figure~\ref{fig:ringpd}. 
The effect of the defect is to cause a disturbance which results in an increase in the
density in front of the defect and a decrease behind.  The density
profile as seen from the defect particle, decays as a power law with
distance from the defect towards its asymptotic value $\rho$:
\begin{eqnarray}
\label{ring:PL+}
\rho_{+j} &\simeq& \bar\rho + \left( \frac{\bar\rho(1-\bar\rho)}{\pi j}
\right)^{1/2} \;.\\
\label{ring:PL-}
\rho_{-j} &\simeq& \bar\rho - \left( \frac{\bar\rho(1-\bar\rho)}{\pi j}
\right)^{1/2}\;.
\end{eqnarray}
We note that this is essentially the same profile as that obtained for
the ASEP with open boundaries in the maximal current phase.

Since the bulk density far away from the defect is
fixed at $\rho$ in the periodic system, we do not have low- and
high-density phases, as such, in this case.  However, the characteristic
density profiles of these phases, in which one has exponential decays
to the left and right of the defect particle respectively, are present
here in the phases marked EL${}^{\pm}$ and ER${}^{\pm}$ respectively.
In the phase ER phase  the defect
only causes a disturbance in front of it and the density profile
decays exponentially towards the asymptotic value.
Similarly, in the phase EL phase  the defect
only causes a disturbance behind it.
In the figure the $+$ or $-$ superscript indicates whether the density
increases or decreases as one goes clockwise around the ring.  
The dashed line separating these regions with positive and negative
density gradients has $\alpha+\beta=1$ and, in common with the ASEP
with open boundaries, along this line the density profile is
completely flat.

The main novelty arising from the particle-number conservation 
is the PC (Phase
Coexistence) phase, where $\alpha<1-\rho$, $\beta<\rho$.
Here, there is coexistence between a
region of low-density $\beta$ in front of the defect particle and
high-density $1-\alpha$ behind the defect particle.  
At the point where
the two regions meet there is a shock in the density.
The shock is localised  at  a position a distance $xN$ in front of
the defect particle, where $x$ is given by $\rho = \beta x +
(1-\alpha)x$ \cite{Derrida96,Mallick96}.  
The interesting point is that the shock exists in an entire region of
the $\alpha$--$\beta$ plane as opposed to the open boundary condition
case where phase coexistence only occurs along the line 
$\alpha=\beta <1/2$.

To understand the PC phase from the analysis, we note that there are
two competing contributions to $F_N(u)$ from the pole in (\ref{eff})
at $z_\alpha$ and the pole at $z_\beta$. Using $u=u_p$ and
(\ref{rhoa},\ref{rhob}) these correspond to densities $\beta$ and
$1-\alpha$ respectively.  Finally we remark that in all phases the
current of first-class particles $J=\rho(1-\rho)$ \emph{except} the PC
phase where $J=\rho(\alpha-\beta)+\beta(1-\alpha)$.

Finally, we mention that other defect particle dynamics
have been studied \cite{AHR98,AHR99,Jafarpour00,Jafarpour00b,Sasamoto00a,RSS00}
and we will discuss these models further in Section~\ref{2species}.

\subsection{Some  properties of the second-class particle case}
The second-class particle case $\alpha=\beta=1$ was subject of a
detailed study by Derrida, Janowsky, Lebowitz and Speer\cite{DJLS93}.
In addition to the case of a single second-class particle,
results were obtained
for several second-class particles and a finite density of
second-class particles (as far as we are aware these computations
have not yet been
extended to the case of general $\alpha$ and $\beta$).  Here we
briefly summarise some of their findings.

For the case of two second-class particles the probability of finding
the particles a distance $r$ apart was computed and shown to decay as
a $1/r^{3/2}$ power law.  Thus there is an effective attractive
interaction and the two particles form a weak bound-state in which,
despite the attractive interaction, the mean separation of the two
particles diverges. The computations of \cite{DJLS93} also considered
a finite density, $\rho_2$ of second-class particles and showed that
in the limit $\rho_2\to 0$ the environment around a second-class
particle is different from case of a single second-class particle.
Although in this limit the global density is zero there 
must be  a clustering of the second-class particles into a bound state.

A finite system with first and second-class particles cannot support
two regions of different densities. However via an astute construction
\cite{DJLS93,FKS91}
one can study shocks on such a system. The idea is that a system with
finite densities $\rho_1$, $\rho_2$ of first and second-class
particles can be viewed as a one-species system in two different
ways. Treating the second-class particles as holes gives a
TASEP with density $\rho_1$ whereas ignoring the difference between
first and second-class particles, so that exchanges between them are
immaterial, yields a TASEP with density $\rho_1+\rho_2$.  Then one can
construct shock profiles as follows: consider the density profile as
seen from some second-class particle where in front of this
second-class particle both first and second-class particles are
treated as particles and behind the second-class particle the
second-class particles are treated as holes.  Then letting the system
size go to infinity yields a profile where $\rho_i$ tends to $\rho_1 +
\rho_2$ for $i$ large and $ \rho_{-i}$ tends to $\rho_1$ for $i$
large. Using this  approach shock profiles were obtained. Later
these profiles were recovered from a calculation directly on the infinite
system (see Section~\ref{definf}).


\section{The partially asymmetric exclusion process with open
  boundaries}
\label{pasep}
\begin{figure}
\begin{center}
\includegraphics[scale=0.9]{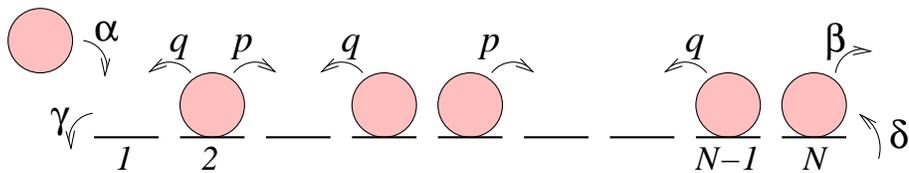}
\end{center}
\caption{\label{fig:pasep} The full parameter space of the partially
  asymmetric exclusion process with open boundaries.  Particles can
  hop in either direction on the lattice, and can enter and leave the
  system at both ends, all at different rates.}
\end{figure}

In this section we consider a generalisation of the ASEP to the
situation where particles can hop both to the left and to the right,
with different rates.  We refer to this variant as the partially
asymmetric exclusion process (PASEP).  The dynamics are as shown in
Figure~\ref{fig:pasep}, and one has particles on the lattice hopping to
the left and right with rates $p$ and $q$ respectively, as long as the
hard-core exclusion constraint is maintained at all times.  
Furthermore, particles
enter the system at the left boundary at rate $\alpha$, as
before, but can also leave from the left boundary at rate $\gamma$.
Likewise, particles leave the right boundary at rate $\beta$, as
before, but also enter there at rate $\delta$.  We shall once again
define the unit of time by setting $p=1$.

Physical interest in this more general model comes from at least two
quarters.  First, there is a mapping between exclusion models and
certain nonequilibrium surface growth models \cite{MRSB86} (note,
these are distinct from the equilibrium surface models discussed in
Sec.~\ref{surf}).  At large lengthscales, the shape of the interface
is governed by a stochastic partial differential equation, the KPZ
equation (see e.g. \cite{Krug97} for a review), that has a nonlinear
term proportional to $1-q$.  The presence or otherwise of this
nonlinearity governs in turn the universal properties of the
interface.  Thus, by taking $q\to 1$ in the PASEP we should learn
something of the crossover between universality classes \cite{DM97}.
The second source of interest lies in the special case where
$\gamma=\delta=0$ and $q>1$.  Then, the bias in the random walks
performed by the particles opposes the current imposed by the boundary
conditions.  This gives rise to a new \emph{reverse-bias} phase not
seen in the ASEP.  Furthermore, the matrix algebra that is used to
solve the PASEP also arises in the context of ballistic reaction
dynamics and has been used to calculate the late-time density decay in
the system \cite{BEK00}.

In Section~\ref{calcZ}  we discussed the three main 
approaches to calculation with matrix products. 
For the PASEP with open boundaries 
the most natural approach turns out to be 
diagonalisation of an explicit representation.
Indeed, to our knowledge, results have
thus far been obtained only through this approach
\cite{Sasamoto99,BECE00,Sasamoto00b,USW04,UW05}.

\subsection{Quadratic algebra}
As for the ASEP, the  stationary distribution is given through products of two
matrices $D$ and $E$ that represent particles and holes respectively.
Again, the product is contracted with a pair of vectors $\bra{W}$ and
$\ket{V}$ to obtain the desired scalar weight, as in (\ref{int:mp}).
The relations that allow the matrix products to be reduced are
\begin{eqnarray}
\label{pasep:DE}
DE - qED &=& D + E \\
\label{pasep:DV}
(\beta D - \delta E) \ket{V} &=& \ket{V} \\
\label{pasep:EW}
\bra{W} (\alpha E - \gamma D) &=& \bra{W} \;.
\end{eqnarray}
These can be proved to solve the stationary master equation either by
considering terms that cancel between domains, as in
Sec.~\ref{RRproof}, or using a slight modification of the purely
algebraic approach of Sec.~\ref{algproof}.  We shall run over the
latter briefly here.

As in Sec.~\ref{algproof}, we decompose the terms in the master
equation into a sum over terms relating to interactions at the
boundaries, and between pairs of sites in the bulk.  That is,
\begin{equation}
\label{pasep:master}
\frac{\rmd}{\rmd t} f(\tau_1, \ldots, \tau_N) = \left( \hat{h}_L +
\sum_{i=1}^{N-1} \hat{h}_{i,i+1} + \hat{h}_R \right) f(\tau_1, \ldots,
\tau_N)
\end{equation}
where $\hat{h}_{i,i+1}$ generate the terms arising from particle hops in the
bulk, as in (\ref{int:gen}), and $\hat{h}_L$ and $\hat{h}_R$ perform the
corresponding duties for interactions at the left and right boundaries
as in (\ref{int:l},\ref{int:r}).

As in Sec.~\ref{RRRproof} we seek to telescope this sum, and we do so
by assuming that the terms in the master equation generated by the
$\hat{h}$ operators can be written using auxiliary matrices as
\begin{eqnarray}
\label{pasep:tilde1}
\hat{h}_{i,i+1} \bra{W} \cdots X_{\tau_{i}} X_{\tau_{i+1}} \cdots
\ket{V} = \bra{W} \cdots [ \tilde{X}_{\tau_i} X_{\tau_{i+1}} -
X_{\tau_{i}} \tilde{X}_{\tau_{i+1}} ] \cdots \ket{V} \\
\hat{h}_L \bra{W} X_{\tau_1} \cdots \ket{V} = - \bra{W}
\tilde{X}_{\tau_1} \cdots \ket{V} \\
\hat{h}_R \bra{W} \cdots X_{\tau_N} \ket{V} = \bra{W} \cdots
\tilde{X}_{\tau_N} \ket{V} \;.
\label{pasep:tilde3}
\end{eqnarray}
Matching these right-hand sides up with the terms generated by the
respective operators  in the master equation yields 
\begin{eqnarray}
0 &=&\tilde{D}D - D \tilde{D} =\tilde{E}E - E \tilde{E} \\
DE - q ED &=& \tilde{E} D - E \tilde{D} = -  \tilde{D}E +  D\tilde{E}  \\ 
\bra{W} (\gamma D - \alpha E ) &=& - \bra{W} \tilde{E}= \bra{W} \tilde{D} 
\\
(\delta E - \beta D) &=& \tilde{D} \ket{V}= - \tilde{E} \ket{V} \;.
\end{eqnarray}
Again, the scalar choices $\tilde{D}=-1, \tilde{E}=1$ yield the
consistent set of relations \PRR.

It turns out that mathematical structure of the solution is much the
same for the case with nonzero $\gamma$ and $\delta$ as that with
$\gamma=\delta=0$.  However, in the former case the various boundary
rates combine with one another in a complicated way, so we shall take
$\gamma=\delta=0$ in the analysis that follows for ease of
presentation.  At the end of this section we shall explain how nonzero
$\gamma$ and $\delta$ were catered for in the works of
\cite{USW04,UW05}.

\subsection{Calculation of $Z_N$ using  an explicit representation}
\label{alsalam}

As we observed in Section~\ref{mpa} of the Introduction, an argument
was presented in \cite{DEHP93} for the $D$ and $E$ matrices to have,
in general, infinite dimension, at least for the case $q=0$.  This is
also the case for general $q$, and one can find a number of different
representations that satisfy the relations \PRR.  The first of these
that we shall consider has
\begin{eqnarray}
\label{pasep:D1}
D &=& \frac{1}{1-q} \left( \begin{array}{ccccc}
1+b & \sqrt{c_0} & 0 & 0 & \cdots\\
0 & 1+bq & \sqrt{c_1} & 0 &\\
0 & 0 & 1+bq^2 & \sqrt{c_2} & \\
0 & 0 & 0 & 1+bq^3 &  \\
\vdots & & & & \ddots
\end{array} \right) \\
\label{pasep:E1}
E &=& \frac{1}{1-q} \left( \begin{array}{ccccc}
1+a & 0 & 0 & 0 & \cdots\\
\sqrt{c_0} & 1+aq & 0 & 0 & \\
0 & \sqrt{c_1} & 1+aq^2 & 0 &\\
0 & 0 & \sqrt{c_2} & 1+aq^3 &\\
\vdots & & & & \ddots
\end{array} \right)
\end{eqnarray}
where the parameters $a$,$b$,$c_n$ are given by
\begin{equation}
\label{pasep:ab}
a = \frac{1-q}{\alpha} - 1 \;,\quad
b = \frac{1-q}{\beta} - 1 
\end{equation}
and
\begin{equation}
c_n = (1-q^{n+1}) (1-ab q^n) \;.
\end{equation}
In this case, the appropriate choice for the boundary vectors is
\begin{equation}
\bra{W_1} = \bra{0} \quad\mbox{and}\quad
\ket{V_1} = \ket{0}
\end{equation}
where we have introduced a set of basis vectors
\begin{equation}
\ket{n} = ( \underbrace{0,\cdots,0}_{n},  1, 0, 0, \cdots, )^{T}
\end{equation}
for $n=0, 1, 2, \ldots$ and their corresponding duals $\bra{n}$ by
transposition.  One can check explicitly that the reduction relations
\PRR\ are satisfied.

One way to evaluate the normalisation using this representation is to
diagonalise the matrix $C=D+E$, as was done 
in 
Section~\ref{diag} for the case $q=0$.  In analogy with (\ref{Cevec})
we parametrise the eigenvalues of $C$ by
 $2(1+\cos \theta)/(1-q)$ where $0< \theta <2\pi$ so that
\begin{equation}
C\ket{\cos \theta} = \frac{2(1+\cos \theta)}{1-q} \ket{\cos \theta}.
\end{equation}
Writing 
\begin{equation}
\label{pasep:vecfun}
\ket{\cos \theta } = \sum_{n=0}^{\infty} f_n(\cos \theta) \ket{n}
\end{equation}
implies from (\ref{pasep:D1}) and (\ref{pasep:E1})
that
\begin{eqnarray}
\fl \lefteqn{ \frac{2(1+\cos \theta)}{1-q} f_n(\cos \theta) =}  \\
&&\frac{1}{1-q} \bigg\{ \sqrt{c_{n-1}} f_{n-1}(\cos \theta)
+ [2+(a+b)q^n] f_n(\cos \theta) +  \sqrt{c_n} f_{n+1}(\cos \theta) \bigg\} \;.
\nonumber
\end{eqnarray}
Putting 
\begin{equation}
f_n(\cos \theta) = \frac{P_n(\cos \theta)}{\prod_{k=0}^{n-1} \sqrt{c_k}}\;,
\end{equation}
this three-term recursion simplifies to
\begin{equation}
\label{pasep:Hrec}
\fl 2 \cos \theta  P_n(\cos \theta) = (1-q^n)(1-abq^{n-1}) P_{n-1}(\cos \theta) + (a+b)q^n P_n(\cos \theta) +
P_{n+1}(\cos \theta) \;.
\end{equation}
Given the boundary conditions $P_{-1}(\cos \theta)=0, P_0(\cos \theta)=1$, this recursion
defines the set of Al-Salam-Chihara polynomials \cite{GR04}.  The
relevance of these polynomials to the PASEP was noticed by Sasamoto
\cite{Sasamoto99}.

Although these polynomials are rather complicated, we need to know
very little about them to evaluate the normalisation
$Z_N=\bra{W}C^N\ket{V}$, other than that they satisfy an orthogonality
relation \cite{Sasamoto99,GR04}
\begin{equation}
\label{pasep:ortho}
\fl \frac{1}{2\pi} \frac{(q,ab;q)_\infty}{(q, ab;q)_n} \int_0^\pi \rmd
 \theta \frac{(\rme^{2i\theta}, \rme^{-2i\theta};
 q)_\infty}{(a\rme^{i\theta}, a\rme^{-i\theta}, b\rme^{i\theta},
 b\rme^{-i\theta}; q)_\infty} P_n(\cos\theta) P_m(\cos\theta) =
 \delta_{n,m},
\end{equation}
which holds when $|a|<1, |b|<1$ and $|q|<1$.  Here we have invoked the
notation
\begin{equation}
(a_1, a_2, \ldots, a_m; q)_n = \prod_{k=0}^{n-1} (1-a_1q^k)(1-a_2q^k)
  \cdots (1-a_mq^k)
\end{equation}
since the quantity on the right-hand side is ubiquitous in the theory
of $q$-series, of which the Al-Salam-Chihara polynomials are an
example \cite{GR04}.  Using this orthogonality relation, we can
construct a representation of the identity
\begin{equation}
\label{pasep:ASC1}
\One = \frac{(q,ab;q)_\infty}{2\pi} \int_0^\pi \rmd \theta
 \frac{(\rme^{2i\theta}, \rme^{-2i\theta};
 q)_\infty}{(a\rme^{i\theta}, a\rme^{-i\theta}, b\rme^{i\theta},
 b\rme^{-i\theta}; q)_\infty} \ket{\cos\theta} \bra{\cos\theta}
\end{equation}
where $\bra{\cos\theta}$ is simply the 
transpose of $\ket{\cos\theta}$.

An integral representation for the normalisation $Z$ is then obtained
as follows.  First, one incorporates this identity into the definition
of $Z$ via $Z_N=\bra{W}C^N \One \ket{V}$.  This yields
\begin{equation}
\fl Z_N = \frac{(q,ab;q)_\infty}{2\pi}\int_{0}^{\pi} \rmd\theta
 \frac{(\rme^{2i\theta}, \rme^{-2i\theta};
 q)_\infty}{(a\rme^{i\theta}, a\rme^{-i\theta}, b\rme^{i\theta},
 b\rme^{-i\theta}; q)_\infty} \bra{W}C^N \ket{\cos\theta}
 \braket{\cos\theta}{V} \;.
\end{equation}
We then use the fact that, by construction,
$C\ket{\cos\theta}=2(1+\cos\theta)/(1-q)$ to obtain
\begin{eqnarray}
\fl Z = \frac{(q,ab;q)_\infty}{2\pi}\int_{0}^{\pi} \rmd\theta
 \frac{(\rme^{2i\theta}, \rme^{-2i\theta};
 q)_\infty}{(a\rme^{i\theta}, a\rme^{-i\theta}, b\rme^{i\theta},
 b\rme^{-i\theta}; q)_\infty} \braket{W}{\cos\theta}
 \left[\frac{2(1+\cos\theta)}{1-q}\right]^N \braket{\cos\theta}{V}
 \;. \nonumber\\
\end{eqnarray}
Finally we note that $\braket{W}{\cos\theta}=\braket{0}{\cos\theta}=1$
and likewise that $\braket{\cos\theta}{V}=1$.  We find then an
integral representation of the normalisation
\begin{equation}
\label{pasep:Zint}
Z_N = \frac{(q,ab;q)_\infty}{2\pi}\int_{0}^{\pi} \rmd\theta
 \frac{(\rme^{2i\theta}, \rme^{-2i\theta};
 q)_\infty}{(a\rme^{i\theta}, a\rme^{-i\theta}, b\rme^{i\theta},
 b\rme^{-i\theta}; q)_\infty}
 \left[\frac{2(1+\cos\theta)}{1-q}\right]^N \;.
\end{equation}
In the limit $q\to0$ this can be written as
\begin{equation}
\label{pasep:Z0}
\fl Z_N = \frac{1-ab}{\pi} \int_{-\pi}^{\pi} \rmd\theta
\frac{\sin^2\theta}{(1+a^2-2a\cos\theta)(1+b^2-2b\cos\theta)}
\left[2(1+\cos\theta)\right]^N
\end{equation}
which can be shown to agree with the finite sum (\ref{open:Z2}) as it
should---see Appendix~\ref{intsum}. In \cite{BECE00} it was also shown
that (\ref{pasep:Zint}) is equivalent to a finite sum expression for
general $q$.  However, the resulting expression is rather cumbersome,
and so in  Section~\ref{pasymp} we will concentrate on the procedure 
to extract the asymptotics of the normalisation directly from
the integral representation (\ref{pasep:Zint}).

\subsubsection{Relation to Motzkin Paths}

The representation given as (\ref{pasep:D1}) and (\ref{pasep:E1}) has
an interpretation in terms of random walks and surfaces, similar to
that described in Section~\ref{surf}.  To see this, let the vector
$\ket{n}$ represent a height $n$ above some origin.  A matrix element
$\bra{m} U \ket{n}$, where $U$ is a product of $\ell$ $D$ and $E$
matrices can then be interpreted as the weight of a set of paths of
horizontal length $\ell$ connecting a point at height $m$ (at the
left-hand end of the surface) to a point at height $n$ (at the
right-hand end of the surface).  The particular choice of boundary
vectors $\bra{W_1} = \bra{0}$ and $\ket{V_1} = \ket{0}$ thus
corresponds to a set of paths that begin and end at the origin.

\begin{figure}
\begin{center}
\includegraphics[scale=0.65]{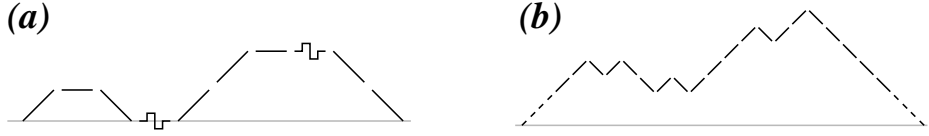}
\end{center}
\caption{\label{fig:motzkin} (a) A bicoloured Motzkin path.  Each
  step  may increase or decrease the
  height of the surface by one unit, or leave it unchanged.  These
  latter horizontal segments come in two \emph{colours}, denoted here
  by the flat and crenellated segments.  (b) The same path mapped onto
  a Dyck path, wherein each Motzkin segment becomes two Dyck segments.
  Two additional segments (shown dotted) are needed at both ends to
  prevent the Dyck path from crossing the origin.}
\end{figure}

The bidiagonal forms of the $D$ and $E$ matrices imply that a step of
the surface must be flat, a diagonal up-step or a diagonal down-step.
Furthermore, the fact that the matrices are semi-infinite 
means that the paths may never go below the origin.  Finally, the
matrix product $Z_N = \bra{0} (D+E)^N \ket{0}$ describes the ensemble
of all possible such paths of length $N$, and has up- and down-step
pairs connecting height $n$ and $n+1$ weighted by $c_n$, and two
different types (or ``colours'') of horizontal segments, coming in
with weights $1+aq^n$ and $1+bq^n$, where again $n$ is the height of
the segment above the origin: see Figure~\ref{fig:motzkin}.  These paths
sometimes appear in the combinatorial literature as \emph{bicoloured
Motzkin paths}.

We remark that each bicoloured Motzkin path can be mapped uniquely to
a Dyck path, which has only diagonal up- and down-steps as described
in Section~\ref{surf}.  This is achieved by mapping each segment of a
Motzkin path to \emph{two} consecutive segments of a Dyck path.
Specifically, diagonal up- and down-steps are mapped to a pair of up-
and down-steps respectively, whilst the two colours of horizontal steps
are mapped to up-down and down-up pairs.  In order that the resulting
path contacts the origin only at its two ends (i.e., \emph{is} a Dyck
path) two additional up-steps are needed at the left end, and
similarly two down-steps at the right--see Figure~\ref{fig:motzkin}.
Thus a Motzkin path of length $N$ corresponds to a Dyck path of length
$2(N+2)$.  Although a similar treatment of the PASEP thermodynamics to
that described for the ASEP in Section~\ref{surf} should in principle
be possible using this approach, to our knowledge it has not yet been
achieved.

\subsection{Asymptotics of the normalisation from the integral
  representation}
\label{pasymp}

\begin{figure}
\begin{center}
\includegraphics[scale=0.6]{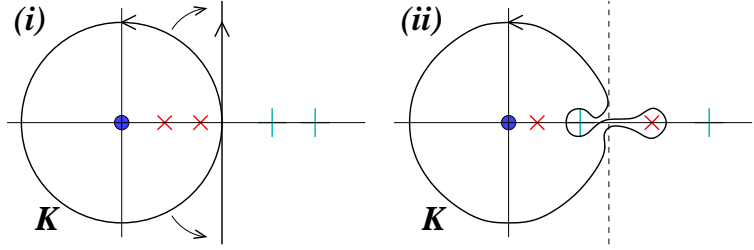}
\end{center}
\caption{\label{fig:contour} Integration contour $K$ for the
  representation (\ref{pasep:Zcont}).  The $\times$ symbols show the
  first of each of the sequences of poles $z=a,qa,q^2a,\ldots,$ and
  $z=a,qb,q^b,\ldots$ that are required to be inside $K$, and the $+$
  symbols those that must be outside.  (i) When $|a|<1$ and $|b|<1$,
  $K$ can be continuously deformed into the contour for the
  saddle-point integration (the vertical line) as shown.  (ii) When
  $|a|>1$, this deformation is no longer possible, and it is necessary
  to add to the saddle-point result the difference between the
  residues at $z=a$ and $z=1/a$.}
\end{figure}

To analyse the asymptotics of the normalisation from its integral
representation (\ref{pasep:Zint}) we first rewrite it as the contour
integral
\begin{equation}
\label{pasep:Zcont}
Z_N = \frac{(q,ab;q)_\infty}{4\pi i}\int_K \frac{\rmd z}{z}
 \frac{(z^2, z^{-2}; q)_\infty}{(az, a/z, bz, b/z; q)_\infty}
 \left[\frac{2+z+z^{-1}}{1-q}\right]^N \;.
\end{equation}
where $K$ is the positively oriented unit circle centred on the
origin.  Recall that the orthogonality relation (\ref{pasep:ortho})
was valid only when $|a|<1$ and $|b|<1$.  In this case the contour $K$
encloses two sequences of poles at $z=a,qa,q^2b, \ldots$ and
$z=b,qb,q^b,\ldots$ along with a number of singularities at the
origin, as shown in Figure~\ref{fig:contour}(i).  We can continue the
function $Z_N$ to $a>1$ and/or $b>1$ by ensuring the contour $K$ is
deformed still to enclose these poles, and further the exclude those
at $z=1/a,q/a,q^2/a, \ldots$ and $z=1/b,q/b,q^2/b, \ldots$ as shown in
Figure~\ref{fig:contour}(ii).

In the former case, where $|a|<1$ and $|b|<1$, the contour $K$ can be
continuously deformed without passing through any poles to coincide
with the line that passes parallel to the imaginary axis through the
saddle point in the integrand at $z=1$.  See
Figure\ref{fig:contour}(i).  Then the normalisation is well approximated
for large $N$ by the expression obtained from applying the
saddle-point method. This is \cite{Sasamoto99,BECE00}
\begin{equation}
\label{pasep:Zsp}
Z_N \sim \frac{4}{\sqrt{\pi}} \frac{(q;q)_\infty^3
  (ab;q)_\infty}{(a;q)_\infty^2 (b;q)_\infty^2} \frac{4^N}{N^{3/2}} \;.
\end{equation}

Let us now consider the case $a>1>qa$, $|b|<1$ where the deformed contour
shown in Figure~\ref{fig:contour}(ii) cannot be continuously deformed
into that used in the saddle-point method.  One must therefore correct
the result by applying the residue theorem to the poles at $z=a$ and
$z=1/a$ so that the former is included in the integration contour (as
required) and the latter excluded.  This procedure gives a correction
\begin{equation}
\label{pasep:Za}
Z_N \sim \frac{(1/a^2;q)_\infty}{(b/a,q)_\infty} (2+a+1/a)^N
\end{equation}
which, for sufficiently large $N$, overwhelms the contribution from
saddle-point expression (\ref{pasep:Zsp}) because $2+a+1/a> 4$ when
$a>1$. Therefore, the  normalisation is well-approximated by the
contribution (\ref{pasep:Za}) from the pair of poles at $z=a$ and
$z=1/a$.  As $a$ is further increased, more and more poles need---in
principle---to be accounted for in the deformation of the contour.
However, one can verify that all further corrections grow less fast
than $(2+a+1/a)^N$ and so (\ref{pasep:Za}) remains the leading
contribution to the normalisation.

\begin{figure}
\begin{center}
\includegraphics[scale=0.8]{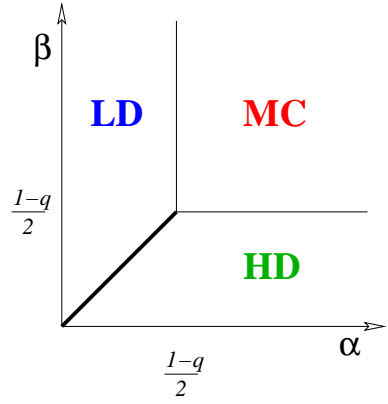}
\end{center}
\caption{\label{fig:pasepJpd} Phases of the PASEP over which the
  particle current is analytic.}
\end{figure}

When $|a|<1$ but $b>1$, we arrive at the same expression
(\ref{pasep:Za}) for the normalisation but with $a$ and $b$ exchanged.
When both $a$ and $b$ are larger than $1$, both of these contributions
are present.  Here, (\ref{pasep:Za}) dominates when $a>b$, and the
corresponding expression with $a \leftrightarrow b$ when $a<b$.  Thus,
for the PASEP we have the same three phases as for the ASEP.  In each
of these the current can be determined using the matrix expression
\begin{equation}
\label{pasep:J}
J = \frac{\bra{W}C^{i-1}(DE - qED) C^{N-i-1}\ket{V}}{Z_N} =
\frac{Z_{N-1}}{Z_N} 
\end{equation}
where the reduction relation (\ref{pasep:DE}) has been used to yield
the same ratio of normalisations that gave the current for the ASEP.
Rewriting the inequalities on $a$ and $b$ given above in terms of
$\alpha$ and $\beta$ we find the phase diagram shown in
Figure~\ref{fig:pasepJpd} with the corresponding currents given in
Table~\ref{tab:pasep}.  Notice the similarities with
Figure~\ref{fig:asepmfpd} and Table~\ref{tab:asepmf}.

\begin{table}
\begin{center}
\begin{tabular}{c|c|c}
Region & Phase & Current $J$ \\\hline
$\alpha < \beta, \alpha < \frac{1-q}{2}$ & Low-density (LD) &
$\frac{\alpha(1-q-\alpha)}{1-q}$ \\
$\beta < \alpha, \beta < \frac{1-q}{2}$ & High-density (HD) &
$\frac{\beta(1-q-\beta)}{1-q}$  \\
$\alpha > \frac{1}{2}, \beta > \frac{1-q}{2}$ & Maximal current (MC) &
$\frac{1-q}{4}$  
\end{tabular}
\end{center}
\caption{\label{tab:pasep} Asymptotic currents in the various phases
  of the PASEP.}
\end{table}

\subsection{Density profiles from the $q$-deformed harmonic oscillator algebra}

Whilst the matrix representation given in (\ref{pasep:D1}) and
(\ref{pasep:E1}) provides a quick route to the normalisation
(\ref{pasep:Zint}), it is less suited to the task of calculating
density profiles.  Here, progress is made by using a representation
that involves the $q$-deformed harmonic oscillator algebra, a
well-studied mathematical object for which many results are known
\cite{Biedenhahn89,Macfarlane89}.

Central to this approach is a pair of creation and annihilation
operators $\hat{a}^\dagger$ and $\hat{a}$ that satisfy a
$q$-commutation relation
\begin{equation}
\label{pasep:qcom}
\hat{a} \hat{a}^\dagger - q \hat{a}^\dagger \hat{a} = 1 - q
\end{equation}
and act on basis vectors according to
\begin{eqnarray}
\label{pasep:adag}
\hat{a}^\dagger \ket{n} &=& \sqrt{1 - q^{n+1}} \ket{n+1} \\
\label{pasep:a}
\hat{a} \ket{n} &=& \sqrt{1 - q^{n}} \ket{n-1} \;.
\end{eqnarray}

If we take
\begin{equation}
\label{pasep:DEaa}
D = \frac{1}{1-q} \left( \One + \hat{a} \right) \quad\mbox{and}\quad
E = \frac{1}{1-q} \left( \One + \hat{a}^\dagger \right)
\end{equation}
we find from (\ref{pasep:qcom}) that the reduction relation
(\ref{pasep:DE}) is satisfied. 
Thus this representation looks like
\begin{eqnarray}
\label{pasep:D2}
D &=& \frac{1}{1-q} \left( \begin{array}{ccccc}
1 & \sqrt{1-q} & 0 & 0 & \cdots\\
0 & 1 & \sqrt{1-q^2} & 0 &\\
0 & 0 & 1 & \sqrt{1-q^3} & \\
0 & 0 & 0 & 1  &  \\
\vdots & & & & \ddots
\end{array} \right) \\
\label{pasep:E2}
E &=& \frac{1}{1-q} \left( \begin{array}{ccccc}
1 & 0 & 0 & 0 & \cdots\\
\sqrt{1-q} & 1 & 0 & 0 & \\
0 & \sqrt{1-q^2} & 1 & 0 &\\
0 & 0 & \sqrt{1-q^3} & 1  &\\
\vdots & & & & \ddots
\end{array} \right)
\end{eqnarray}
 In order that (\ref{pasep:DV}) and
(\ref{pasep:EW}) are also respected  (bearing in mind $\gamma=\delta=0$)
we  require
\begin{equation}
\label{pasep:WVnn}
\braket{W}{n} = \kappa \frac{a^n}{\sqrt{(q;q)_n}} \quad\mbox{and}\quad
\braket{n}{V} = \kappa \frac{b^n}{\sqrt{(q;q)_n}}
\end{equation}
where the parameters $a$ and $b$ are as previously given in
(\ref{pasep:ab}). Thus this representation is the generalisation
to $q \neq 0$ of that used in Section~\ref{diag}.
 The constant $\kappa$ appearing in
(\ref{pasep:WVnn}) is fixed by the convention that $\braket{V}{W} = 1$.
That is, we require
\begin{equation}
\frac{1}{\kappa^2} = \sum_{n=0}^{\infty} \frac{(ab)^n}{(q;q)_n} \;.
\label{kappa}
\end{equation}
This sum, which converges when $|ab|<1$, features prominently in the
literature on $q$-series as it is the $q$-analogue of the exponential
function \cite{GR04}.  When it converges, it can be expressed as an infinite
product and $\kappa^2 = (ab;q)_\infty$.

Diagonalisation of the matrix $C=D+E$ proceeds in much the same way as
described above in Subsection~\ref{alsalam}.  Here, the eigenfunctions
are a vector of functions
\begin{equation}
\ket{\cos\theta} = \sum_{n=0}^{\infty}
\frac{H_n(\cos\theta)}{\sqrt{(q;q)_n}} \ket{n}
\end{equation}
with eigenvalue $\lambda(\cos\theta)=2(1+\cos\theta)/(1-q)$ as before.  The
polynomials $H_n(\cos\theta)$ are found to satisfy the three-term recurrence
\begin{equation}
2\cos\theta H_n(\cos\theta) = (1-q^n) H_{n-1}(\cos\theta) +
H_{n+1}(\cos\theta)\quad n\geq 0
\end{equation}
using the same approach as previously.  Taking $H_0=1$, allows the
functions $H_n(\cos\theta)$ to be identified as $q$-Hermite
polynomials, the properties of which are discussed in
\cite{AAR00,GR04,EB02}.  Of particular importance is the $q$-exponential
generating function of these polynomials, since this allows
computation of the scalar products $\braket{W}{\cos\theta}$ and
$\braket{\cos\theta}{V}$ that appear during the calculation of the
normalisation \cite{Sasamoto99,BECE00}.  One finds for the former
\begin{equation}
\braket{W}{\cos\theta} = \kappa \sum_{n=0}^{\infty} \frac{a^n
  H_n(\cos\theta)}{(q;q)_n} =
  \frac{1}{(a\rme^{i\theta},a\rme^{-i\theta};q)_\infty}
\end{equation}
and the corresponding expression with $a\to b$ for the latter.  With
the knowledge of the orthogonality relation satisfied by the polynomials
\cite{AAR00,GR04} one can write down the identity
\begin{equation}
\label{pasep:HP2}
\One = \frac{(q;q)_\infty}{2\pi} \int_0^\pi \rmd\theta
(\rme^{2i\theta}, \rme^{-2i\theta}; q)_\infty
\ket{\cos\theta}\bra{\cos\theta}
\end{equation}
which is the final part of the jigsaw that is an expression
for the normalisation.  One can verify that the same integral
(\ref{pasep:Zint}) is obtained using this alternative representation.

The key benefit offered by this approach is that expressions for the
difference in density between neighbouring pairs of sites
\begin{equation}
\label{pasep:Delta}
\Delta_i \equiv \rho_{i} - \rho_{i+1} = \frac{\bra{W} C^{i-1} [DC -
    CD] C^{N-i-2} \ket{V}}{Z_N}
\end{equation}
involve the matrix $DC-CD=DE-ED$ which is diagonal in this representation.
This one sees from (\ref{pasep:DEaa}) and the definitions
(\ref{pasep:adag}) and (\ref{pasep:a}).  Specifically,
\begin{eqnarray}
\bra{n} [DE - ED] \ket{m} &=& \frac{\bra{n} [\hat{a} \hat{a}^\dagger -
    \hat{a}^\dagger \hat{a} ] \ket{m}}{1-q} \\ &=&
\frac{\left[(1-q^{n+1}) - (1-q^n)\right]}{1-q} \delta_{n,m} = q^n
\delta_{n,m} \;.
\end{eqnarray}
To express the density gradient (\ref{pasep:Delta}) in terms of
integrals over the $q$-Hermite polynomials, one inserts two of the
identities (\ref{pasep:HP2}), one in front of the combination
$[DC-ED]$ and one behind.  This gives rise to a double integral which
is analysed for large $N$ and $i$ in much the same way as discussed
above in Subsection~\ref{pasymp} but with additional complexity
stemming from the fact that one has the deformation of two contours to
contend with.

\begin{figure}
\begin{center}
\includegraphics[scale=0.8]{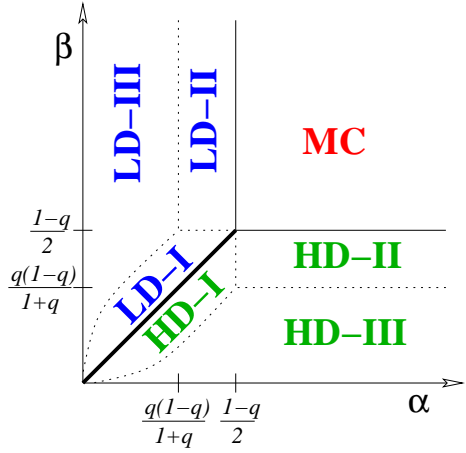}
\end{center}
\caption{\label{fig:paseppd} Full phase diagram of the PASEP showing
  additional sub-phases that have different density decay forms.}
\end{figure}

The detailed analysis of this double integral was conducted in
\cite{Sasamoto00b}.  As with the ASEP, one finds additional sub-phases
relating to different boundary profiles within  the
high-density and low-density phases.
The full phase diagram is as
shown in Figure~\ref{fig:paseppd} and the density profiles one obtains
summarised as follow.
\begin{itemize}
\item \textbf{Maximal-current phase} $\alpha>\frac{1-q}{2}$,
  $\beta>\frac{1-q}{2}$.  Once again, one has the universal power-law
  density profile seen both in the ASEP with open boundaries and on a
  ring for sufficiently large $\alpha$ and $\beta$.  That is, near the
  right boundary one has the decay
\begin{equation}
\rho_{N+1-j} \sim \frac{1}{2} \left( 1 - \frac{1}{\sqrt{\pi j}} \right)
\end{equation}
  and a similar decay from the left obtained by invoking the particle
  hole symmetry that was also present in the ASEP, viz,
  $\rho_{i}(\alpha,\beta) = 1 - \rho_{N+1-i}(\beta,\alpha)$.
\item \textbf{Low-density phases} $\alpha < \frac{1-q}{2},
  \alpha<\beta$.  In all of the low-density phases there is a region
  of constant density $\rho = \frac{\alpha}{1-q} < \frac{1}{2}$ that
  extends from the left boundary into the bulk.  Near the right
  boundary the decay takes three different forms which define three
  sub-phases.
\begin{itemize}
\item \textbf{LD-I} $\alpha<\beta<\frac{\alpha}{q+a},
  \beta<\frac{1-q}{2}$.  Here there is an exponential decay
  with a characteristic length $\xi$ given by
\begin{equation}
\xi^{-1} = \ln \left[ \frac{\beta(1-q-\beta)}{\alpha(1-q-\alpha)}
  \right] \;.
\end{equation}
This phase is present in the totally symmetric case $q=0$ (see
Figure~\ref{fig:aseppd}).
\item \textbf{LD-II} $\frac{q(1-q)}{1+q} < \alpha < \frac{1-q}{2},
  \beta > \frac{1-q}{2}$.  This phase, which is also present when
  $q=0$, has an exponential decay characterised by
\begin{equation}
\xi^{-1} = \ln \left[ \frac{(1-q)^2}{4\alpha(1-q-\alpha)} \right]
\end{equation}
which is additionally modulated by a power-law decay with exponent
$\frac{3}{2}$, as in the totally asymmetric case.
\item \textbf{LD-III} $\beta>\frac{\alpha}{q+a},
  \alpha<\frac{q(1-q)}{1+q}$.  When $q>0$, a new phase appears that
  has a purely exponential decay with a characteristic length
\begin{equation}
\xi^{-1} = \ln \left[ \frac{q}{(q+\alpha)^2} \right] \;.
\end{equation}
\end{itemize}
\item \textbf{High-density phases} $\beta < \frac{1-q}{2},
  \beta<\alpha$. The properties of the high-density phases are
  obtained from the particle-hole symmetry.  These all have an
  exponential decay at the left boundary to a constant bulk density
  that extends to the right boundary.
\end{itemize}

\subsection{Reverse bias phase}

The foregoing analysis applies to the case $q<1$, since this is a
restriction on the representations of the identity (\ref{pasep:ASC1})
and (\ref{pasep:HP2}) that were required in the analysis.  As we
previously noted, the case where $q>1$ is of interest since then the
particles in the bulk prefer to hop to the left, even though the
boundary conditions enforce a current that flows in the opposite
direction.

Although the $q$-Hermite polynomials have an extension to $q>1$
\cite{Askey89} it is unclear how to handle the resulting integral
representation of the normalisation and other matrix product
expressions \cite{BECE00}.  An alternative approach, followed in
\cite{BECE00}, is to note that, if one were actually to evaluate the
integral (\ref{pasep:Zint}) by performing, say, a Laurent expansion,
one would end up with a sum containing a finite number of terms.  This
is because this integral is another way to express the result of a
direct reordering using the reduction relations \PRR, a procedure that
is known to terminate.  Such an expression is then valid for any $q$,
not just $q<1$ as was required to derive the integral representation in
the first place.  In other words, we can access the asymptotics for
$q>1$ by first fixing $N$ at some finite value with $q<1$, put $q>1$
in the result and only then at the very end take $N\to\infty$.

It is possible, using a clutch of identities from the theory of
$q$-series, to find the finite sum implied by (\ref{pasep:Zint})
\cite{BECE00}.  We do not quote this formula---a complicated mixture
of ordinary and $q$-combinatorial factors---here.  Rather, we simply
state that the leading term in the normalisation for $q>1$ is, for
large $N$ \cite{BECE00},
\begin{equation}
Z_N \sim A(q,\alpha,\beta) \left( \frac{1}{q-1} \sqrt{\frac{(q-1+\alpha)(q-1+\beta)}{\alpha\beta}} \right)^N
q^{\frac{1}{4}N^2}
\end{equation}
where $A(q,\alpha,\beta)$ is a function that is independent of $N$.
Note that this result holds for all values of $\alpha$ and $\beta$:
there is only one reverse-bias phase.  For (\ref{pasep:J}) we see that
in this phase, the current vanishes asymptotically as
\begin{equation}
J \sim \left( \frac{\alpha\beta (q-1)^2}{(q-1+\alpha)(q-1+\beta)}
\right)^{\frac{1}{2}} q^{\frac{1}{4} - \frac{1}{2}N} \;.
\end{equation}

Although the density profile in the reverse bias phase has not been
calculated explicitly, one can conjecture its form using a simple
argument.  First we expect most of the particles to be in a domain by
the left boundary, and most of the vacant sites by the right.  Now, if
the interface between these particle- and hole-rich regions were
closer to the right boundary than the left, a particle at the
interface will escape from the right boundary in a shorter time than a
hole from the left.  Therefore, in this instance the interface will
tend to drift to the left.  On the other hand, if the interface were
closer to the left boundary it would, by the same argument, drift to
the right.  Therefore, the lattice will typically be half full in the
steady state, a feature noticed in computer simulations for small
systems.

\subsection{Entry and exit at both boundaries}

We finally briefly mention the case of the PASEP where particles may
enter and leave the system at both boundaries, as shown in
Figure~\ref{fig:pasep}.  A representation similar to that used in
Subsection~\ref{alsalam} was established in \cite{USW04}, where it was
also shown that the functions appearing in the eigenvector of the
matrix $C$ are a complete set of Askey-Wilson polynomials
\cite{AW85,GR04}.  These are a set of very general polynomials that
contain both the Al-Salam-Chihara and $q$-Hermite polynomials
encountered above as special cases.  They have four parameters related
to the transition rates $\alpha$, $\beta$, $\gamma$ and $\delta$ as
follows
\begin{eqnarray}
a_{\pm} &=& \frac{1}{2\alpha} \left[ (1-q-\alpha+\gamma) \pm
  \sqrt{(1-q-\alpha+\gamma)^2 + 4\alpha\gamma} \right] \\
b_{\pm} &=& \frac{1}{2\beta} \left[ (1-q-\beta+\delta) \pm
  \sqrt{(1-q-\beta+\delta)^2 + 4\beta\delta} \right] \;.
\end{eqnarray}
As in Subsection~\ref{alsalam}, the boundary vectors are
$\bra{W}=\bra{0}$ and $\ket{V}=\ket{0}$, and the polynomials satisfy
an orthogonality relation similar to (\ref{pasep:ortho}), but with
additional factors appearing to accommodate the new parameters
\cite{AW85,GR04}.  These carry through to the expression one
eventually obtains for the normalisation
\begin{eqnarray}
\fl Z_N =
\frac{(q,a_{+}a_{-},b_{+}b_{-},a_{+}b_{+},a_{+}b_{-},a_{-}b_{+},a_{-}b_{-};q)_\infty}{2\pi(a_{+}b_{+}a_{-}b_{-},q)_\infty}
\times {} \nonumber\\
\fl \int_{0}^{\pi} \rmd\theta
 \frac{(\rme^{2i\theta}, \rme^{-2i\theta};
 q)_\infty}{(a_{+}\rme^{i\theta}, a_{+}\rme^{-i\theta}, b_{+}\rme^{i\theta},
 b_{+}\rme^{-i\theta}, a_{-}\rme^{i\theta}, a_{-}\rme^{-i\theta}, b_{-}\rme^{i\theta},
 b_{-}\rme^{-i\theta}; q)_\infty}
 \left[\frac{2(1+\cos\theta)}{1-q}\right]^N \;.
\end{eqnarray}
One recognises a structure similar to (\ref{pasep:Zint}) and the
asymptotic analysis via a contour integral detailed in \cite{USW04} is
basically the same as described in Subsection~\ref{pasymp}.  When
$q<1$, one finds the three familiar phases: a maximal-current phase
when $a_{+}>1, b_{+}>1$; a low-density-phase when $a_{+}>1$ and
$a_{+}>b_{+}$; and a high-density phase when $b_{+}>1$ and
$b_{+}>a_{+}$.  When $q>1$ the same three phases are manifested as the
model is invariant under the exchanges $(\alpha,\beta,\gamma,\delta,q)
\leftrightarrow (q^{-1}\delta, q^{-1}\gamma, q^{-1}\beta,
q^{-1}\alpha, q^{-1})$, at least when all four boundary rates are
nonzero.  Since the physical properties of these phases are very
similar to their counterparts in the ASEP with open boundaries, we do
not discuss them further here.  Further details, including an analysis
of the bulk densities through the introduction of a fugacity as
described in Subsection~(\ref{gce}), are given in \cite{USW04}.  As
further shown in \cite{UW05}, the eigenvalue spectrum of the matrix
$C$ can be used to determine density profiles and correlation
functions.  These show once again the universal power-law decay in the
maximal current phase, and various exponential decays in the high- and
low-density phases.


\section{Macroscopic density profiles for open boundary ASEP}
\label{free}

In this review we have so far mostly concentrated on how the current
and density profiles in driven diffusive systems are calculated from
the matrix product expressions for the steady state.  Since one has to
hand the entire stationary distribution of microstates, it is in
principle possible to calculate any macroscopic quantity of interest.
For example, the existence of a matrix product solution has recently
admitted the calculation of the Gibbs-Shannon entropy in the symmetric
exclusion process with open boundaries \cite{DLS07}.  In this section
we take a closer look at a quantity that has received a great deal of
recent attention in the context of open driven-diffusive systems,
namely free energy functionals which characterise the probability that
a deviation from the most likely density profile is seen in a
macroscopically large (but finite) system.

To be more precise, we consider macrostates corresponding to a density
profile $\rho(x)$ where $x$ lies in the (one-dimensional) interval
$[0,1]$ which contains some large number $N$ of lattice sites.  As we
shall see explicitly below, each macroscopic profile $\rho(x)$ can be
realised in number of ways that grows exponentially with the number of
microscopic degrees of freedom $N$.  This exponential growth leads to
the key property of statistical mechanical systems that in the
thermodynamic limit, $N\to\infty$, one particular macrostate---let us
call this $\bar{\rho}(x)$---is very much more likely to be realised
than any other (except possibly at a phase transition).  That is, the
probability $P[\rho]$ of seeing a particular macrostate behaves for
large $N$ as
\begin{equation}
P[\rho] \sim \exp\left( - N {\cal F}[\rho] \right)
\end{equation}
where ${\cal F}[\rho]$ is a functional of the density profile that
vanishes when $\rho(x) = \bar{\rho}(x)$.  In an equilibrium system,
${\cal F}[\rho]$ is essentially the free energy of the macrostate,
relative to the most likely profile.  Of course, in a nonequilibrium
steady state, the probability of seeing a deviation from the most like
profile can be measured, and so by analogy with the equilibrium case,
we refer to the object ${\cal F}[\rho]$ that results as a free energy
functional.  Progress in calculating such free energy functionals and
related quantities has been summarised in the lecture notes of Derrida
\cite{Derrida07}.

\subsection{Free energy functional for the symmetric simple exclusion
  process (SSEP)}
\label{hardway}

In this subsection, we will sketch the derivation of the free energy
functional for the nonequilibrium symmetric simple exclusion process
(SSEP) with open boundaries that has been achieved through the matrix
product solution for the steady-state distribution \cite{DLS01,DLS02}.
This version of the model has symmetric hopping of a single species at
unit rate in the bulk, entry at the left and right boundaries at rate
$\alpha$ and $\delta$ respectively, and exit at the left and right
boundaries at rate $\beta$ and $\gamma$.
In other words, we have
the dynamics illustrated in Figure~\ref{fig:pasep} of
Section~\ref{pasep} with the bias $q=1$. 

In order to illustrate the procedure by which the free energy
functional was first calculated in \cite{DLS01,DLS02} from the matrix
product expressions, we shall follow it here for the much simpler case
in which the model parameters are chosen so as to give rise to an
\emph{equilibrium} steady state.  To find this set of parameters, we
employ a Kolmogorov criterion \cite{Kelly79,Mukamel00}.  This states
that if, for every loop $\Cee_1\to\Cee_2\to\cdots\to\Cee_n\to\Cee_1$
in configuration space, the equality
\begin{eqnarray}
\label{free:kol}
\fl W(\Cee_1\to\Cee_2) W(\Cee_2\to\Cee_3) \cdots W(\Cee_{n-1}\to\Cee_n)
W(\Cee_n \to \Cee_1) = {}\nonumber\\ 
W(\Cee_n \to \Cee_{n-1}) W(\Cee_{n-1}\to\Cee_{n-2}) \cdots
W(\Cee_2\to\Cee_1) W(\Cee_1\to\Cee_n)
\end{eqnarray}
holds, the steady state of the system is an equilibrium state for
which the detailed balance relation
\begin{equation}
\label{free:db}
\frac{f(\Cee)}{f(\Cee')} = \frac{W(\Cee'\to\Cee)}{W(\Cee\to\Cee')}
\end{equation}
also holds for every pair of configurations $\Cee$ and $\Cee'$ between
which transitions occur at nonzero rate.  Here we find that
(\ref{free:kol}) is automatically satisfied unless, in total, some
nonzero number $m$ particles exits at one boundary and re-enters at
the other to return to the starting configuration going one way round
the loop.  Then, the ratio of the two products in (\ref{free:kol}) is
$(\alpha\beta/\gamma\delta)^m$.  Therefore, to realise an equilibrium
steady state one must have $\alpha\beta=\gamma\delta$.

Since the bulk hopping occurs at the same rate in both directions, the
detailed balance relation (\ref{free:db}) implies that all
configurations that have $M$ particles on the lattice are equally
likely in the steady state.  By considering a pair of configurations
that differ by the addition or removal of a single particle at the
left boundary, one finds that if $f_M$ is the steady-state weight of
any one of the $M$-particle configurations,
\begin{equation}
f_{M+1} = \frac{\alpha}{\gamma} f_{M} \;.
\end{equation}
Hence, the probability of seeing a particular configuration $\tau_1,
\tau_2, \ldots, \tau_N$ (where, as previously, $\tau_i=1$ if site $i$
is occupied, and zero otherwise) is given by the product
\begin{equation}
P(\tau_1, \tau_2, \ldots, \tau_N) = \prod_{i=1}^{N} \frac{
(\frac{\alpha}{\gamma})^{\tau_i}}{1+\frac{\alpha}{\gamma}} \;.
\end{equation}

Once the underlying distribution of microstates has been established,
the free energy functional is obtained by taking an appropriate
combination of thermodynamic and continuum limits.  To this end, we
divide the system into a number $n$ of boxes, box $j$ containing $N_j
= y_j N$ sites and $M_j = \rho_j N_j$ particles.  The $j$ pairs $(y_j,
\rho_j)$ can thus be used to specify a macrostate of the system in the
limit $N\to\infty$.  To find the probability of this macrostate we
note that
\begin{equation}
\label{free:boxes}
P(N_1, M_1; N_2, M_2; \cdots; N_j, M_j) = \prod_{i=1}^{j} P_{\rm
  box}(N_i, M_i)
\end{equation}
where $P_{\rm box}(N, M)$ is the probability that $N$ particles are in
a box of size $M$ in the steady state.  This is given by
\begin{eqnarray}
\fl \lefteqn{P_{\rm box}(N, M) = {N \choose M}
\frac{(\frac{\alpha}{\gamma})^M}{(1+\frac{\alpha}{\gamma})^N}} \\ 
\fl&\sim& \exp\left( - N \left[ \ln \left( 1 + \frac{\alpha}{\gamma}
\right) - \frac{M}{N} \ln \frac{\alpha}{\gamma} + \frac{M}{N} \ln
\frac{M}{N} + \left( 1 - \frac{M}{N} \right) \ln \left(1 - \frac{M}{N}
\right) \right] \right)
\end{eqnarray}
where the second expression, valid for large $N$, has been obtained
using Stirling's approximation.  Inserting this into
(\ref{free:boxes}), and taking the thermodynamic limit, one finds that
\begin{equation}
\label{free:therm}
\fl - \lim_{N\to\infty} \frac{1}{N} \ln P(y_1, \rho_1; \cdots; y_j, \rho_j) =
  \sum_{i=1}^{j} y_i \left[ \rho_i \ln
  \frac{\rho_i}{\frac{\alpha}{\gamma+\alpha}} + (1 - \rho_i) \ln
  \frac{1 - \rho_i}{1 - \frac{\alpha}{\gamma+\alpha}} \right]
\end{equation}
in terms of the intensive box sizes $y_j$ and densities $\rho_j$.  The
continuum limit is now straightforward: we take $y_i = \delta x_i\to 0
$ and $j\to\infty$ where $\delta x_i$ is some portion of the interval
$x \in [0,1]$ such that $\sum_{i} \delta x_i = 1$ and the density
profile $\rho(x)$ becomes a function of $x$.  This procedure yields
\begin{equation}
\label{free:eqm}
{\cal F}^{\rm eq}[\rho] = \int_0^1 \rmd x \left[ \rho(x) \ln
  \frac{\rho(x)}{\bar{\rho}} + (1 - \rho(x)) \ln
  \frac{1-\rho(x)}{1-\bar{\rho}} \right]
\end{equation}
for the free energy functional, where we have introduced the parameter
$\bar{\rho} = \frac{\alpha}{\alpha+\gamma}$.  It is easily verified
that ${\cal F}[\rho(x)]$ vanishes if $\rho(x)=\bar{\rho}\;\;\forall x \in
[0,1]$, indicating that the flat profile is the most likely profile in
the thermodynamic limit as one would expect at equilibrium.  We also
note that the free energy functional is local in the density field
$\rho(x)$; that is, one can express (\ref{free:eqm}) as an integral
over a free energy density $f(x)$ that depends only on the local
properties of the density field $\rho(x)$ at the point $x$ (and
nowhere else).

When the equilibrium constraint on the boundary rates
$\alpha\beta=\gamma\delta$ is relaxed, this local property of the free
energy density no longer holds.  The reason for this is that when the
system is divided into boxes, the joint distribution of box sizes and
occupancy does not factorise as in (\ref{free:boxes}).  That is, there are
long-range correlations in the density profiles that are implied by
the matrix-product structure of the steady-state weights.  This makes
evaluating the free energy functional by taking the appropriate
combination of thermodynamic and continuum limits a much more
complicated exercise. We therefore omit the technical details of the
calculation here, referring the reader to \cite{DLS02} for a full
account.

The way in which the free energy functional is obtained in
\cite{DLS01,DLS02} is to introduce a pair of fugacities for each box,
one of which controls its mean size $y_i$ and the other its density
$\rho_i$.  This procedure is similar to the use of the grand canonical
ensemble described in Section~\ref{gce}, albeit with a much larger
number of fugacities, and yields the generating function of the joint
distribution that extends to the nonequilibrium case the expression
(\ref{free:boxes}).  The values of the fugacities are then fixed by
finding the saddle-point of this generating function. It
turns out that the resulting expression for the free energy functional,
\begin{equation}
\label{free:ssep}
{\cal F}[\rho] = \int_0^1 \rmd x \left[ \rho(x) \ln
  \frac{\rho(x)}{\sigma(x)} + (1 - \rho(x)) \ln \frac{1 - \rho(x)}{1-\sigma(x)} +
  \ln \frac{\sigma'(x)}{\sigma(1)-\sigma(0)} \right] \;,
\end{equation}
involves a companion function $\sigma(x)$ which is related to the
fugacities described above, and that is fixed by the saddle-point
analysis as the function that maximises the functional ${\cal
F[\rho]}$ for a given profile $\rho(x)$.  One can verify, by asking
for a vanishing functional derivative $\frac{\delta {\cal F}}{\delta
\sigma (x)}$, that this function must satisfy the nonlinear differential
equation
\begin{equation}
\label{free:comp}
\sigma(x) + \sigma(x)(1-\sigma(x)) \frac{\sigma''(x)}{\sigma'(x)^2} = \rho(x) \;.
\end{equation}
Additionally, the detailed analysis presented in \cite{DLS01,DLS02}
shows that $\sigma(x)$ must further be a monotonic function, and satisfy
the boundary conditions
\begin{equation}
\sigma(0) = \frac{\alpha}{\alpha+\gamma} \quad\mbox{and}\quad
\sigma(1) = \frac{\delta}{\delta+\beta} \;.
\end{equation}
We note that the monotonicity requirement on $\sigma(x)$ ensures that
the argument of the final logarithm appearing in (\ref{free:ssep}) is
positive, and further that the value of the companion function
$\sigma(x)$ at some point $x$ will in general depend on the
\emph{entire} profile $\rho(x)$ through the differential equation
(\ref{free:comp})---the free energy density is non-local, as
previously claimed.

\subsection{Effective local thermal equilibrium and additivity principle}
\label{easyway}
The companion function $\sigma(x)$ that appears in the nonequilibrium
free energy functional (\ref{free:ssep}) is somewhat mysterious at a
first encounter.  However, a physical meaning can be ascribed to this
function, starting from the observation that the values imposed at the
boundaries coincide with the mean densities of the left and right
boundary sites in the steady state, and hence also the densities of
the particle reservoirs at each end of the system.  In the bulk, one
finds that an infinitesimal segment of the system located at a point
$x \in (0,1)$ can be thought of as being  in local equilibrium with a reservoir at density
$\sigma(x)$.

To see this, it is convenient first to generalise the functional
(\ref{free:ssep}) to a density profile $\rho(x)$ defined on an
arbitrary interval $x \in [a,b]$ in such a way that if the profile is
translated and scaled so that it covers a different interval $x \in
[a',b']$, the free energy per unit length is unchanged.  The
functional that has this property and coincides with (\ref{free:ssep})
for the case $a=0,b=1$ is
\begin{equation}
\label{free:ssepg}
\fl {\cal F}_{[a,b]}([\rho]; \sigma(a), \sigma(b)) = \int_a^b
\rmd x \left[ \rho(x) \ln \frac{\rho(x)}{\sigma(x)} + (1 - \rho(x)) \ln
\frac{1 - \rho(x)}{1-\sigma(x)} + \ln \frac{(b-a)\sigma'(x)}{\sigma(b)-\sigma(a)}
\right] \;.
\end{equation}
In the limit of an infinitesimal interval, $b \to a$, this expression
approaches the corresponding expression for the equilibrium system
\begin{equation}
\label{free:eqmg}
{\cal F}^{\rm eq}_{[a,b]}([\rho]; \bar{\rho}) = \int_a^b
\rmd x \left[ \rho(x) \ln \frac{\rho(x)}{\bar{\rho}} + (1 - \rho(x)) \ln
\frac{1 - \rho(x)}{1-\bar{\rho}}
\right]
\end{equation}
if we identify $\sigma(a)$ with $\bar{\rho}$, the density of the
reservoirs coupled to the equilibrium system.

The observation that an effective local thermal equilibrium applies does
not itself allow one to recover the full expression (\ref{free:ssepg})
for the free energy functional since \textit{a priori} one does not
know what densities the effective intermediate reservoirs should
have.  However, it was noticed in \cite{DLS02} that one can construct
the free energy functional through an additivity principle that
involves the modified free energy
\begin{equation}
\label{free:mod}
{\cal H}_{[a,b]}([\rho]; \sigma(a), \sigma(b)) = {\cal
 F}_{[a,b]}([\rho]; \rho_a,\rho_b) + (b-a) \ln J(\sigma(a), \sigma(b))
\end{equation}
where $J(\sigma(a), \sigma(b))$ is the steady-state current through a
system of length $b-a$ coupled to a reservoir of density $\sigma(a)$
at the left boundary, and of density $\sigma(b)$ at the right.  For
the SSEP, this current is
\begin{equation}
J = \frac{\sigma(b) - \sigma(a)}{b - a} \;,
\end{equation}
where we have assumed that $\sigma(b) > \sigma(a)$.  (Since the
dynamics are symmetric, the case $\sigma(a)> \sigma(b)$ can be treated
by making the replacement $x \to 1-x$.)  The additivity principle
given in \cite{DLS02} relates the free energy functional for a system
to that of two subsystems created by inserting a reservoir at a point
$y \in (a,b)$.  It reads
\begin{equation}
\label{free:add}
\fl {\cal H}_{[a,b]}([\rho]; \sigma(a), \sigma(b)) = \max_{\sigma(y)} 
\left\{ {\cal H}_{[a,y]}([\rho]; \sigma(a), \sigma(y)) + {\cal
  H}_{[y,b]}([\rho]; \sigma(y), \sigma(b)) \right\} \;.
\end{equation}
That is, $\sigma(y)$ is the density one should choose for the
reservoir placed at the point $y$ so that the combined free energy of
the two subsystems is maximised.  One can verify by recursively
subdividing the subsystems created by inserting intermediate
reservoirs, and by assuming that each infinitesimal segment created
in this way is in a local equilibrium with its boundary reservoirs,
that the expression (\ref{free:ssep}) results.  The fact that one is
looking for a maximum in (\ref{free:add}) further implies that the
reservoir densities $\sigma(x)$ must satisfy the differential equation
(\ref{free:comp}).  Thus the additivity principle (\ref{free:add}) in
tandem with the effective local equilibrium property and the free energy
functional given by (\ref{free:comp}) are
equivalent.

\subsection{Properties and applications of the free energy functional
  for the SSEP}
\label{SSEPprops}

It is worth reiterating  a few properties exhibited by the free energy
functional (\ref{free:ssep}) noted in \cite{DLS02}.  First, as is required, the equilibrium
version (\ref{free:eqm}) is recovered in the limit where the right
boundary reservoir density $\sigma(1)$ approaches that of the left
$\sigma(0)$.  To see this, we write (without loss of generality) for
the effective reservoir density in the bulk
\begin{equation}
\sigma(x) = \sigma(0) + [ \sigma(1) - \sigma(0) ] x + \delta\sigma(x) \;.
\end{equation}
If one substitutes this expression into the differential equation
(\ref{free:comp}) that governs $\sigma(x)$, one finds that the
function $\delta\sigma(x)$ must be proportional to
$[\sigma(1)-\sigma(0)]^2$ if $\sigma(x)$ is to remain bounded as
$\sigma(1) \to \sigma(0)$.  In turn, this implies that $\sigma(x) \to
\sigma(0) = \sigma(1)$ in this limit, yielding the desired equilibrium
free energy functional, with $\bar{\rho}$ equal to both boundary
reservoir densities.

It is also straightforward to show that the most probable profile
coincides with the stationary profile, which for the SSEP is simply
the linear function $\bar{\rho}(x) = \sigma(0) + [\sigma(1)-\sigma(0)]
x$.  This is achieved by asking for the functional derivative of
(\ref{free:ssep}) with respect to $\rho(x)$ to vanish.  That is,
\begin{equation}
\label{free:opt}
\frac{\delta {\cal F}[\rho]}{\delta \rho(x)} = \ln \left[
  \frac{\rho(x)}{\sigma(x)} \, \frac{1-\sigma(x)}{1-\rho(x)}
  \right] = 0 \;,
\end{equation}
and hence we must have $\sigma(x)=\rho(x)$.  The differential equation
(\ref{free:comp}) then implies that the second derivative of
$\sigma(x)$ must vanish, which (along with the boundary conditions)
results in the optimal $\sigma(x)$ and $\rho(x)$ coinciding with the
linear stationary profile.  It is easy to see that then
(\ref{free:ssep}) is zero, indicating that this profile appears with
probability $1$ in the thermodynamic limit; it can also be shown
\cite{DLS02} that any other macroscopic profile appears with a
probability exponentially small in  the number of lattice sites.

It is interesting to compare the magnitude of the free energy
functional for a given profile $\rho(x)$ with that for an equilibrium
system that has the same stationary profile.  This can be realised by
coupling a one-dimensional chain of sites along its length to a series
of particle reservoirs that have their chemical potentials tuned in
such a way that the desired equilibrium density $\bar{\rho}(x)$ in a
small interval $x \in [a,b]$ is attained.  Since the equilibrium free
energy density is local, the free energy functional for the whole
system is obtained by summing (\ref{free:eqmg}) over all these small
intervals.  This gives
\begin{equation}
{\cal F}^{\rm eq}[\rho] = \int_0^1 \rmd x \left[ \rho(x) \ln
  \frac{\rho(x)}{\bar{\rho}(x)} + (1 - \rho(x)) \ln
  \frac{1-\rho(x)}{1-\bar{\rho}(x)} \right] \;.
\label{free:eq}
\end{equation}
Now (\ref{free:ssep}) reduces to the equilibrium expression
(\ref{free:eq}) when $\sigma(x) = \bar{\rho}(x)$, the equilibrium
profile (which is linear).  However, the nonequilibrium free energy is
obtained by maximising the expression (\ref{free:ssep}) with respect
to $\sigma$ and generally the maximum is not realised by $\sigma(x) =
\bar{\rho}(x)$. Therefore ${\cal F}[\rho] \ge {\cal F}^{\rm eq}[\rho]$
which means that the probability of witnessing a particular
fluctuation away from the optimal profile for a system driven out of
equilibrium is suppressed compared to that for an equilibrium system
with the same optimal profile.  This is not always the case however: a
similar analysis for the asymmetric simple exclusion process (see
below) shows that fluctuations can also be enhanced relative to the
equilibrium state.

A final further application of the free energy functional is to
explore the optimal profiles in the nonequilibrium system after
imposing additional global constraints.  For example, one can ask for
the most likely profile given that the overall density $\bar{\rho} =
\int_0^1 \rmd x \rho(x)$ is fixed.  It turns out \cite{DLS02} that
the the most likely profile has an exponential form and, unless the
overall imposed density happens to be equal to the equilibrium density
(in which limit the linear profile is recovered), the density 
is discontinuous at the 
boundaries $x=0,1$.

\subsection{Free energy functional for the partially asymmetric simple
  exclusion process (PASEP)}

Free energy functionals have also been derived for the partially
asymmetric exclusion process (PASEP) in the case where entry and exit
of particles occurs at the left and right boundaries respectively, and
the bulk bias is to the right (i.e., the forward-bias regime, $q<1$)
\cite{DLS02b,DLS03}.  In principle, one could follow through the procedure
outlined above in Subsection~\ref{hardway}, where one starts with the
full stationary distribution of microstates and takes the continuum
limit.  It turns out to be more straightforward instead to use the
microscopic distribution to extend the additivity principle discussed
in Subsection~\ref{easyway}.  Then, by the effective local equilibrium
property, one can construct the full free energy functional as before.

When the bulk dynamics are asymmetric, it turns out that two versions
of the additivity principle come into play.  One applies when the left
boundary reservoir density $\sigma(0)$ is less than that at the right;
the other when the reverse is true.  In both cases, the additivity
relation involves the modified free energy defined by
Equation~(\ref{free:mod}), but where now the current is that found (for
example) by an application of the extremal current principle discussed
in Section~\ref{asep}.  That is,
\begin{equation}
J(\sigma,\sigma') = \left\{ \begin{array}{ll}
\displaystyle \min_{\rho \in [\sigma, \sigma']} \rho(1-\rho) & \sigma \le \sigma' \\
\displaystyle \max_{\rho \in [\sigma', \sigma]} \rho(1-\rho) & \sigma \ge \sigma'
\end{array} \right. \;.
\end{equation}

When $\sigma(0) \ge \sigma(1)$, the additivity formula is the same as
for the symmetric case, Equation~(\ref{free:add}) (see \cite{DLS03} for
details of the calculation).  Using the local equilibrium property,
one then finds the free energy functional to be
\begin{eqnarray}
\label{free:asepd}
\fl {\cal F}_{[a,b]}([\rho]; \sigma(a), \sigma(b)) = -(b-a) \ln
J(\sigma(0),\sigma(1)) + {}\nonumber\\
\!\!\! \max_{\sigma(x)} \int_a^b \rmd x \left[ \rho(x) \ln (\rho(x)[1-\sigma(x)]) + (1
- \rho(x)) \ln [(1 - \rho(x))\sigma(x)] \right] \;.
\end{eqnarray}
Again, the companion function $\sigma(x)$ that gives the effective reservoir
densities in the bulk must match the actual reservoir densities at the
boundaries, and must also be a nonincreasing function.

When $\sigma(0) \le \sigma(1)$, the additivity formula takes on the
different form
\begin{equation}
\fl
{\cal H}_{[a,b]}([\rho]; \sigma(a), \sigma(b)) = \min_{\rho_y \in
  \{\sigma(a), \sigma(b)\}} \left\{ {\cal H}_{[a,y]}([\rho];
  \sigma(a), \sigma(y)) + {\cal H}_{[y,b]}([\rho]; \sigma(y),
  \sigma(b)) \right\} \;,
\end{equation}
where again we refer the reader to \cite{DLS03} for the details of the
calculation.  The fact that the intermediate reservoir inserted at the
point $y$ always takes on a density that is equal to that of one of
the boundary reservoirs implies that effective reservoir density is
constant on one side of the point $y$.  Further subdivision on that
side then has no effect.  On the other side the situation is the same
as for the previous subdivision: on one side of the division, the
density will be constant and equal to that of the appropriate boundary
reservoir, whereas on the other further subdivisions will be required.
The upshot of this is that the reservoir density function $\sigma(x)$ will
be a step function, taking the value $\sigma(x)=\sigma(a)$ for $a<x<y$ and
$\sigma(x)=\sigma(b)$ for $y<x<b$ for some value $y$.  This value will be such
that the overall free energy is minimised, that is
\begin{eqnarray}
\fl {\cal F}_{[a,b]}([\rho]; \sigma(a), \sigma(b)) = -(b-a) \ln
J(\sigma(a), \sigma(b)) + {}\nonumber\\
\!\!\! \min_{a\le y\le b} \Bigg\{\int_a^y \rmd x \left[ \rho(x) \ln (\rho(x)[1-\sigma(a)]) + (1
- \rho(x)) \ln [(1 - \rho(x))\sigma(a)] \right] + {}\nonumber\\
\hspace{3em} \int_y^b \rmd x \left[ \rho(x) \ln (\rho(x)[1-\sigma(b)]) + (1
- \rho(x)) \ln [(1 - \rho(x))\sigma(b)] \right] \Bigg\} \;.
\end{eqnarray}
Note that despite considerable cosmetic differences between this
formula and (\ref{free:asepd}), the two expressions are in fact very
similar: in both cases, one needs to find the set of reservoir
densities $\sigma(x)$ that leads to an extremum of the same joint
functional of $\rho(x)$ and $\sigma(x)$.  The fact that in one case,
the additivity principle involves a maximum and in another a minimum
has its origins in the nature of the saddle-point which provides the
relevant asymptotics \cite{Derrida07}.

We also note that the bias parameter $q$ does not enter explicitly
into these equations, only implicitly through the relationship between
the boundary reservoir densities $\sigma(a)$ and $\sigma(b)$, and the
microscopic transition rates $\alpha, \beta, \gamma, \delta$ and $q$.
Specifically
\begin{equation}
\sigma(a) = \frac{\alpha}{1-q} \quad \mbox{and} \quad
\sigma(b) = 1- \frac{\beta}{1-q} \;,
\end{equation}
and so $q$ affects phase boundaries in the $\alpha$-$\beta$ plane, as
was seen in Section~\ref{pasep}.  The $q$-independence of the free
energy functionals does not contradict the rich structure seen in the
density profiles in Section~\ref{pasep}, since the deviations from the
bulk density at the boundaries vanish under rescaling in the
$N\to\infty$ limit.  This lack of $q$-dependence also implies that
(\ref{free:ssep}) is not obtained as $q\to 1$: in other words, the
limits $N\to\infty$ and $q\to 1$ do not commute (as we have already
seen).  The free energy functional
in the weakly asymmetric limit $1-q = \lambda/N$
has been computed explicitly using the matrix product approach
and on this scale a $q$ (or $\lambda$) dependence
is apparent \cite{ED04b}.

Closer scrutiny of these free energy functionals for the PASEP
\cite{DLS03} shows similar properties to those seen for the SSEP above
in Subsection~\ref{SSEPprops}.  For example, one finds that the
stationary profile (here, a constant profile after rescaling in the
thermodynamical limit) is also the most likely profile, except along
the boundary between the high- and low-density phases along which---as
has been discussed before---there is a superposition of shocks with
the shock location distributed uniformly across the system.  Then, any
one of these shock profiles is found to minimise the free energy
functional.  Again, one can compare the relative size of a fluctuation
away from the most likely profile in the nonequilibrium system, and an
equilibrium system coupled to a reservoir with a spatially-varying
chemical potential.  In the case where $\sigma(a) > \sigma(b)$, it is
found that (as for the SSEP), such fluctuations are suppressed;
however when $\sigma(a)<\sigma(b)$ the opposition between the boundary
densities and the bulk bias results in these fluctuations being
enhanced.

A final interesting feature of the PASEP is that density fluctuations in the
maximal current phase are non-Gaussian.  Evidence for this is provided
by the existence of a discontinuity at a density of $\rho=\frac{1}{2}$
in the probability of witnessing a density $\rho$ in a box located
somewhere in the bulk of the system.  That the distribution is indeed
non-Gaussian is confirmed by explicit calculations for the totally
asymmetric case ($q=0$) \cite{DLS03,DEL04} that exploit the
relationship to equilibrium surface models outlined in
Subsection~\ref{surf}.

\subsection{Remarks}

In this section we have seen that once one has found a function that
is additive when two subsystems are connected together via a reservoir
with an appropriate density, the distribution of density profile
macrostates can be constructed given the existence of a local
equilibrium property.  This is an appealing approach, and one hopes
that it can be applied to a much wider range of systems than exclusion
processes.  The difficulty is, however, that one does not know
\textit{a priori} what form the additivity principle should take:
here, we have relied on the complete knowledge of the underlying
distribution of microstates to construct it.  Nevertheless, it is
worth remarking that the free energy functional for the symmetric
exclusion process (\ref{free:ssep}) has been obtained in a purely
macroscopic formalism \cite{BDSGJLL02}.  In this approach it is
assumed that, in the combined thermodynamic and continuum limit, the
macroscopic density profile evolves deterministically as a consequence
of the law of large numbers.  Then, it can be shown that the
probability of witnessing a deviation $\rho(x)$ from the most likely
profile $\bar{\rho}(x)$ in the steady state is given by a functional
of the most likely trajectory through phase space that begins at time
$t=-\infty$ at stationarity $\bar{\rho}(x)$ and is constrained to
reach $\rho(x)$ at $t=0$ \cite{BDSGJLL01}.


\section{Two-species Models with quadratic algebra}
\label{2species}

So far we have seen exact matrix product solutions for the steady
state of two classes of model systems with nonequilibrium dynamics:
the open boundary ASEP and PASEP---various aspects of which were
discussed in detail in Sections~\ref{asep}, \ref{open}, \ref{pasep}
and \ref{free}---which had a single species of particles hopping on a
one-dimensional lattice with open boundaries; a two-species model with
periodic boundary conditions, which we examined in Section~\ref{ring}.
In this section we are going to search for $2$-species exclusion
models that can be solved using the matrix product approach.  For
clarity, we reiterate that in this work the number of species $n$
relates to the number of particle species, excluding vacancies.

It is not known how to perform an exhaustive search of \emph{all}
possible matrix product steady states, but it is possible to search a
restricted subset where the matrix products involved can be
systematically {\em reduced} using expressions like \RR\ and \RRR.  To this
end, let us recall the proof of the reduction relations that applied
for exclusion models on the ring geometry given in
Subsection~\ref{RRRproof}.  This proof concerned stationary weights
that were given by a trace of matrices
\begin{equation}
\label{2s:f}
f(\tau_1, \tau_2, \ldots, \tau_N) = \tr ( X_{\tau_1} X_{\tau_2} \cdots
X_{\tau_N} )
\end{equation}
where the variable $\tau_i=0,1,2,\ldots, n$ indicates the species of
particle occupying site $i$ ($\tau_i=0$ denotes a vacancy) and
$X_{\tau_i}$ is the corresponding matrix whose forms is to be
determined.  We showed that when $W(\tau_i \tau_{i+1} \to \tau_{i+1}
\tau_{i})$ is the rate at which particles on sites $i$ and $i+1$
exchange places, if  one can find auxiliary matrices $\tilde{X}_i$
such that
\begin{equation}
\label{2s:m}
\fl W(\tau_{i+1}\tau_i \to \tau_i\tau_{i+1}) X_{i+1} X_i \nonumber -
W(\tau_i\tau_{i+1} \to \tau_{i+1}\tau_i) X_{i} X_{i+1} = \tilde{X}_i
X_{i+1} - X_i \tilde{X}_{i+1}
\end{equation}
is satisfied, the weights given by (\ref{2s:f}) are stationary.

If we are to obtain matrix reduction relations, we require that the
auxiliaries $\tilde{X}_i$ are scalars (rather than matrices).  We must
also insist that these reduction relations describe an associative
algebra, i.e., that no matter what order the reduction relations are
applied, one always ends up with the same sum of irreducible strings.
The various ways in which this can be be achieved  were formalised, classified and catalogued by
Isaev, Pyatov and Rittenberg \cite{IPR01}.
In this section, we
briefly outline the classification scheme for the
the case of two
particle species, $n=2$,
and show what physical
dynamics the various possibilities correspond to.  We then move on to
discuss geometries other than the ring.  

\subsection{Classification of Isaev, Pyatov and Rittenberg for
  two-species models}
\label{IPR}
In this section we will mostly use the notation established above,
where vacancies are denoted by $0$ and particles by $1$ and $2$.
Occasionally, it will be helpful to state model dynamics using a
natural ``charge representation'', where particles of species 1 and
vacancies are relabelled as positive and negative charges and
particles of species 2 as vacancies:
\begin{equation}
1 \mapsto + \qquad 2 \mapsto 0 \qquad  0 \mapsto - \; .
\label{chargerep}
\end{equation}
Note that our notation differs from that used by \cite{IPR01}.  We
will also use  a shorthand for the transition rates
\begin{equation}
w_{\tau \tau'} \equiv W(\tau \tau'\to \tau' \tau)
\end{equation}
and, since we are assuming the auxiliary matrices are scalars, we will
write them as $\tilde{X}_\tau = x_\tau$.

We first restrict ourselves to the case $w_{12}>0$, $w_{20} >0$ and
$w_{12} > 0$, i.e. there are exchanges in at least one direction
between particles and vacancies and between the two species of
particles.  Then, the relations (\ref{2s:m}) become
\begin{eqnarray}
w_{10} X_1 X_0 - w_{01} X_0 X_1 &=& x_0 X_1 - x_1 X_0 \nonumber \\
w_{20} X_2 X_0 - w_{02} X_0 X_2 &=& x_0 X_2 - x_2 X_0 \nonumber \\
w_{12} X_1 X_2 - w_{21} X_2 X_1 &=& x_2 X_1 - x_1 X_2 
\label{sqa}
\end{eqnarray}

It remains to check whether the relations (\ref{sqa}) are consistent.
The approach of Isaev, Pyatov and Rittenberg \cite{IPR01} (see also
that of Karimipour \cite{Karimipour99}) is to generalise the reduction
process leading to (\ref{int:reord}).  That is, one seeks to use
(\ref{sqa}) to reduce any product of matrices to a sum of irreducible
strings.  First we fix the order of irreducible strings as $X_0^l
X_2^m X_1^n$.  Then we require that reducing an arbitrary product $U$
of the matrices $X_i$ to a sum of irreducible strings
\begin{equation}
U =  \sum_{l,m,n} a_{l,m,n} X_0^l X_2^m X_1^n
\end{equation}
is independent of the order in which the reduction is carried out
using the elementary rules (\ref{sqa}).  Reduction rules which satisfy
this requirement are referred to by Isaev, Pyatov and Rittenberg as
\emph{PBW-type algebras} \cite{Bergman78}.

For the two-species case the requirement 
is that the two possible reductions of the
product $X_1 X_ 2 X_0$, illustrated schematically below, give
equivalent results.  \unitlength=6mm
\begin{picture}(17,4)
\put(1.5,1.5){$  X_1 X_2 X_0$}
\put(4.5,2){\vector(1,1){1}}
\put(4.5,1.4){\vector(1,-1){1}}
\put(6,3){$X_1 X_0 X_2$}
\put(6,0){$X_2 X_1 X_0$}
\put(9,3.2){\vector(1,0){1}}
\put(9,0.2){\vector(1,0){1}}
\put(10.5,3){$X_0 X_1 X_2$}
\put(10.5,0){$X_2 X_0 X_1$}
\put(13.5,3){\vector(1,-1){1}}
\put(13.5,0.4){\vector(1,1){1}}
\put(15,1.5){$X_0 X_2 X_1$}
\end{picture}
\vspace*{2ex}

\noindent Calculating explicitly using (\ref{sqa}) yields \cite{IPR01}
\begin{eqnarray}
\nonumber
&&
x_1 \, w_{0 2} ( 
w_{12}-w_{21}-w_{10}+w_{01}) X_0 X_2  +
x_2 \, w_{0 1} ( 
w_{12}-w_{21}+w_{20}-w_{02} ) X_0 X_1
\\
\nonumber
&&+\ x_0 \, w_{2 1} ( 
w_{20}-w_{02}-w_{10}+w_{01} ) X_2 X_1
+ x_1 x_2 ( w_{12}-w_{21}+ w_{20}-w_{10}  ) \, X_0
\\
\label{rel1}
&&
+\ x_1 x_0 ( w_{2 1} - w_{0 2} ) \, X_2
+ x_2 x_0 (w_{10}-w_{12} -w_{20}+w_{02}) \, X_1\ =\ 0
\; ,
\end{eqnarray}
which  results in the following six  conditions on the
hopping rates $w_{ij}$ from the requirement that each of the
six terms in the above equation vanish:
\begin{eqnarray}
\label{PBW1}
x_1 \, w_{0 2} ( 
w_{12}-w_{21}-w_{10}+w_{01}) =0\;,
\\
\label{PBW2}
x_2 \, w_{0 1} ( 
w_{12}-w_{21}+w_{20}-w_{02} ) =0\;,
\\
\label{PBW3}
x_0 \, w_{2 1} ( 
w_{20}-w_{02}-w_{10}+w_{01} )=0\;,
\\
x_1 x_2 ( w_{12}-w_{21}+ w_{20}-w_{10}  )=0\;,
\label{PBW4} 
\\
\label{PBW5}
 x_1 x_0 ( w_{2 1} - w_{0 2} )=0\;,
\\
\label{PBW6}
x_2 x_0 (w_{10}-w_{12} -w_{20}+w_{02})  = 0\;.
\end{eqnarray}
The solutions of these equations were classified in \cite{IPR01}. Here
we summarise the various nontrivial solutions which are physically
relevant and the corresponding models which have been previously
studied in the literature. The classification hinges  on how
many of the $x_i$ are zero.

\subsubsection{Class A (includes Karimipour's model of overtaking)} 

The first class of solutions has none of the $x_i=0$.  There are then
two possible solutions to (\ref{PBW1})--(\ref{PBW6}).

\begin{itemize}
\item \textbf{Solution AI}\quad $w_{ij} = w\quad \forall i,j$. Physically, this
corresponds to symmetric exclusion with two species which, although
labelled distinctly, have identical dynamics. 
Matrices  are easily found by setting $X_2 = (x_2/x_1)X_1$ which reduces 
(\ref{sqa}) to the  single species  condition
$\ds w(X_1 X_0 - X_0 X_1)  = x_0 X_1 - x_1 X_0$.
On the ring this system
has a simple steady state where all allowed configurations are
equally likely.
\item \textbf{Solution AII}\quad $w_{ij}= v_i-v_j >0$ for $i<j$ and $w_{ij}=
0$ for $i>j$.  This corresponds to a model where the hop rates are
totally asymmetric and of the form
\begin{equation}
1\,0 \mathop{\rightarrow}\limits^{v_1}\, 0\,1 \qquad
2\,0 \mathop{\rightarrow}\limits^{v_2}\, 0\,2 \qquad
1\,2 \mathop{\rightarrow}\limits^{v_1-v_2}\, 2\,1 \;.
\label{AIIdyn}
\end{equation}
That is, since $v_1>v_2$, a particle of species 1 is faster than a
particle of species 2 and overtakes with rate $v_1-v_2$.  The
corresponding matrix algebra is
\begin{eqnarray}
v_1 X_1 X_0 &=& x_0 X_1 - x_1 X_0 \nonumber\\
v_2 X_2 X_0 &=& x_0 X_2 - x_2 X_0 \nonumber\\
(v_1-v_2) X_1 X_2 &=& x_2 X_1 - x_1 X_2 \;.
\label{AIIalg}
\end{eqnarray}
 This solution of (\ref{PBW1})--(\ref{PBW6})
was first noted and studied by Karimipour \cite{Karimipour99} and
followed up in \cite{KK00}.  As it happens, this model can be
generalised to more than two species, and as such will be discussed in
its full generality in Section~\ref{sec:Karimipour}.
Representations are given in Appendix~\ref{appreps} Equation (\ref{Karimrep}).
\end{itemize}

\subsubsection{Class B (includes asymmetric second-class particle dynamics)}

This  second class of solutions has one of $x_i=0$.  The first two
families of solutions within this class are obtained by taking
$x_2=0$, and then choosing the hop rates to satisfy (\ref{PBW1}),
(\ref{PBW3}) and (\ref{PBW5}) in two different ways.

\begin{itemize}
\item \textbf{Solution BI}\quad  $w_{12}=w_{20} = p_2$,  $w_{21}=w_{02}=q_2$
, $w_{10}=p_1$ and $w_{01}=q_1$ with $p_2-q_2=p_1-q_1$.  This
  corresponds to a model with the hop rates
\begin{equation}
1\,0 \mathop{\rightleftharpoons}\limits^{p_1}_{q_1}\, 0\,1 \qquad
2\,0 \mathop{\rightleftharpoons}\limits^{p_2}_{q_2}\, 0\,2 \qquad
1\,2 \mathop{\rightleftharpoons}\limits^{p_2}_{q_2}\, 2\,1 
\label{BIdyn}
\end{equation}
or, in the charge representation,
\begin{equation}
+\,- \mathop{\rightleftharpoons}\limits^{p_1}_{q_1}\, -\,+ \qquad
0\,- \mathop{\rightleftharpoons}\limits^{p_2}_{q_2}\, -\,0 \qquad
+\,0 \mathop{\rightleftharpoons}\limits^{p_2}_{q_2}\, 0\,+   \qquad.
\end{equation}
This model is thus an asymmetric generalisation of the second-class
particle of Section~\ref{ring} and was first used by \cite{DJLS93}
for the special case $q_2=q_1$, $p_2=p_1$.  The corresponding matrix
algebra is
\begin{eqnarray}
p_1 X_1 X_0 - q_1 X_0 X_1 &=& x_0 X_1 -x_1 X_0 \nonumber\\
p_2 X_2 X_0 - q_2 X_0 X_2 &=& x_0 X_2 \nonumber\\
p_2 X_1 X_2 - q_2 X_2 X_1 &=& -x_1 X_2 
\label{BIalg}
\end{eqnarray}
A representation of this algebra is given in Appendix~\ref{appreps}.
\item \textbf{Solution BII} $w_{02}=w_{21}=0$, $w_{12} =\alpha$,
  $w_{10}=p$, $w_{01}=q$ and $w_{12}=\beta$.  The corresponding model
  has hop rates
\begin{equation}
1\,0 \mathop{\rightleftharpoons}\limits^{p}_{q}\, 0\,1 \qquad
2\,0 \mathop{\rightarrow}\limits^{\alpha}\, 0\,2 \qquad
1\,2 \mathop{\rightarrow}\limits^{\beta}\, 2\,1  
\end{equation}
or, in the charge representation,
\begin{equation}
+\,- \mathop{\rightleftharpoons}\limits^{p}_{q}\, -\,+ \qquad
0\,- \mathop{\rightarrow}\limits^{\alpha}\, -\,0 \qquad
+\,0 \mathop{\rightarrow}\limits^{\beta}\, 0\,+   \qquad .
\label{BIIdyn}
\end{equation}
This model is another generalisation of the second-class problem first
studied with $p=1$, $q=0$ for a single species two particle
\cite{Mallick96} and later generalised in \cite{AHR98}.  The matrix
algebra is
\begin{eqnarray}
p X_1 X_0 - q X_0 X_1 &=& x_0 X_1 -x_1 X_0 \nonumber\\
\alpha X_2 X_0  &=& x_0 X_2 \nonumber\\
\beta  X_1 X_2 &=& -x_1 X_2 \;.
\label{BIIalg}
\end{eqnarray}
This algebra can be mapped onto that used to solve the PASEP with open
boundaries, \PRR, if one takes 
\begin{equation}
X_2 = \ket{V}\bra{W}
\end{equation}
and  $x_0=1$, $x_1=-1$.  
Thus one can make use of results from Section~\ref{pasep},
along with techniques for models on the ring described in
Section~\ref{ring} to analyse various cases of this model \cite{AHR98,AHR99,
Jafarpour00,Jafarpour00b,Sasamoto00a,RSS00}.
\end{itemize}

Choosing $x_1$ or $x_3$ to be zero results in the same solution as BII
under relabelling particles, and one additional solution when
$x_1=0$.  This is
\begin{itemize}
\item \textbf{Solution BIII}\quad $w_{01}=w_{21}=0$, $w_{10}=\alpha$,
  $w_{21}=\beta$, $w_{20}=p$, $w_{02}=q$ and $p-q=\alpha-\beta$.  The
  corresponding model has hop rates
\begin{equation}
1\,0 \mathop{\rightarrow}\limits^{\alpha}\, 0\,1 \qquad
2\,0 \mathop{\rightleftharpoons}\limits^{p}_{q}\, 0\,2 \qquad
1\,2 \mathop{\rightarrow}\limits^{\beta}\, 2\,1 
\end{equation}
and the matrix algebra is
\begin{eqnarray}
\alpha X_1 X_0   &=& x_0 X_1  \nonumber\\
p X_2 X_0-q X_0 X_2  &=& x_0 X_2 - x_2 X_0 \nonumber\\
\beta  X_1 X_2 &=& x_2 X_1  \;.
\label{BIIIalg}
\end{eqnarray}
However on the ring this results in a steady state in which all
allowed configurations are equally likely: as can be seen from
(\ref{BIIIalg}) in any periodic string of matrices, all $X_2$ and
$X_0$ can be eliminated through the first and third relations.
\end{itemize}

\subsubsection{Class C (includes non-overtaking two-species dynamics)}

The third class of solutions is obtained by taking two of the scalar
quantities $x_i$ equal to zero.  Equations (\ref{PBW4})--(\ref{PBW6})
and two of (\ref{PBW1})--(\ref{PBW3}) are then automatically
satisfied.  The equation that remains has a structure that is
independent of which of the $x_i$ are taken to be nonzero, so we take
$x_1=x_2=0$ which leaves us to satisfy (\ref{PBW3}).  This can be done
in two ways.

\begin{itemize}
\item \textbf{Solution CI} $w_{10}=p_1$, $w_{01}=q_1$, $w_{20}=p_2$, 
$w_{02}=q_2$ and $p_1-q_1 = p_2-q_2$.  This corresponds to a model in
  which all six exchanges may take place
\begin{equation}
1\,0 \mathop{\rightleftharpoons}\limits^{p_1}_{q_1}\, 0\,1 \qquad
2\,0 \mathop{\rightleftharpoons}\limits^{p_2}_{q_2}\, 0\,2 \qquad
1\,2 \mathop{\rightleftharpoons}\limits^{w_{12}}_{w_{21}}\, 2\,1 
\end{equation}
and that has the matrix algebra
\begin{eqnarray}
p_1 X_1 X_0 - q_1 X_0 X_1 &=& x_0 X_1  \nonumber\\
p_2 X_2 X_0 - q_2 X_0 X_2 &=& x_0 X_2 \nonumber\\
w_{12} X_1 X_2 - w_{21} X_2 X_1 &=& 0  \;.
\label{CIalg}
\end{eqnarray}
Unfortunately this algebra is not useful in describing a physical
system with periodic boundary conditions.  To see this one can take
for example a representation where $X_0 = \One$ and $x_0 = p_1-q_1=
p_2-q_2$ which satisfies the first two relations of
(\ref{CIalg}). However the third relation of (\ref{CIalg}) has the
form of a vanishing deformed commutator
\begin{equation}
X_1 X_2 - r X_2 X_1 = 0 \;.
\end{equation}
Thus if we commute say a matrix $X_1$ all the way around the ring we
will end up with the same matrix product multiplied by a factor
$r^{N_2}$ where $N_2$ is the number of species 2 particles on the ring
and $r= w_{21}/w_{12}$.    Thus, only for $w_{21}= w_{12}$ ($r=1$) can we use this
algebra on a periodic system; alternatively, it can be used for
general $r$ on a closed segment as we discuss in Section~\ref{sec:closeg}.
\item \textbf{Solution CII} Equation (\ref{PBW3}) can also be
satisfied by taking $w_{12} =w_{21}=0$ and unrestricted choices for the
remaining rates $w_{10}=p_1$, $w_{01}=q_1$, $w_{20}=p_2$, 
$w_{02}=q_2$.  This has similar dynamics to the previous model, but with
no exchange of species 1 and 2
\begin{equation} 
1\,0 \mathop{\rightleftharpoons}\limits^{p_1}_{q_1}\, 0\,1 \qquad
2\,0 \mathop{\rightleftharpoons}\limits^{p_2}_{q_2}\, 0\,2
\end{equation}
The matrix algebra is similar to (\ref{CIalg}) but with the third relation absent.
In contrast to the previous case, this algebra can be used to describe
the above dynamics on a ring \cite{Evans96}.  This model 
and its generalisation to multiple species will be discussed further in
Section~\ref{sec:multispecies}.  
\end{itemize}

\subsubsection{Class D (includes the cyclically symmetric ABC model)}
\label{sec:ABC}

In this case, we take all $x_1=0$ which leaves all six hop rates free.
\begin{equation}
1\,0 \mathop{\rightleftharpoons}\limits^{w_{10}}_{w_{01}}\, 0\,1 \qquad
2\,0 \mathop{\rightleftharpoons}\limits^{w_{20}}_{w_{02}}\, 0\,2 \qquad
1\,2 \mathop{\rightleftharpoons}\limits^{w_{12}}_{w_{21}}\, 2\,1 
\end{equation}
The algebra is a set of deformed commutators
\begin{eqnarray}
w_{10} X_1 X_0 - w_{01} X_0 X_1 &=& 0 \\
w_{20} X_2 X_0 - w_{02} X_0 X_2 &=& 0 \\
w_{12} X_1 X_2 - w_{21} X_2 X_1 &=& 0 \;.\label{Dalg}
\end{eqnarray}
Explicit forms for $X_0$, $X_1$ and $X_2$ are discussed in
Section~\ref{sec:closeg}. On a periodic system,
as with case CI discussed above, we require that a matrix product is
left invariant after commuting one of the matrices once around the
ring.  The condition for this is that
\begin{equation}
\left[\frac{w_{01}}{w_{10}}\right]^{n_0}
\left[\frac{w_{21}}{w_{12}}\right]^{n_2}
=
\left[\frac{w_{02}}{w_{20}}\right]^{n_0}
\left[\frac{w_{12}}{w_{21}}\right]^{n_1}
=
\left[\frac{w_{10}}{w_{01}}\right]^{n_1}
\left[\frac{w_{20}}{w_{02}}\right]^{n_2}
= 1
\label{Ddbcond}
\end{equation}
where $n_i$ is the number of particles of species $i$.  These
conditions are satisfied for example by the ABC model
\cite{EKKM98a,EKKM98b} in the special case where all particle numbers
are equal $n_0=n_1=n_2$. The fact that the rhs of (\ref{Dalg}) are all zero implies that to use these algebraic relations, detailed balance must hold in the steady state. Therefore the condition
(\ref{Ddbcond}) is actually the condition for 
detailed balance to hold. The corresponding energy turns out to be an interesting long-range function \cite{EKKM98b}.

In the ABC model, the particles and vacancies are relabelled as
\begin{equation}
1 \mapsto A \quad 2 \mapsto B \quad 0 \mapsto C
\end{equation}
and have dynamics
\begin{equation}
A\,B \mathop{\rightleftharpoons}\limits^{q}_{1}\, B\,A \qquad
B\,C \mathop{\rightleftharpoons}\limits^{q}_{1}\, C\,B \qquad
C\,A \mathop{\rightleftharpoons}\limits^{q}_{1}\,  A\,C 
\end{equation}
where we take $q<1$.  This corresponds to the choice of rates
$w_{12}=w_{20}=w_{01}=q$ and $w_{21}=w_{02}=w_{10}=1$.  Note that the
dynamics are invariant under cyclic permutation of the three particle
types, and that $A$ particles tend to move to the left of $B$
particles, $B$ particles tend to move to the left of $C$ particles and
$C$ particles tend to move to the left of $A$ particles.  In the
steady state of this model this results in a strong phase separation
into three domains of $A$, $B$ and $C$ in the order $ABC$.  In the
limit $N \to \infty$ with $q<1$ fixed, the domains are pure in that
far away from the boundary of the domain the probability of finding a
particle of a different species (to that of the domain) tends to zero.
That is, particles from a neighbouring domain may penetrate only a
finite distance.  In the case where we have equal numbers of each
species condition (\ref{Ddbcond}) is satisfied and one can use the
matrix product to calculate the steady state exactly \cite{EKKM98b}.
This model also exhibits an interesting phase transition in the weakly
asymmetric limit where $q$ varies with system size $N$ as $q= \exp(
-\beta/N)$, thus approaching unity in the thermodynamic limit.  Then,
according to the value of $\beta$, the steady state can either order
into three domains which are rich in $A$ $B$ and $C$ (but are not pure
domains) or a disordered phase where the particles are typically in a
disordered configuration \cite{CDE03}.


\subsection{Geometries other than the ring}

In the previous subsection, we summarised a classification of all the
possible sets of dynamics for which a matrix product state with scalar
auxiliaries (i.e., one that has a set of reduction relations) exists
with a unique decomposition into irreducible strings of matrices.  We
also assessed whether these matrix product states could be used for
models on the ring.  We now extend this enquiry to other
one-dimensional geometries that can be constructed.

\subsubsection{Open boundary conditions}
\label{sec:extop}
We first consider open boundary conditions, i.e., those where
particles can enter and leave at the boundaries.  The most general
dynamics is to have rates
$\alpha_{ij}$ at which a particle of species $i$ (or a vacancy if $i=0$)
is transformed into a
particle of species $j$ at site $1$ and rates $\beta_{ij}$ at which a
particle of species $i$ is transformed into a particle of species $j$
(or a vacancy of $j=0$)
at site $N$.  If we are to have statistical weights of the form
(\ref{int:mp}), i.e.,
\begin{equation}
f(\tau_1, \ldots, \tau_N) = \bra{W} X_{\tau_1} X_{\tau_2} \cdots
X_{\tau_N} \ket{V}
\end{equation}
we then obtain additional conditions involving the matrices and
vectors $\bra{W}$ and $\ket{V}$.  These may be written as
\begin{eqnarray}
\sum_{i} \bra{W} \alpha_{ij}X_i =  -\bra{W} x_j \label{alphaij}
 \\
\sum_i \beta_{ij}X_i \ket{W}  =   x_j \ket{W} \label{betaij}
\end{eqnarray}
where  $\alpha_{ii}= - \sum_{j\neq i} \alpha_{ij}$
and $\beta_{ii}= - \sum_{j\neq i} \beta_{ij}$.
Since $\sum_j \alpha_{ij}=\sum_j \beta_{ij} =0$ we require
\begin{equation}
\sum_i x_i =0 \label{xsum}
\end{equation}

In the open boundary case, typically, in addition to the conditions for
the solution classes of Section~\ref{IPR}, one has  further
constraints on the boundary rates to allow conditions
(\ref{alphaij},\ref{betaij}) to be satisfied.  We do not attempt to
catalogue all solutions here, rather we point out which of the
solution classes of Section~\ref{IPR} may, in principle, be used in an
open system and reference some examples.
Some general considerations of
the open boundary conditions for which one has a matrix product
state using  the quadratic algebra (\ref{sqa}) are discussed
in the two species cases in \cite{ADR98}.

\begin{itemize}
\item \textbf{Class A}\quad Solution AI, which
corresponds to symmetric exclusion of two differently labelled
species with identical bulk dynamics, can be
used in the open boundary case where the two species are
injected and extracted with different rates, under certain conditions.
Solution AII can be combined with certain open boundary conditions that were
 found by Karimipour \cite{KK00}. We shall specify these
  boundary interactions and discuss this model more thoroughly in
  Section~\ref{sec:multispecies}.
\item \textbf{Class B}\quad Both solutions BI and BII can be used in
  models with open boundary conditions .
   One case that has been studied is the dynamics
  implied by solution BII with $q=0$ \cite{EFGM95a,EFGM95b,DE96,Arita06}.  
  This is
  sometimes referred to as the bridge model because one has two
  particle species with opposite velocities that slow down to exchange
  places when they meet, rather like cars on a narrow bridge.
  Unfortunately, the full phase diagram of the bridge model---and in
  particular an interesting broken symmetry region---cannot be
  described by a quadratic algebra (i.e., one where the
  auxiliaries are scalars and reduction relations are implied on the steady-state weights). The BII $q\neq 0$ case has also been studied \cite{Kolomeisky97,Uchiyama07}.
\item \textbf{Class C}\quad Recall that Class C comprises solutions
  where only one of the $x_i$ are nonzero.  Hence, it is not possible
  to satisfy that sum rule (\ref{xsum}), and hence the dynamics
  implied by these solutions cannot be solved in an open system using
  a simple matrix product algebra.
\item \textbf{Class D}\quad Algebras within class D can only
be used in models with open boundaries under the special conditions of detailed balance, which generally does not hold.  On the other hand,
as we discuss below, detailed balance \emph{does} apply in
  conservative models with closed boundaries and thus such algebras
  are relevant there---see below.
\end{itemize}

\subsubsection{Closed segment}
\label{sec:closeg}

A closed segment is a one-dimensional lattice of size $N$ with {\em
closed} boundary conditions. That is, particles cannot enter or leave
at the left boundary or right boundaries (sites 1 and $N$).  For the
moment, let us consider conserving models where particles are neither
created nor destroyed in the bulk: the only moves that are allowed are
those where particles or vacancies on neighbouring sites exchange
places.  It is straightforward to show using a Kolmogorov criterion
(\ref{free:kol}) that detailed balance is satisfied in the steady
state of any such model.  The reason for this is that in order to
create a loop in configuration space, every exchange of a pair must be
accompanied by an exchange in the opposite direction (as long as the
dynamics is not totally asymmetric).  Hence the forward loop contains
the same set of exchanges as the reverse, and (\ref{free:kol}) is
satisfied. Some quadratic algebras for closed segments and segments
closed at one end and open at the other were discussed in
\cite{ADR98}.

\begin{figure}
\begin{center}
\includegraphics[scale=0.8]{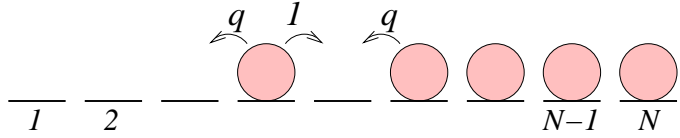}
\end{center}
\caption{\label{fig:pasep-closed} Typical configuration in the steady
  state of the partially asymmetric exclusion process on a closed
  segment of $N$ sites with $q<1$.}
\end{figure}

The simplest example of this type is the partially asymmetric
exclusion process (PASEP) on the closed segment i.e. just one species
of particle \cite{SS94}.  The appropriate matrix algebra in this case
is
\begin{equation}
DE - qED =0\;.
\label{qdefcom}
\end{equation}
Taking $q<1$ the configuration with the highest weight will be the
configuration with all particles stacked up to the right end of the
lattice as shown in Figure~\ref{fig:pasep-closed}. The weight of any
configuration can then be obtained by moving particles to the left
from the maximal weight configuration and multiplying the weight by a
factor $q$ at each move as implied by (\ref{qdefcom}).  Then the
statistical weight of any configuration $\tau_1, \tau_2, \ldots,
\tau_N$ will be given by
\begin{equation}
f(\tau_1, \tau_2, \ldots, \tau_N)  = q^{-\sum_k k \tau_k }\;.
\label{closegw}
\end{equation}
and  the maximal weight is
\[ q^{-\sum_{k=N-M+1}^{N} k}
=  q^{-  {N+1 \choose 2}  + {N-M+1 \choose 2} }\;.
\]
The steady state weight (\ref{closegw}) is in fact a Boltzmann weight
with an energy function ${\cal H} \propto | \ln q| \sum_k k \tau_k$
and the dynamics respects detailed balance with respect to this
weight.

Let us now turn to Class D models, where the matrix algebra takes
the form of three deformed commutators (\ref{Dalg}).
An  explicit form for the operators $X_1$,$X_2$,$X_3$
involves tensor products of matrices which obey  deformed commutation 
relations
\begin{eqnarray}
X_1 = D \otimes E(\omega_{21}/\omega_{12})\otimes  \One \\
X_2 = \One  \otimes D \otimes E(\omega_{02}/\omega_{20})  \\
X_0 = E(\omega_{10}/\omega_{01}) \otimes\One \otimes  D 
\end{eqnarray}
where 
\begin{equation}
D E(r) - r E(r) D =0\;.
\end{equation}
So, for example, 
\begin{eqnarray}
\lefteqn{w_{10} X_1 X_0 - w_{01} X_0 X_1 =}\nonumber \\
&&\left[w_{10} D E(\omega_{10}/\omega_{01}) - w_{01}  E(\omega_{10}/\omega_{01})D \right]  \otimes E(\omega_{21}/\omega_{12})\otimes D  =0 \;.
\end{eqnarray}

To obtain statistical weights here, we need to prescribe a suitable
contraction operation i.e. appropriate  boundary vectors. Using the representation of the deformed commutator algebra in Appendix~\ref{appreps} (\ref{defcomrep})
we see that if there are $n_1$ species 1 particles, $n_2$ species 2 particles
and  $n_0$  vacancies then
\begin{equation}
\bra{W} = \bra{0}\otimes\bra{0}\otimes\bra{0}
\qquad
\ket{V} = \ket{n_1}\otimes \ket{n_2}\otimes \ket{n_0}
\end{equation}
will give a nonzero contraction for those configurations with the correct number of each species. In particular for the reference configuration the weight is
\begin{equation}
\bra{W} X_0^{n_0} X_2^{n_2} X_1^{n_1} \ket{V}
= \left( \frac{\omega_{21}}{\omega_{12}}\right)^{n_2}\;.
\end{equation}

\subsubsection{A single defect on an infinite system}  \label{definf}
In order to investigate the structure of
shocks, Derrida, Lebowitz and Speer studied a single second-class
particle on an infinite system\cite{DLS97}. The idea was to describe the
stationary measure as seen from the second-class particle by a matrix
product.  The model considered was the partially asymmetric
generalisation with dynamics
\begin{equation}
1\,0 \mathop{\rightleftharpoons}\limits^{p}_{q}\, 0\,1 \qquad
2\,0 \mathop{\rightleftharpoons}\limits^{p}_{q}\, 0\,2 \qquad
1\,2 \mathop{\rightleftharpoons}\limits^{p}_{q}\, 2\,1 \qquad.
\end{equation}

Taking the second-class particle to be at the origin and considering a
window of $m$ sites to the left and $n$ sites to the right of the
second-class the stationary probabilities (note, not weights) are
written as
\begin{equation}
P(\tau_{-m}, \ldots, \tau_{-1}, \tau_{1}, \ldots, \tau_{n}) = \bra{W}
\left[ \prod_{i=-m}^{-1} X_{\tau_i}\right] A \left[ \prod _{j=1}^n
X_{\tau_j}\right] \ket{V}
\end{equation}
where $A$ is once more the matrix corresponding to a second-class
particle.  In order for this form to hold for a window of arbitrary
size i.e. all values of $n$, $m = 0\ldots \infty$ one requires that
\begin{eqnarray}
\bra{W} C &=& \bra{W}\nonumber \\
C \ket{V} &=& \ket{V}
\label{CW}
\end{eqnarray}
where $C=D+E$.  To ensure that the probabilities are correctly
normalised we further require
\begin{equation}
\bra{W} A \ket{V} =1\;.
\end{equation}

The bulk dynamics falls within the solution class BI, and so the
matrix algebra (\ref{BIalg}) should be used with $p_1=p_2=p$ and
$q_1=q_2=p$.  This particular application has the unusual property
that the choice of $x_0$ and $x_1$ has physical consequences.  This is
because $x_0$ and $x_1$ cannot be scaled out of equations (\ref{CW}).
The correct choice is made by insisting that the desired asymptotic
densities $\rho_{+}$ far to the right of the second-class particle ($n
\to \infty$) and $\rho_{-}$ far to the left ($m \to \infty$) are
obtained.  It turns out that one should take
\begin{eqnarray}
x_0 = (p-q)(1-\rho_-)(1-\rho_+)\\
x_1  = -(p-q)\rho_-\rho_+
\end{eqnarray}
Choosing $\rho_+ > \rho_-$ corresponds to a shock profile as soon from
the second-class particle with the second-class particle tracking the
position of the shock.  The structure of the shock was analysed, in
particular the decay to the asymptotic values is exponential.  An
interesting result is that the characteristic length of the decay
becomes independent of the asymmetry when
\begin{equation}
\frac{p}{q} > \left( \frac{\rho_+(1-\rho_-)}{\rho_-(1-\rho_+)} \right)^{1/2}
\end{equation}

To obtain a representation of the required matrices and vectors, one
can use (\ref{BIrep1})--(\ref{BIrep0}) given in Appendix~\ref{appreps}
for solution class BI.  Then, using (\ref{CW}), one can construct
$\bra{W}$, $\ket{V}$ to be eigenvectors $D+E$ with eigenvalue 1. The
expressions for the boundary vectors are, however, quite complicated
in this representation.  An alternative approach, used in
\cite{DLS97}, exploits an infinite-dimensional (rather than
semi-infinite) representation.  Finite dimensional representations
along special curves were studied in \cite{Speer97}.

\subsection{Non-conserving two-species models with quadratic algebra}
\label{nonconservative}

In the foregoing, we have assumed that the bulk dynamics conserve
particle numbers.  In this section we show how the matrix algebras
which were encountered can be adapted to models that have
non-conservative dynamics.  The basic idea is to augment the dynamics
with additional processes that move between different sectors (i.e.,
configurations with a particular number of particles).  These moves
will be constructed in such a way that detailed balance between
sectors is satisfied, even though detailed balance does not hold within
a sector.  As such, one can realise any prescribed distribution of
sectors: we choose weights that correspond to the grand canonical
ensemble that was introduced in Section~\ref{ring} as a means to study
systems with fixed particle numbers.  We first discuss this method in
detail, and then give two concrete examples.

\subsubsection{Non-conserving dynamics which generate a grand canonical
  ensemble}

As we have just described, the aim is to construct some dynamics such
that the statistical weight of a configuration with $M$ particles (one
a designated species) on an $N$-site ring is
\begin{equation}
\label{fens}
f(\tau_1, \tau_2, \ldots, \tau_N; M) = u^M \tr( X_{\tau_1}
X_{\tau_2} \cdots X_{\tau_N} )
\end{equation}
where $X_{\tau}$ are the matrices that give the steady state for some
conservative process, and $u$ is a fugacity that controls the mean
particle number $M$.  Let us denote transition rates for this
conservative process as $W_0(\Cee\to\Cee')$, and the additional rates
that serve to increase or decrease $M$ by one as
$W_{\pm}(\Cee\to\Cee')$.  The statistical weights must then satisfy
the master equation
\begin{eqnarray}
\fl \sum_{\Cee'} \left[ f(\Cee';M) W_0(\Cee' \to \Cee) - f(\Cee;M)
  W_0(\Cee \to \Cee') \right] + {} \nonumber\\
\fl\qquad \sum_{\Cee'} \left[ f(\Cee';M-1) W_{+}(\Cee' \to \Cee) - f(\Cee;M)
  W_{-}(\Cee \to \Cee') \right] + {} \nonumber\\
\sum_{\Cee'} \left[ f(\Cee';M+1) W_{-}(\Cee' \to \Cee) - f(\Cee;M)
  W_{+}(\Cee \to \Cee') \right] = 0 \;.
\end{eqnarray}
Now, the first summation is what appears in the master equation for
the conservative process, and since all the weights for fixed $M$ are
proportional to $u^M$, this first summation vanishes.  The second
and third summations can be made to vanish if we choose
\begin{equation}
\label{dbens}
\frac{W_{-}(\Cee \to \Cee')}{W_{+}(\Cee'\to\Cee)} = \frac{f(\Cee';
  M-1)}{f(\Cee; M)} = \frac{1}{u} \frac{ \tr(
X_{\tau_1'} X_{\tau_2'} \cdots X_{\tau_N'}) }{ \tr( X_{\tau_1}
X_{\tau_2} \cdots X_{\tau_N} ) }
\end{equation}
where $\tau_i$ and $\tau_i'$ are the site occupation variables
specifying to configurations $\Cee$ and $\Cee'$ that have $M$ and
$M-1$ particles respectively.  As we can see, this takes the form of a
detailed balance relation (\ref{free:db}).  The two examples that
follow serve to demonstrate.

\subsubsection{Grand-canonical dynamics for the ASEP with a second-class
  particle}\label{gcmat}

Recall that in Section~\ref{ring} we considered a version of the ASEP
with a second-class particle, i.e., a model with dynamics
\begin{equation}
1\,0 \mathop{\rightarrow}\limits^{1}\, 0\,1 \qquad
2\,0 \mathop{\rightarrow}\limits^{\alpha} \, 0\,2 \qquad
1\,2 \mathop{\rightarrow}\limits^{\beta} \, 2\,1 \qquad \;.
\label{plusmin}
\end{equation}
This falls within the family of solutions denoted BII above.  In the
analysis of this model we introduced a fugacity $u$ as a trick to
expediate calculation of statistical weights, tuning its value to
yield the desired mean particle density on the ring in the canonical
ensemble.

This ensemble can also be generated using physical dynamics as we have
just described.  This is achieved by noting that the algebra
Let $M$ count the number of positive charges (species $1$ particles).
We note that the algebra (\ref{BIIalg}) implies
\begin{equation}
\frac{\tr( \cdots X_2 X_0 \cdots)}{\tr(\cdots X_1 X_2 \cdots)} =
- \frac{\beta}{\alpha} \frac{x_0}{x_1} \;.
\end{equation}
Hence from (\ref{dbens}), we see that the grand canonical ensemble is
realised physically if we introduce two non-conserving processes
\begin{eqnarray}
+\,0 \mathop{\rightleftharpoons}\limits^{\delta}_{\gamma} 0\,- 
\label{noncon1}
\end{eqnarray}
where the rates satisfy
\begin{equation}
\frac{\delta}{\gamma} = - \frac{\beta}{\alpha u} \frac{x_0}{x_1} \;.
\end{equation}
We remark that $\delta$ and $\gamma$ as free parameters that enforce a
particular fugacity $u$ in the grand canonical ensemble generated by
these non-conservative dynamics.
This particular implementation
of a model with non-conserving dynamics
was used in \cite{ELMM04} to generate
a grand canonical ensemble dynamically.

\subsubsection{A model in which no particle numbers are conserved}

A second non-conserving generalisation of the ASEP on a ring with a
second-class particle is best described using the charge
representation. For simplicity,  we  take the dynamics of BII with $p=1$ and
$q=0$ and take $\alpha=\beta=1$. The case $q \neq 0$ was studied in \cite{EKLM02}.
Thus, the conserving dynamics are
\begin{equation}
+\,0 \mathop{\rightarrow}\limits^{1} \, 0\,+ \qquad
0\,- \mathop{\rightarrow}\limits^{1} \, -\,0 \qquad
+\,- \mathop{\rightarrow}\limits^{1} \, -\,+ \qquad  \;.
\end{equation}
To these dynamics we seek to add the following non-conserving moves
\begin{equation}
\label{noncon2} +\,0 \,
\mathop{\rightleftharpoons}\limits^{w}_{1}\, 0\,0 \qquad 0\,- \,
\mathop{\rightleftharpoons}\limits^{w}_{1} \, 0\,0 \;.
\end{equation}
in such a way that the BII matrix algebra (\ref{BIIalg}) can be
exploited.  To achieve this, we require  matrices such that
(\ref{dbens}) is satisfied for the
non-conserving moves.  That is, 
\begin{equation}
w = \frac{1}{u} \frac{\tr( \cdots X_{0} X_{0} \cdots)}{\tr ( \cdots X_{+} X_{0}
  \cdots )} = \frac{1}{u} \frac{\tr( \cdots X_{0} X_{0} \cdots)}{\tr ( \cdots X_{0} X_{-}
  \cdots )}
\end{equation}
where $u$ is a fugacity counting the total number of positive and
negative charges.  To make contact with more familiar problems, let us
put $X_{+}=D$, $X_{-}=E$ and $X_0=A$.  We see that the previous
equation is  satisfied if
\begin{equation}
u = \frac{1}{w} \quad \mbox{and}\quad A^2 = A \;,
\end{equation}
i.e., we require $A$ to be a projector.  This is 
achieved using the matrices for the second-class particle problem of
Section~\ref{ring}, (\ref{ring:DE}--\ref{ring:A})
i.e. $A = \ket{V}\bra{W}$,
 where $D\ket{V}=\ket{V}$, $\bra{W}E = \bra{W}$ and
$\braket{W}{V}=1$.  The statistical weight of a configuration is then
given by
\begin{equation}
f(\tau_1, \tau_2, \ldots, \tau_N) = w^{M} \tr( X_{\tau_1} X_{\tau_2}
\cdots \tau_N)
\label{wweight}
\end{equation}
where $M$ is the number of vacancies in the configuration.  

In \cite{EKLM02}
it was shown that an interesting  phase transition 
arises as $w$ is varied.  For $w < 1$ there is a tendency
to create positive and negative particles and this ultimately leads to
a vanishing steady-state density of vacancies.  On the other hand for
$w>1$ there is a tendency to eliminate positive and negative charges
at the left and right boundaries respectively of domains of
vacancies. Thus domains of vacancies tend to grow and in the steady
state this results in a finite density of vacancies.  The transition
is remarkable in that it is a phase transition in a periodic one
dimensional system with local dynamics without an absorbing state.
Here, we shall use the generating function technique of
Section~\ref{gfa} to quickly obtain the asymptotics of the partition
function and hence demonstrate the phase transition.

First, notice from (\ref{noncon2})
that configurations with no vacancies are dynamically inaccessible.
Therefore all configurations have at least one   vacancy and we write the partition function as
\begin{equation}
Z_N = \tr\left[ wA (D + E + w A)^{N-1}\right]
= w \bra{W} (D + E + w A)^{N-1}\ket{V}
\end{equation}
(An unimportant subtlety is that
we have ignored the degeneracy factor in placing a
given configuration of particles   on the periodic lattice;
this factor is bounded from above by $N$ and hence does not contribute
to the exponential part of $Z_N$ which controls  the macroscopic physics, as will be seen below.)

We introduce the generating function
\begin{equation}
F(z) = \sum_{N=1}^\infty  z^N Z_N = w z \bra{W} \frac{1}{1-zG}\ket{V}
\end{equation}
where we define $G= D+E + w A$.
Now, we may write
\begin{equation}
1-zG = (1-\eta D -\nu A)(1-\eta E -\nu A)
\end{equation}
where 
\begin{equation}
z= \eta(1-\eta)\qquad  zw = 2\nu(1-\eta) - \nu^2\;. \label{znu}
\end{equation}
Then using the generating function technique and (\ref{znu}) we obtain
\begin{eqnarray}
F(z)&=& 
 w z \bra{W} \frac{1}{(1-\eta E - \nu A)}\, \frac{1}{(1-\eta D - \nu A)}\ket{V}
\\
&=& \frac{wz}{(1-\eta - \nu)^2}
= \frac{wz}{(1-\eta)^2 -  wz}\;.
\end{eqnarray}

The singularities of $F(z)$ are a square root singularity at $z=1/4$
(coming from $\eta(z) = (1- \sqrt{1-4z})/2$) and a pole at $z=
w/(1+w)^2$ which only exists if $w >1$.  Using the results of
Appendix~\ref{gf} one quickly 
obtains the asymptotic behaviour of $Z_N$
\begin{eqnarray}
\mbox{if}\quad w<1 \quad Z_N \sim \frac{1}{\pi^{1/2}} \frac{w}{(1-w)^2}
\frac{4^N}{N^{3/2}}\\
\mbox{if}\quad w>1 \quad Z_N \sim  \frac{w-1}{w+1}
\frac{(w+1)^{2N}}{w^N}\;.
\end{eqnarray}

Due to the form (\ref{wweight}) of the weights, $\theta$,
the density of vacancies, is given by
\begin{equation}
\theta\equiv \lim_{N\to \infty } \frac{\overline{M}}{N}=  
\lim_{N\to \infty} \frac{w}{N}\frac{\partial \ln {\cal Z}_N}{\partial
w}\;, \label{eq:theta1}
\end{equation}
where $\overline{M}$ is the average number of vacancies in the system.
Therefore we obtain
\begin{eqnarray}
\theta &=&0 \quad\mbox{for}\quad w < 1\\
\theta &=&  \frac{w -1}{w+1} \quad\mbox{for}\quad w > 1
\end{eqnarray}
demonstrating the phase transition in the density of vacancies  at $w=1$.
\section{Multispecies Models with quadratic algebra}
\label{sec:multispecies}

In the previous section we identified all possible two-species
dynamics that have a steady state of matrix product form with a quadratic
algebra i.e. obeying  relations of the type given by (\ref{sqa}).
We deferred the discussion of two families of dynamics, as both can be
generalised in a straightforward way to arbitrarily many species of
particles.  We now examine these cases in more detail.
Further discussions of multi-species quadratic algebras
are given in \cite{IPR01,Aneva02,Aneva03,PT02,Twarock03}.

\subsection{ASEP with disordered hopping rates}
\label{pwdis}
We first consider a variant of the ASEP on the ring in which each
particle has a different forward and reverse hop rate (and there is no overtaking)\cite{BFL96,Evans96}. This is a generalisation of the two species model,
case CII of Section~\ref{2species}.
There are $M$ particles and $N-M$ vacancies on the ring of size $N$.
We  give
each of the $M$ particles an index $\mu$, and introduce
for each $\mu$ a pair of rates $p_\mu$ and $q_\mu$ so that the dynamics of
particle $\mu$ are
\begin{equation}
\mu \,0 
\mathop{\rightleftharpoons}\limits^{p_\mu}_{q_\mu}\, 0\, \mu  \;.
\end{equation}
The corresponding matrix algebra is a generalisation of class CII of
Section~\ref{2species} to $M$ particle species.  It takes the form
\begin{equation}
p_\mu D_\mu E - q_\mu E D_\mu = D_\mu
\label{msalg1}
\end{equation}
where we have adopted the notation $X_0 =E$ and $X_\mu = D_\mu$. One
possible representation of these matrices is given in
Appendix~\ref{appreps} as Equation~(\ref{DmuEalg}).  However, 
it is simplest to perform calculations not with a
representation, but directly from the algebra (\ref{msalg1}).

On a periodic system let the steady state weight of a
configuration specified by $\{ n_\mu\}$ where $n_\mu$ is the number
of empty sites in front of particle $\mu$, be $f_N(n_1,\ldots,
n_M)$. The matrix product form for $f_N$ is
\begin{equation}
 f_N( n_1,n_2,\ldots, n_M)=\mbox{Tr} \left[
D_1\ E^{n_1}\ D_2\ E^{n_2}\ \ldots D_M\ E^{n_M} \right]
\label{msmatf1}
\end{equation}
Using (\ref{msalg1}) for $\mu=M$ in (\ref{msmatf1}) gives
the relation 
$$
f_N( n_1,\ldots, n_M) - \frac{q_M}{p_M} f_N( n_1,\ldots,n_{M-1}+1, n_M-1)=
\frac{1}{p_M} f_{N-1}( n_1,\ldots,n_{M-1}, n_M-1).
$$
The procedure is continued  using, in sequence,  (\ref{msalg1})
 with $\mu=M-1,M-2 \ldots 1$ to obtain
$$
\left[1- \prod_{i=0}^{M-1}\frac{q_{M-i}}{p_{M-i}}
                          \right] f_N( n_1,\ldots, n_M)
 =                         \left[ \sum_{i=0}^{M-1} \frac{1}{p_{M-i}}
                           \prod_{j=M+1-i}^{M} \frac{q_j}{p_j}\right]
                           f_{N-1}( n_1,\ldots, n_M-1)
$$ The effect of this manipulation has been to commute a hole
initially in front of the particle $M$ backwards one full turn around
the ring.  The result is that the weight of a configuration of size
$N$ is expressed as a multiple of the weight of a configuration of
size $N-1$ with one hole fewer  in front of particle $M$.  Repeating the
commutation procedure for a hole initially in front of particles
$M-1,M-2,\ldots 1$ implies that the weights are
\begin{equation}
\fl f_N(\left\{ n_\mu \right\}) = \prod_{\mu=1}^{M} g_{\mu}^{n_{\mu}}f_M(0,\ldots,0)\;\;\;
\mbox{where}\;\;\; g_{\mu} =\left[
                         \sum_{i=0}^{M-1} \frac{1}{p_{\mu-i}}
                           \prod_{j=\mu+1-i}^{\mu} \frac{q_j}{p_j}
                           \right]
\ \left[1-\prod_{k=1}^{M}\frac{q_{k}}{p_{k}}
                          \right]^{-1} \; 
\label{diss}
\end{equation}
and we can take $f_M(0,\ldots,0)=1$.

\subsubsection{Bose-Einstein Condensation}
\label{BEC}
An interesting phenomenon that may occur in this system is that of
condensation. In that case the steady state is dominated by
configurations where one particle has an extensive value for
$n_\mu$ \cite{Evans96,KF96}. 
This is easiest to understand when $q_\mu= 0$ i.e. we
consider only forward hops. 
Then (\ref{diss}) reduces to $g_\mu = 1/p_\mu$ and we may write
\begin{equation}
f_N( \left\{ n_\mu \right\}) =
\exp \left( - \sum_{\mu=1}^M n_\mu \epsilon_\mu \right)\quad\mbox{where}
\quad \epsilon_\mu = \ln p_\mu\;.
\label{BEf}
\end{equation}
The weight (\ref{BEf}) clearly has the form of the weight of an ideal
Bose gas with $kT =1$. Here the `bosons' are vacancies and the Bose
states correspond to particles with the energy of the state determined
by the hop rate of the particle. The equivalent of the density of
states for the Bose system is the distribution of particle hop rates
which we denote $\sigma(p)$.  If there is a minimum hop rate $p_{\rm min}$
this will correspond to the ground state of the Bose system. Then, if
for low hop rates the distribution of particle hop rates follows
\begin{equation}
\sigma(p) \sim (p-p_{\rm min})^\gamma\quad\mbox{with}\quad \gamma>0\;,
\end{equation}
Bose condensation will occur for high enough vacancy density or
equivalently {\em low} enough particle density.  When condensation
occurs the slowest particle (the one with the minimum hop rate $p_{\rm
min}$) has an extensive number of vacancies in front of it whereas the
rest of the particles will have gaps in front of them comprising some
finite number of vacancies. Thus a `traffic jam' forms behind the
slowest particle.  In \cite{EH05} it is shown that the ASEP on a
periodic system may be mapped onto a Zero Range Process and the
condensation transition is fully discussed in the context of the Zero
Range Process.

Recently, further progress has been made in understanding
the case where $q_\mu\neq0$ by using extreme values statistics and renormalisation arguments \cite{JSI05a}.

\subsection{Karimipour's model of overtaking dynamics}
\label{sec:Karimipour}

This is the generalisation of the two-species case AII, which had the
two particle species hopping with different speeds, and the faster
ones overtaking the slower.  The many-species generalisation 
allows an arbitrary number of particle species labelled
$\mu = 1\ldots P$.
A particle of species $\mu$ has  a forward hop rate
or `velocity'  $v_{\mu}$,
and adjacent
particles $\mu$ and $\nu$ exchange places with rate $v_\mu - v_\nu$ if
$v_\mu > v_\nu$, i.e., 
\begin{eqnarray}
\mu \,0 \mathop{\rightarrow}\limits^{v_\mu}\, 0\,\mu \nonumber \\
\mu \,\nu \mathop{\rightarrow}\limits^{v_\mu-v_\nu}\, \nu \,\mu
\quad\mbox{if}\quad v_\mu >v_\nu \;.
\end{eqnarray}
The model provides
an elegant generalisation of the open boundary TASEP
and  its algebra.
The generalisation of (\ref{AIIalg}) to the multispecies case is
\begin{eqnarray}
v_\mu X_\mu X_0 &=& x_0 X_\mu - x_\mu X_0 \quad \mu >0 \nonumber\\
(v_\mu-v_\nu) X_\mu X_\nu &=& x_\nu X_\mu - x_\mu X_\nu \quad  v_\mu > v_\nu 
\;.
\label{AIIalgm}
\end{eqnarray}
Writing $X_0 = E$, $X_\mu = D_\mu/P$
and making the convenient choice $x_0=1$, $x_\mu= - v_\mu/P$ for $\mu >0$
yields
\begin{eqnarray}
D_\mu E  &=& \frac{1}{v_\mu} D_\mu +E\nonumber\\
D_\mu D_\nu &=&  \frac{v_\mu}{v_\mu-v_\nu}D_\nu- \frac{v_\nu}{v_\mu-v_\nu}D_\mu
\quad v_\mu >v_\nu\;.
\label{Kalg}
\end{eqnarray}
On a periodic lattice the relations (\ref{Kalg}) are the only ones
which need be satisfied and this can be done by choosing a scalar
representation $E= 1/\epsilon$ and $D_\mu = \frac{v_\mu}{v_\mu -
\epsilon}$ where $\epsilon$ is chosen to be less than the lowest hop
rate.  Thus on a ring all allowed configurations are equally likely.

More interestingly, the relations (\ref{msalg1}) may  be used for an open
system with suitable boundary conditions.
As discussed in Section~\ref{sec:extop}
we require $\sum_{\mu=0}^P x_\mu =0$ which with the above choice of $x_0$
and $x_\mu$ implies 
\begin{equation}
\frac{1}{P}  \sum_{\mu=1}^P  v_\mu = 1\;.
\label{avvd}
\end{equation}
At the left hand boundary
a particle of species $\mu$
is  injected with rate $\alpha_\mu$ if the first site is empty.
The choice $\alpha_\mu= \alpha v_\mu/P$ reduces (\ref{alphaij})
to
\begin{equation}
\bra{W} E  = \frac{1}{\alpha} \bra{W}
\label{Ka}
\end{equation}
At the left hand boundary a particle of species $\mu$ leaves the system with rate $\beta_\mu$. The choice $\beta_\mu = v_\mu +\beta -1$ ensures that explicit representations of the matrices can  be constructed (see Appendix~\ref{appreps})
and (\ref{betaij}) becomes 
\begin{eqnarray}
D_\mu \ket{V} = \frac{v_\mu}{v_\mu -(1-\beta)} \ket{V}
\label{Kb}
\end{eqnarray}
Relations (\ref{Kalg}), (\ref{Kb}) and (\ref{Ka}) provide an elegant
generalisation of the usual ASEP relations \RR\ which are recovered
when we have just one species of particle with hop rate $v_1=1$.  In
general, an infinite dimensional representation of (\ref{Kalg}) is
needed to satisfy the boundary conditions (\ref{Kb}); see, for
example, (\ref{Karimrep}) in Appendix~\ref{appreps}.

The fully disordered system is realised when each particle that enters
has a velocity drawn from some distribution $\sigma(v)$, with support
$[v_{\rm min},\infty]$ so that the lowest velocity is $v_{\rm min}$,
and (\ref{avvd}) is replaced by
\begin{equation}
\int_{v_{\rm min}}^\infty  \rmd v\,  \sigma(v)   v  = 1\;.
\label{avvc}
\end{equation}
Thus, when site one is vacant a particle enters with rate $\alpha$ and
its velocity $v$ is assigned according to the distribution
$\sigma(v)v$ giving an effective rate $\alpha(v) = \sigma(v) v
\alpha$.  Therefore we can think of a reservoir of particles with
velocities distributed according to $\sigma(v)$ each attempting to
enter with rate $v\alpha$.  A particle reaching site $N$ exits with
rate $\beta(v) = v+ \beta -1$. Then we can replace $X_\mu$ in the
discrete case by $X(v)$ in (\ref{AIIalgm}). Letting $X_0 = E$ and
$X(v) = \sigma(v) D(v)$, $x_0=1$ and $x(v) = - \sigma(v) v$ recovers
the same algebra (\ref{Kalg}), (\ref{Kb}) and (\ref{Ka}) with $D_\nu$
replaced by $D(v)$, e.g.,
\begin{equation}
D(v)D(v') = \frac{v}{v-v'}D(v')-
\frac{v'}{v-v'}D(v)\quad\mbox{for}\quad v>v'\;.
\end{equation}

Karimipour \cite{Karimipour99b,KK00} studied, amongst other things,
the phase diagram of the
open boundary model and found the structure depends on the
distribution $\sigma(v)$ through the parameter
\begin{equation}
l[\sigma] = \frac{1}{v_{\rm min}^2} - \left\langle \frac{v}{(v-v_{\rm min})^2} \right\rangle
\end{equation}
which characterises the form of $\sigma(v)$ near the lowest hop rate
$v_{\rm min}$.  If $l[\sigma] <0$ the three phases (low-density, high-density
and maximal current) of the open boundary ASEP remain although the
phase boundaries depend on parameters such as $l[\sigma]$ and $v_{\rm min}$.
However if $l[\sigma] \geq 0$ the high-density phase is suppressed and
only the low-density and maximal current phases exist.

\section{More complicated matrix product states}
\label{formal}

Recall that in Section~\ref{algproof} we set out a general
cancellation scheme for matrices that guarantees a stationary solution
of a master equation for a process on the ring.  This involved a set
of ordinary matrices $X_{\tau}$ and their auxiliaries
$\tilde{X}_{\tau}$.  By restricting to the case of scalar auxiliaries,
we showed, in the case of two particle species in
Section~\ref{2species}, that it was possible to determine all possible
sets of dynamics that give rise to a matrix product steady state.  In
this section, we shall consider the more complicated case where the
auxiliaries are matrices or more complicated operators.

Here it is not possible---as far as we are aware---to perform an
exhaustive search for solutions.  However, it has been shown that if
one has a one-dimensional lattice with $n$ particle species, open
boundary conditions and arbitrary nearest-neighbour dynamics in the
bulk, there does exist a matrix product solution, involving
auxiliaries that in general will not be scalars, obeying the algebraic
relations set out in Section~\ref{algproof}.  In the next subsection
we review this existence proof which is due to Krebs and Sandow
\cite{KS97}.  Unfortunately, this proof does not lead to any
convenient reduction relations, like \RR\ for the ASEP, that might be
in operation, Also, the proof is not constructive in that it requires
the steady state to already be known in order to construct explicit
matrices. Rather, the proof demonstrates that there are no internal
inconsistencies in the cancellation mechanism of
Section~\ref{algproof}.

In the last two subsections we discuss two models of which we are
aware that give examples of a matrix product state with matrix auxiliaries.

\subsection{Existence of a matrix product solution for models with
  open boundaries}

The existence proof of a matrix product state \cite{KS97} (see also \cite{RS97}
for the discrete time case) applies to
models with open boundary conditions and arbitrary nearest-neighbour
interactions.  Let us restate the master equation for a general process
in the form introduced in 
Section~\ref{algproof}.  It reads
\begin{equation}
\label{formal:me}
\fl \frac{\rmd}{\rmd t} f(\tau_1, \ldots, \tau_N) \equiv \hat{H} f(\tau_1,
\ldots, \tau_N) = \left( \hat{h}_L + \sum_{i=1}^{N-1} \hat{h}_{i,i+1}
+ \hat{h}_R \right) f(\tau_1, \ldots, \tau_N) \;,
\end{equation}
where the operator $\hat{h}_{i,i+1}$ applied to the function of $N$
state variables $\tau_i$ generates the gain and loss terms arising
from interactions between neighbouring pairs of sites in the bulk, and
$\hat{h}_L$ and $\hat{h}_R$ do the same at the boundaries.  Here we
shall consider models that have $n$ distinct particle species in
addition to vacancies: i.e., the occupation variables take the values
$\tau_i=0,1,\ldots,n$.  The matrix product
expressions
\begin{equation}
\label{formal:mp}
f(\tau_1, \tau_2, \cdots, \tau_N) = \bra{W} X_{\tau_1} X_{\tau_2}
\cdots X_{\tau_N} \ket{V}
\end{equation}
provide the steady state solution to (\ref{formal:me}) for some set of
matrices $X_{\tau}$ and vectors $\bra{W}$ and $\ket{V}$ if one can
find a second set of auxiliary matrices $\tilde{X}_{\tau}$ such that
the relations
\begin{eqnarray}
\label{formal:tilde1}
\hat{h}_{i,i+1} \bra{W} \cdots X_{\tau_{i}} X_{\tau_{i+1}} \cdots
\ket{V} = \bra{W} \cdots [ \tilde{X}_{\tau_i} X_{\tau_{i+1}} -
X_{\tau_{i}} \tilde{X}_{\tau_{i+1}} ] \cdots \ket{V} \\
\label{formal:tilde2}
\hat{h}_L \bra{W} X_{\tau_1} \cdots \ket{V} = - \bra{W}
\tilde{X}_{\tau_1} \cdots \ket{V} \\
\hat{h}_R \bra{W} \cdots X_{\tau_N} \ket{V} = \bra{W} \cdots
\tilde{X}_{\tau_N} \ket{V}
\label{formal:tilde3}
\end{eqnarray}
hold.  Then, one obtains a zero right-hand side of the master equation
(\ref{formal:me})  via a pairwise cancellation of terms
coming from (\ref{formal:mp})--(\ref{formal:tilde3}).
Thus, relations (\ref{int:h3}--\ref{int:hr3})
for the ASEP generalise to
\begin{eqnarray}
\label{ft1}
\hat{h}_{i,i+1} X_{\tau_{i}} X_{\tau_{i+1}}  =  \tilde{X}_{\tau_i} X_{\tau_{i+1}} -
X_{\tau_{i}} \tilde{X}_{\tau_{i+1}}  \\
\label{ft2}
\hat{h}_L \bra{W} X_{\tau_1} = - \bra{W}\tilde{X}_{\tau_1} \\
\hat{h}_R  X_{\tau_N} \ket{V} = \tilde{X}_{\tau_N} \ket{V}\;.
\label{ft3}
\end{eqnarray}
Relations (\ref{ft1}--\ref{ft2}) give the general form for a
cancellation mechanism involving matrix (or possibly tensor)
auxiliaries.  Such a cancellation scheme has been proposed in various
contexts \cite{DEM95,Schutz96,SS95} including a generalisation to
longer (but finite) range interactions \cite{KS99}.  Some consequences
of this scheme have been explored in \cite{Karimipour00}.

The existence of the matrices $X_{\tau}$ and $\tilde{X}_{\tau}$ 
and vectors $\bra{W}$ and $\ket{V}$ appearing in (\ref{ft1}--\ref{ft2})
is proved by constructing them within
an explicit representation.  This representation has basis vectors
that correspond to configurations of the rightmost $N-k+1$ sites of
the $N$-site lattice.  These we denote as $\ket{\tau_k \tau_{k+1}
\cdots \tau_{N}}$.  The vector $\ket{V}$ is then ascribed the role of
a vacuum state, and the matrices $X_{\tau}$ that of creation operators
in such a way that
\begin{equation}
\label{formal:X}
X_{\tau} \ket{V} = \ket{\tau} \quad \mbox{and} \quad X_{\tau}
\ket{\tau_{k+1} \tau_{k+2} \cdots \tau_{N}} = 
\ket{\tau \tau_{k+1} \tau_{k+2} \cdots \tau_{N}} \;.
\end{equation}
One then defines the vector $\bra{W}$ via the scalar products
\begin{equation}
\label{formal:sp}
\braket{W}{\tau_1 \tau_2 \ldots \tau_N} = f(\tau_1, \tau_2, \cdots,
\tau_N)
\end{equation}
so that (\ref{formal:mp}) gives the desired statistical weights.

The auxiliary matrices $\tilde{X}_\tau$ are defined as
\begin{equation}
\label{formal:Xtilde}
\tilde{X}_\tau = \left( \sum_{j=1}^{N-1} \hat{h}_{j,j+1} + \hat{h}_R \right)
X_\tau
\end{equation}
where the operators $\hat{h}_{j,j+1}$ and $\hat{h}_R$ are extended to
the full space of all sub-configurations $\ket{\tau_k \tau_k+1 \cdots
\tau_N}$ as follows.  First, the bulk operator is defined in terms of
the microscopic transition rates $W(\Cee \to \Cee')$ as
\begin{equation}
\fl \bra{\tau_k \tau_{k+1} \cdots \tau_{i}' \tau_{i+1}' \cdots \tau_N}
\hat{h}_{i,i+1} \ket{\tau_k \tau_{k+1} \cdots \tau_{i} \tau_{i+1}
\cdots \tau_N} = W(\tau_i \tau_{i+1} \to \tau_{i}' \tau_{i+1}')
\end{equation}
when $i \ge k$ and $(\tau_i',\tau_{i+1}') \ne (\tau_i, \tau_{i+1})$,
and
\begin{equation}
\fl \bra{\tau_k \tau_{k+1} \cdots \tau_{i} \tau_{i+1} \cdots \tau_N}
\hat{h}_{i,i+1} \ket{\tau_k \tau_{k+1} \cdots \tau_{i} \tau_{i+1}
\cdots \tau_N} = - \sum_{\tau_i', \tau_{i+1}'} W(\tau_i \tau_{i+1} \to
\tau_{i}' \tau_{i+1}')
\end{equation}
under the same condition that $i \ge k$.  All other elements of these
operators are zero.  Similarly, at the right boundary we have
\begin{equation}
\bra{\tau_k \cdots \tau_N'} \hat{h}_R \ket{\tau_k \cdots \tau_N} =
W(\tau_N \to \tau_N')
\end{equation}
when $\tau_N \ne \tau_N'$, and
\begin{equation}
\bra{\tau_k \cdots \tau_N} \hat{h}_R \ket{\tau_k \cdots \tau_N} =
-\sum_{\tau_N'} W(\tau_N \to \tau_N') \;.
\end{equation}
With these definitions established, the relations
(\ref{formal:tilde1})--(\ref{formal:tilde3}) follow 
after some straightforward manipulations\cite{KS97}.

Let us take stock of this construction. We have seen that not only
does the existence of a set of vectors and matrices that satisfy
(\ref{formal:tilde1})--(\ref{formal:tilde3}) imply a stationary
solution of matrix product form of the master equation
(\ref{formal:me}), but also that such a set of vectors and matrices
can always be constructed {\em once} one knows the weights
(\ref{formal:sp}).  As we previously saw in Section~\ref{pasep} for
the PASEP, one can find a number of different representations of the
matrices $D$ and $E$ (which correspond to $X_1$ and $X_0$ in the more
general setting) even when the auxiliary matrices $\tilde{D}$ and
$\tilde{E}$ are the same.  In the representation detailed above the
auxiliaries are not scalars but rather complicated objects.  Hence we
see that (i) there may be many choices of both the matrices $X_{\tau}$
and their auxiliaries that correspond to the stationary solution of a
single master equation; (ii) matrix reduction relations, like those
found for the ASEP \RR, constitute only sufficient conditions on
$X_{\tau}$, $\bra{W}$ and $\ket{V}$, since valid representations where
these relations do not hold can be found (that used in this section
provides an example); and (iii) the construction of the auxiliary
matrices through the generators of the stochastic process in
(\ref{formal:Xtilde}) does not necessarily imply the existence of any
convenient reduction relations for the $X$ matrices that allow, for
example, efficient computation of statistical properties.

\subsubsection{Existence of a matrix product state for models on the ring}
As we saw in Section~\ref{ring}
the matrix product state on the ring involves using a trace operation.
The algebraic relations
to be satisfied are given in the general case  by (\ref{ft1}).
However, as we saw in Section~\ref{2species}, sometimes it happens that, 
although the algebraic relations 
are consistent, the rotational invariance of the
trace operation leads to global constraints on particle numbers.  For
example, in the ABC model discussed in \ref{sec:ABC} we found that the
matrix product state was consistent only for models in which the
numbers of each particle species was the same.  It is essentially this
property of models with periodic boundary conditions which has
precluded the development of an existence proof for a matrix product
state parallel to that given above.  These difficulties are discussed
in more detail in \cite{Krebs00}.

\subsection{Non-conserving models with finite-dimensional representations}

We now discuss a specific reaction-diffusion model whose steady state
can be represented in matrix product form, but where the auxiliaries
$\tilde{X}_\tau$ are matrices.  This model contains one species of
particle and the stochastic processes are diffusion, coagulation and
decoagulation.  These last two updates involve the annihilation or
creation of a particle adjacent to another particle.  If the bulk
rates are chosen in the following way \cite{HSP96}
\begin{equation*}
0 \, 1 
\mathop{\rightleftharpoons}\limits^{q}_{q^{-1}}\, 1\,0 \qquad
1\,1 
\mathop{\rightleftharpoons}\limits^{q}_{\Delta q^{-1}}\, 1\,0 \qquad
1\,1 
\mathop{\rightleftharpoons}\limits^{q^{-1}}_{\Delta q }\, 0\,1 
\end{equation*}
then for suitable boundary conditions the steady state may be written in
matrix product form.

Originally in \cite{HSP96} a closed segment was considered.  Then, one
requires $\hat{D}\ket{V} =\hat{E}\ket{V} = \bra{W} \hat{D} = \bra{W}
\hat{E} =0$ the conditions to be satisfied come from the bulk algebra
\begin{eqnarray*}
\hat{E} E - E \hat{E} &=& 0 \\
\hat{E} D - E \hat{D} &=& q^{-1} D E + q^{-1} DD - (\Delta + 1)q ED \\
\hat{D} E - D \hat{E} &=& q E D  + q  DD - (\Delta + 1)q^{-1} ED \\
\hat{D} D - D \hat{D} &=& \Delta q E D  + \Delta q^{-1}  DD - (q +q^{-1}) ED 
\end{eqnarray*}
A four-dimensional representation of the matrices was found
\begin{eqnarray*}
 E  &=& \left( \begin{array}{cccc}
                 q^{-2}& q^{-2}&0&0\\
                  0&(\Delta+1)^{-1} &(\Delta+1)^{-1}&0\\
                  0&0& 1&q^2\\
                  0&0&0&q^2
                  \end{array}
                   \hspace{0.1in} \right)\, ,
\\[0.5ex]
 D  &=& \left( \begin{array}{cccc}
                 0 & 0 &0&0\\
                  0&\Delta/(\Delta +1) &\Delta/(\Delta+1)&0\\
                  0&0& \Delta &0\\
                  0&0&0& 0
                  \end{array}
                   \hspace{0.1in} \right)\, ,
\\[0.5ex]
 \hat{E}  &=& \left( \begin{array}{cccc}
                   0 &  0 &q^{-1}& -(q-q^{-1})^{-1}\\
                  0&0  &(q-q^{-1})& -q\\
                  0&0  &\Delta (q-q^{-1})& -\Delta q\\
                  0&0&0&0
                  \end{array}
                   \hspace{0.1in} \right)\, ,
\\[0.5ex]
\hat{D}  &=& \left( \begin{array}{cccc}
                 0 & -\Delta q^{-1} &0& 0\\
                  0& -\Delta(q-q^{-1}) &0 &0\\
                  0&0  &-\Delta (q-q^{-1})& \Delta q\\
                  0&0&0& 0
                  \end{array}
                   \hspace{0.1in} \right)\, ,
\end{eqnarray*}
with boundary vectors
\begin{equation}
\bra{W} = (1-q^2, 1, 0, a)\qquad \ket{V}= (b, 0, q^2, q^2-1)^T
\end{equation}
where  $a \neq b$ so that $\braket{W}{V} \neq 0$.

However there is a subtlety with this model on a closed segment which
is that the empty lattice is dynamically inaccessible and itself
comprises an inactive steady state. If we wish to exclude this
configuration we should choose $a$ and $b$ so that $\bra{W} E^N
\ket{V}$ vanishes.  This can be accomplished but in
doing so renders $a$ and $b$ $N$-dependent.

A first order phase transition occurs at $q^2 = 1 + \Delta$.  For $q^2
< 1 + \Delta$ the system is in a low-density phase whereas for $q^2 >
1 + \Delta$ the system is in a high-density phase.  Within the matrix
product form of the steady state the transition can be understood as
the crossing of eigenvalues of $C$.  This may occur here not because
the matrices are of infinite dimension but because they have negative
elements (cf.\ the transfer matrices that arise in equilibrium
statistical mechanics, as described in Section~\ref{ising}).

Various other non-conserving models with finite-dimensional matrix
product states have been identified.  For example, the bulk dynamics
just described has also been studied in the case of an open left
boundary, where particles enter with rate $\alpha$ and leave with rate
$\gamma$ and a closed right boundary by Jafarpour \cite{Jafarpour03}.
Then, one requires $\hat{D}\ket{V} =\hat{E}\ket{V} =0$ and
\begin{eqnarray}
-\bra{W} \hat{D} = \alpha \bra{W} \hat{E} - \gamma \bra{W} \hat{D}\\
-\bra{W} \hat{E} = \gamma \bra{W} \hat{E} - \alpha \bra{W} \hat{D} \;.
\end{eqnarray}
Then if
\begin{equation}
\alpha = \left( \frac{1}{q}-q + \beta \right) \Delta
\end{equation}
one can find two-dimensional representations of the algebra.  Further
generalisations of this class of model have been shown to have
finite-dimensional matrix product states \cite{Jafarpour04}.  A
general systematic approach to determine a necessary condition for the
existence of a finite-dimensional matrix product state has been put
forward by Hieida and Sasamoto \cite{HS04} in which more examples are
given.


\subsection{Several classes of particles}

Second-class particles were discussed in Section~\ref{sscp}
and the matrix product solution to the second-class particle problem
was detailed in Section~\ref{ring}.
The matrix algebras for two-species
generalisations of the second-class particle
arose in the BI and BII solution
classes of the classification scheme of
Section~\ref{2species}.  We now consider a natural generalisation of the second-class
particle to many classes  of particle, labelled $\tau=1,2,\ldots,n$ with
dynamics
\begin{equation}
\tau \,0  \mathop{\rightarrow}\limits^{1}\, 0\,\tau\quad \forall \tau
\quad\mbox{and}\quad
\tau \,\tau'
\mathop{\rightarrow}\limits^{1}\, \tau' \,\tau \quad\mbox{if}\; \tau <
\tau' \;.
\end{equation}
In this case  it turns out one requires more complicated operators and auxiliaries than matrices.

For $n=3$,  the case of three particle classes, the algebraic
relations to be  satisfied, (\ref{ft1}),  become
\begin{equation}
\tilde{X}_{i} X_i - X_{i} \tilde{X}_{i} =0 \;, \; i=0,1,\ldots 3
\label{3sp1}
\end{equation}
and
\begin{eqnarray}
X_{\tau} X_0 &=&  \tilde{X}_0 X_{\tau}  -X_0 \tilde{X}_{\tau}  =
-\tilde{X}_{\tau} X_{0}  +X_{\tau} \tilde{X}_{0}  \;, \;
\tau>0 \nonumber \\
X_1 X_2 &=& \tilde{X}_2 X_1 - X_2 \tilde{X}_1 
= -\tilde{X}_1 X_2 + X_1 \tilde{X}_2 \nonumber \\
X_1 X_3 &=& \tilde{X}_3 X_1 - X_3 \tilde{X}_1 =
-\tilde{X}_1 X_3 + X_1 \tilde{X}_3 \nonumber \\
X_2 X_3 &=& \tilde{X}_3 X_2 - X_3 \tilde{X}_2 =
-\tilde{X}_2 X_3 - X_2 \tilde{X}_3 \;.
\label{3sp}
\end{eqnarray}
The two-class problem, where the last lines of (\ref{3sp})
are absent, 
can be solved  (as we have seen in Section~\ref{RRRproof}) with scalar
choices $\tilde{X}_1 = -1$, $\tilde{X}_0=1$ and $\tilde{X}_2 = 0$.  It
turns out that three-class problem is not a simple generalisation of
the two-class case, but instead a much more complication solution is
needed \cite{MMR99}.  Matrices which satisfy (\ref{3sp1},\ref{3sp}) have been
given as
\begin{eqnarray}
\hspace{-0.in} X_1 &=&  \left( \begin{array}{ccccc}
                  D&0&E&0&.\\
                  0&D&0&E&. \\
                  0&0&D&0&.\\
                  0&0&0&D&.\\
                  . &.&.&.&.  
                   \end{array}
                   \hspace{0.2in} \right),
       \nonumber 
 X_2 =  \left( \begin{array}{cccccc}
               D&-E&0&0&. \\
               0&0&0&0&.  \\
               0&0&0&0&.  \\
               0&0&0&0&.  \\
              . &.&.&.&. \end{array}
                   \hspace{0.2in} \right)
\hspace{0.2in}\\
X_3 &=&   \left( \begin{array}{cccccc}
               E&0&0&0&. \\
               D&0&0&0&.  \\
               0&0&0&0&.  \\
               0&0&0&0&.  \\
               .&.&.&.&. \end{array}
                   \hspace{0.2in} \right),
 X_0 =  \left( \begin{array}{cccccc}
                  E&0&0&0&.\\
                  0&E&0&0&. \\
                  D&0&E&0&. \\
                  0&D&0&E&. \\
                  .&.&.&.&.
                   \end{array}
                   \hspace{0.2in} \right).
\label{mat}
\end{eqnarray}
The objects $D$ and $E$ that appear as elements of these
infinite-dimensional matrices are themselves infinite dimensional
matrices $D$ and $E$ which satisfy the familiar relation $DE = D + E$.
In other words the $X_i$ are rank four tensors, as indeed are the
auxiliaries in this representation.  These were also given by
\cite{MMR99}, and take the form
\begin{eqnarray}
\tilde{X}_1 &=& \left( \begin{array}{ccccc}
                 D/2+\One&0&E/2-\One&0&.\\
                  0&D/2+\One&0&E/2-\One\\
                  0&0&D/2+\One&0&.\\
                  0&0&0&D/2+\One&.\\
                  .&.&.&.&.
                  \end{array}
                   \hspace{0.2in} \right)\, ,
        \nonumber \\
 \tilde{X}_2 &=& \left( \begin{array}{ccccc}
               \One-D/2&\One-E/2&0&0&. \\
               0&0&0&0&.  \\
               0&0&0&0&.  \\
               0&0&0&0&.  \\
              .&.&.&.&. \end{array}
                   \hspace{0.2in} \right)\, ,\nonumber  \\
 \tilde{X}_3 &=& \left( \begin{array}{ccccc}
               E/2-\One&0&0&0&. \\
              \One-D/2&0&0&0&.  \\
               0&0&0&0&.  \\
               0&0&0&0&.  \\
               .&.&.&.&. \end{array}
                   \hspace{0.2in} \right)\, ,   \nonumber \\
 \tilde{X}_0 &=& -\left( \begin{array}{ccccc}
                 E/2+\One&0&0&0&. \\
                  0&E/2+\One&0&0&. \\
                 D/2-\One&0&E/2+\One&0&.\\
                  0&D/2-\One&0&E/2+\One&.\\
                  .&.&.&.&.
                  \end{array}
                   \hspace{0.2in} \right)\, .
\end{eqnarray}
The tensor, or matrix within matrix, form of (\ref{mat})
is closely related to that of the operators used to determine the steady state current 
fluctuations in an open system \cite{DEM95}.

In \cite{MMR99}, some algebraic relations involving \emph{only} the
matrices $X_\tau$ (i.e., not the auxiliaries) were quoted.  These
included relations which had triples and quartets of matrices reducing
to pairs, and others which transformed triples and pairs to other
triples and pairs.  As noted in \cite{MMR99}, the relations given
comprise only a subset of those that would actually be required to
perform a complete reduction of any string of matrices to an
irreducible form.  Hence, a concise statement of the steady state of
this system, like that encoded by the reduction relations \RR\ for the
ASEP, is not  easy.
As such, no calculations beyond simple cases\cite{MMR99} have yet been attempted for the stationary properties of this model.

However recently,  generalising a construction by Angel \cite{Angel06}
for  the two-class problem, Ferrari and Martin \cite{FM05} have shown how to generate the steady state of  the multi-class problem. This construction can then be used to determine operators and auxiliaries to
solve the steady state  for more than three classes \cite{EFM07}.


\section{Discrete-time updating schemes}
\label{discretetime}

So far we have reviewed interacting particle systems involving
continuous time, or equivalently, random sequential dynamics as
discussed in Section~\ref{asep}.  However, this is not necessarily the
most natural choice of dynamics with which to model some physical
systems of interest.  For example, in traffic flow and pedestrian
modelling \cite{CSS00,Helbing01} it is desirable that the microscopic
constituents are able to move simultaneously---
this  often originates from the existence
of a smallest relevant timescale, e.g. reaction times in traffic.
Therefore, in these systems, parallel
dynamics in which all particles are updated in one discrete timestep,
may be used.
In this section we consider the ASEP under three types
of discrete-time dynamics: sublattice, ordered sequential and fully
parallel.

In all these discrete-time cases the steady state weights are given by
the eigenvector with eigenvalue one of the transfer matrix of the
dynamics, which we write in a schematic notation is  
\begin{equation}
\hat{T} f(\tau_1,\ldots, \tau_N) =
f(\tau_1,\ldots, \tau_N)\;.
\label{fulltm}
\end{equation}
The transfer matrix $\hat{T}$ applied to the weight
$f(\tau_1,\ldots, \tau_N)$  generates a sum
of terms, each being the weight of a configuration multiplied by the
transition {\em probability} in one discrete timestep
 from that configuration to $\{\tau_1,\ldots,
\tau_N\}$.

\subsection{Sublattice parallel dynamics}

In sublattice updating the timestep is split into two halves: in the
first half site 1, the even bonds $(2,3)$,$(4,5)\ldots(2L-2,2L-1)$ and
site $2L$ are updated simultaneously; in the second half time step the
odd bonds $(1,2)$,$(3,4)\ldots(2L-1,2L)$ are updated simultaneously.
Note that we require the total number of sites $N=2L$ to be even.

\begin{figure}
\begin{center}
\includegraphics[scale=0.6]{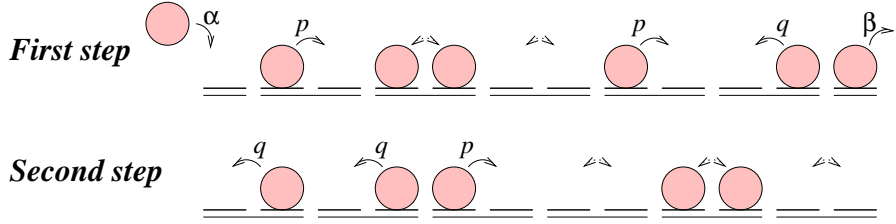}
\end{center}
\caption{\label{fig:sublattice} Typical configuration in the steady
  state of the partially asymmetric exclusion process on a closed
  segment of $N$ sites with $q<1$.}
\end{figure}

In Figure~\ref{fig:sublattice} we show how this type of updating is
applied to partially asymmetric exclusion dynamics with open boundary
conditions.  In an update of site 1, if site 1 is empty a particle
enters with probability $\alpha$.  In an update of a bond $i,i+1$, the
two possible transitions are: if site $i$ is occupied and site $i+1$
is empty the particle moves forward with probability $p$, or else if
site $i+1$ is occupied and site $i$ is empty the particle moves
backward with probability $q$.  In an update of site $N$, if site $N$
is occupied the particle leaves the system with probability $\beta$.
We note that this two-step process avoids the possibility of a
conflict occurring (e.g., two particles attempting to hop into the
same site simultaneously).

The ASEP with sublattice dynamics was first studied in the case of
deterministic bulk dynamics ($p=1$, $q=0$) by Sch\"utz
\cite{Schutz93}. A matrix product solution for this case was found by
Hinrichsen \cite{Hinrichsen96} and a matrix product solution for the
general case of stochastic bulk dynamics was found by Rajewsky et al \cite{RS97,RSSS98}.  We
outline this general case here.

One proceeds by constructing the transfer matrix for the whole
timestep as a product of operators for each half-timestep
\begin{equation}
\hat{T} =  \hat{T_2}\hat{T_1}  
\end{equation}
where
\begin{eqnarray}
T_1 = 
\hat{L} \hat{T}_{23} \hat{T}_{45}\ldots \hat{T}_{2L-2\,2L-1} \hat{R}  \\
T_2 = 
\hat{T}_{12} \hat{T}_{34}\ldots \hat{T}_{2L-1\,2L}  
\end{eqnarray}
$\hat{T}_{i i+1},\hat{L},\hat{R}$ when acting on $f(\tau_1 \dots \tau_N)$
generate the weights of configurations from which 
$\tau_1 \dots \tau_N$ could have been reached by 
a transition associated with that bond or boundary site,
multiplied by the transition probability.

We may write
\begin{eqnarray}
\hat{T}_{i\,i+1} = 
\left(
\begin{array}{cccc}
     1&0&0&0\\
     0&1-q&p&0\\
     0&q&1-p&0\\
     0&0&0&1
\end{array}
\right)\label{Trep}\\[1ex]
\hat{L} = 
\left(
\begin{array}{cc}
     1-\alpha&0\\
     \alpha &1
\end{array}
\right)\qquad
\hat{R} = 
\left(
\begin{array}{cc}
     1&\beta\\
     0 &1-\beta
\end{array}
\right)\label{LRrep}
\end{eqnarray}
where the basis for $\hat{T}$ is $00$,$01$,$10$,$11$
and for $\hat{L}$ and $\hat{R}$ is $0$,$1$.
So, for example,
\begin{eqnarray}
\fl \lefteqn{\hat{T}_{i i+1} f( \cdots \tau_{i-1}\, 1\, 0\, \tau_{i+1}\cdots)
=}\\\
&&  p f( \cdots \tau_{i-1}\,0\, 1\, \tau_{i+1}\cdots)
+ (1-p) f( \cdots \tau_{i-1}\, 1\, 0 \, \tau_{i+1}\cdots)\nonumber\;.
\end{eqnarray}

The matrix product state for sublattice updating takes the form
\begin{equation}
\label{slmp}
f_N(\tau_1, \tau_2, \ldots, \tau_N) = \bra{W} \hat{X}_{\tau_1} X_{\tau_2}
\cdots \hat{X}_{\tau_{2L-1}}X_{\tau_{2L}} \ket{V}
\end{equation}
in which, for $i$ odd, $X_{\tau_i}$ is a matrix $D$ if site $i$ is
occupied by a particle ($\tau_i=1$) or a matrix $E$ otherwise; and
for, $i$ even, $\hat{X}_{\tau_i}$ is a matrix $\hat{D}$ if site $i$ is
occupied by a particle or a matrix $\hat{E}$ otherwise.  Note that
different matrices, hatted and unhatted, are used for the odd and even
sublattices.

The cancellation mechanism is as follows
\begin{eqnarray}
\hat{T}_{i\,i+1} \, \left[ X_{\tau_i} \hat{X}_{\tau_{i+1}} \right] \;=\;
\hat{X}_{\tau_i}  X_{\tau_{i+1}}
\label{slcancel}
\\
\langle W|  \hat{L} \hat{X}_{\tau_1} \;=\; \langle W| X_{\tau_1}\qquad
 \hat{R} X_{\tau_{2L}} |V\rangle\;=\; \hat{X}_{\tau_{2L}} |V\rangle \,.
\label{slbccancel}
\end{eqnarray}
Note that the action of the first half time step transfer matrix $\hat{T}_1$
is to put hatted matrices at the even sites and unhatted matrices on the odd sites, 
then the action of 
$\hat{T}_2$ is to restore them to their original sites.
Thus,  the matrix product
state (\ref{slmp}) is indeed an eigenvector with eigenvalue one of the full transfer matrix (\ref{fulltm}).
Using the  form (\ref{Trep},\ref{LRrep}) of the operators,
the mechanism (\ref{slcancel}, \ref{slbccancel})
implies the following algebraic relations
\begin{eqnarray}
\label{spbulk}
[E,\hat{E}] &=& [D,\hat{D}] \;=\; 0 \nonumber \\
\hat{E}D &=& (1-q)E\hat{D} + p D \hat{E} \\
\hat{D}E &=& (1-p) D \hat{E} + q E \hat{D} \nonumber
\end{eqnarray}
and the boundary conditions
\begin{equation}
\label{spbc}
\begin{array}{c}
   \langle W| \hat{E} (1-\alpha) \;=\; \langle W| E  \\[1mm]
   \langle W| (\alpha \hat{E}+\hat{D}) \;=\; \langle W| D \end{array}
\qquad
\begin{array}{c}
   (1-\beta)D|V\rangle \;=\; \hat{D}|V\rangle \\[1mm]
   (E+\beta D)|V\rangle \;=\; \hat{E}|V\rangle \end{array}\,.
\end{equation}
Remarkably, the same matrices and vectors $D,E$ as in the random sequential
case can be used to solve these relations. To see this we make the ansatz
\cite{RS97}
\begin{equation}
\hat{E} = E + \lambda \One\quad ; \quad
\hat{D} = D - \lambda \One\; ,
\end{equation}
where $\lambda$ is a scalar to be fixed,
and we find that (\ref{spbulk}, \ref{spbc}) reduce to
\begin{eqnarray}
pDE - q ED = \lambda \left[ (1-p) D + (1-q) E\right]
\label{RS1}\\
   D|V\rangle \;=\; \frac{\lambda}{\beta}|V\rangle \quad ; \quad
 \langle W |E  \; =\;  \langle W |\frac{\lambda (1-\alpha)}{\alpha}\;.
\label{RS2}
\end{eqnarray}   
Finally we may set
\begin{equation}
\lambda^2 = \frac{1}{(1-q)(1-p)}
\end{equation}
and  define
\begin{equation}
E' = \lambda(1-q)E \quad ; \quad
D' = \lambda(1-p)D
\end{equation}
which then satisfy
\begin{eqnarray}
pD'E' - q E'D' = D' + E'\\
   D'|V\rangle \;=\; \frac{1}{(1-q)\beta}|V\rangle \quad ; \quad
 \langle W |E'  \; =\;  \langle W |\frac{(1-\alpha)}{\alpha(1-p)}\;.
\end{eqnarray}   
Thus, we have rewritten (\ref{RS1},\ref{RS2}) in terms of matrices
obeying the usual algebra for the PASEP \PRR\ with redefined $\alpha$
and $\beta$.  From this rewriting, the phase diagram can, in principle,
be deduced using the general results of \cite{BEK00}.

\subsection{Ordered sequential dynamics}

Another discrete-time  updating scheme is to update
each site in a fixed sequence in each time step.  Two particularly
obvious choices of sequence are as follows.

\paragraph{Forward-ordered update} Here, the timestep begins by updating
site 1 wherein if site 1 is empty a particle enters with probability
$\alpha$.  Next, the bond $1,2$ is updated such that if site 1 is
occupied and site 2 is empty the particle moves forward with
probability $p$, or else if site 2 is occupied and site 1 is empty the
particle moves backward with probability $q$.  Then the bonds $i,i+1$ for $i =2 \dots N-1$
are similarly updated in order.  The time step concludes with site $N$
being updated wherein if site $N$ is occupied the particle leaves the
system with probability $\beta$.  Note that in the forward sequence it
is possible for a particle to move several steps forward in one
timestep.

\paragraph{Backward-ordered update} In this case, the time step begins
by updating site $N$, then bonds $N-1,N$ to $1,2$ in backward sequence
and finally site $1$, where all the individual updates follow the
usual rules, as described above.  Note that due to particle-hole
symmetry the dynamics of the vacancies in the backward order is the
same as the updating of particles in the forward order. Therefore
there is a symmetry between the steady states for the forward and
backward orders.

\bigskip

Let us focus on the backward-ordered dynamics.  We construct the
transfer matrix for the whole timestep as a product of operators for
each update of the sequence
\begin{equation}
\hat{T} = \hat{L} \hat{T}_{12} \hat{T}_{23}\ldots \hat{T}_{N-1 N} \hat{R}  
\end{equation}
where
$ \hat{L}, \hat{T}_{ii+1},\hat{R}  $ are as in
(\ref{Trep},\ref{LRrep}).

The matrix product solution is of the usual form
\begin{equation}
f_N(\tau_1, \tau_2, \ldots, \tau_N) = \bra{W} {X}_{\tau_1} X_{\tau_2}
\ldots X_{\tau_{N}} \ket{V}
\label{mss}
\end{equation}
The cancellation mechanism is precisely the same as for the sublattice
parallel case (\ref{slcancel}, \ref{slbccancel}).  In the ordered case
the hat matrices do not appear in the steady state weights at the end
of the time step, rather they are auxiliary matrices which appear
during the update procedure and move from right to left through the
lattice.

We conclude that backward ordered updating has the exact same phase
diagram as sublattice updating, although exact expressions for
correlation functions do depend on the details of the
updating\cite{RSSS98,ERS99}. Finally the phase diagram for forward ordered updating can
be obtained from the particle-hole symmetry mentioned above.
Current and density profiles have been calculated in \cite{BPV00,BP01}.

\subsection{Fully parallel dynamics}

In parallel dynamics (sometimes referred to as fully parallel to
distinguish from the sublattice updating) {\em all} bonds and boundary
sites are updated simultaneously.  This dynamics is  considered
the most natural for modelling traffic flow \cite{CSS00}.  In an
update at the bond $(i,i+1)$, if site $i$ is occupied and site $i+1$
is empty the particle moves forward with probability $p$.  Under this
type of updating scheme one cannot include reverse hopping without
introducing the possibility of conflicts occuring, or of a particle
hopping to two places at once.  Also note that it is the occupancies
at the beginning of the timestep which determine the dynamical events.

In this case the matrix product solution is of the usual form
(\ref{mss}) but the algebraic relations have a more complicated
structure which was first elucidated in \cite{ERS99}. There it was
found  that recursion relations between systems of different sizes were
higher than first order: for example the relation
\begin{equation}
\fl f_N(\ldots 0 1 0 0 \ldots)
=(1-p) f_{N-1}(\ldots 0 1 0 \ldots) + f_{N-1}(\ldots 0 0 0 \ldots)
+pf_{N-2}(\ldots 0  0 \ldots)\;,
\label{example_recur}
\end{equation}
which relates the weights of size $N$ to those of size $N-1$ and size $N-2$,
was discovered to hold.
This in turn implies that algebraic relations between the operators
are {\em quartic} rather than quadratic.  For example the rules in the
bulk are %
\begin{eqnarray}
{ EDEE} &=& (1-p){ EDE} + { EEE} + p{ EE}
\label{bulk1}\\
 { EDED} &=&  { EDD} + { EED} + p{ ED}
\label{bulk2}  \\
{ DDEE} &=& (1-p){ DDE} + (1-p){ DEE} + p(1-p){ DE}
 \label{bulk3}\\
{ DDED} &=& { DDD} + (1-p){ DED} + p{ DD}
\label{bulk4}
\end{eqnarray}
and there are other rules which we do not quote here for matrices near to
the boundary (see \cite{ERS99}).  These rules were proved using a
domain approach related to, but more complicated than, that presented in
in Section~\ref{RRproof} \cite{ERS99}. Subsequently, an algebraic
proof similar in spirit to the cancellation mechanism for the ordered
sequential updating case was found \cite{dGN99} but it is still too
involved to present here.

The quartic algebra, (\ref{bulk1})--(\ref{bulk4}), as well as the
other conditions mentioned above, can be reduced to quadratic rules by
making a convenient choice for the operators involved. The trick is to
write
\begin{eqnarray}      
D = \left( \begin{array}{cc}  D_1 & 0  \\ D_2 & 0 \end{array} \right)
\;, \quad
E = \left( \begin{array}{cc}  E_1 & E_2\\  0  & 0 \end{array} \right)
\;,
\label{DE}
\end{eqnarray}
where $D_1$, $D_2$, $E_1$, and $E_2$ are matrices of, in
general, infinite dimension; that is, $D$ and $E$ are written as rank
four tensors with two indices of (possibly) infinite dimension and the
other two indices of dimension two.  Correspondingly, we write
$\langle { W}|$ and $| { V} \rangle$ in the form
 \begin{eqnarray}
\langle { W}| = (\ \langle W_1|,\ \langle W_2|\ )
  \;,\;\;\; 
| { W}\rangle =
 \left( \begin{array}{c}| V_1 \rangle \\ | V_2 \rangle \end{array} \right) \;,
 \label{WV}
\end {eqnarray}
 where $\langle W_1|$, $\langle W_2|$, $| V_1 \rangle$, 
and $| V_2 \rangle $ are
vectors of the same dimension as $D_1$ and $E_1$.  
$D_1$, $E_1$, $\langle W_1|$, and
$|V_1\rangle$ satisfy the quadratic relations
\begin{equation}
D_1 E_1 = (1-p)\left[ D_1 +E_1 +p  \right] \;, \label{D1E1con}
\end{equation}\begin{equation}
D_1 | V_1 \rangle = \frac{p(1-\beta)}{\beta} |V_1\rangle \;\;\;,\;\;\;
\langle W_1|E_1 =  \langle W_1| \frac{p(1-\alpha)}{\alpha}, \label{D1V1}
\end{equation}
 and $D_2$, $E_2$, $\langle W_2|$, and $|V_2\rangle$ satisfy 
\begin{equation}
E_2 D_2 = p \left[ D_1 +E_1 +p  \right], \label{E2D2con}
\end{equation}\begin{equation}
E_2 | V_2 \rangle = p|V_1\rangle \;\;\;,\;\;\;
\langle W_2|D_2 = \langle W_1|p. \label{E2V2}
\end{equation}
and $\langle W_1|$ satisfying (\ref{D1E1con},\ref{D1V1}) are presented
in \cite{ERS99}.  In addition, one can choose $D_2 \propto E_1$, $E_2
\propto D_1$, $\langle W_2|\propto \langle W_1|$, $| V_2 \rangle
\propto |V_1\rangle$ so that (\ref{E2D2con}), (\ref{E2V2}) reduce to
(\ref{D1E1con}), (\ref{D1V1}).  Along the curve
\begin{equation}
(1-\alpha)(1-\beta)=1-p\;,
\label{specialline}
\end{equation} 
scalar representations of $D_1$, $E_1$, $|V_1\rangle$ and $\langle
W_1|$ can be found and ${ D,E}$ are $2\times 2$ operators.

The continuous time limit is recovered by setting $p= \rmd t$,
 replacing $\alpha \to \alpha \rmd t$, $\beta \to \beta \rmd t$ and
 letting the time step $\rmd t \to 0$.  Then
 (\ref{D1E1con},\ref{D1V1}) reduce to the usual ASEP quadratic algebra
 \RR.

\begin{figure}
\label{parpd}
\begin{center}
\includegraphics[scale=0.7]{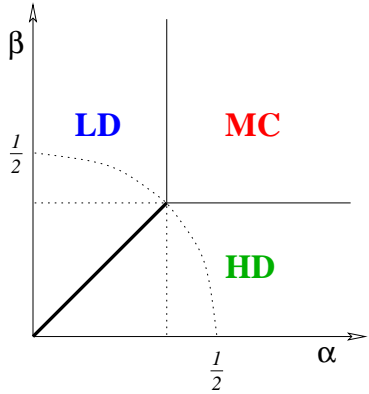}
\caption{Phase diagram for the ASEP with 
parallel  update. Note that $\alpha \leq 1$, $ \beta \leq 1$.
MC is the maximum current
phase, LD and HD are the low and high-density phases, respectively.
The straight dashed lines are the boundaries between 
the low density and  maximal current phases and the high density
and  maximal current phase at $\alpha = 1- \sqrt{1-p}$ and $\beta = 1-
\sqrt{1-p}$ respectively.
The curved dashed line
is the line given by (\ref{specialline}) and intersects the line
$\alpha=\beta$ at $\alpha=\beta=1-\sqrt{1-p}=q$ }
\end{center}
\end{figure}

The phase diagram is presented in Figure~\ref{parpd}.  It appears
similar to the familiar continuous time phase diagram except that the
transition lines from low density to maximal current and high density
to maximal current are at $\alpha = 1- \sqrt{1-p}$ and $\beta = 1-
\sqrt{1-p}$ respectively. In the continuous time limit described above
one recovers $\alpha=1/2$ and $\beta = 1/2$. Also, in the limit
of deterministic bulk  dynamics, $p \to 1$ \cite{TE98} 
the maximal current phase 
disappears from the phase diagram as is also the case in 
other updating schemes in this limit.

\section{Summary and outstanding challenges}
\label{fin}

In this work we have reviewed the physical properties of systems that
are driven out of equilibrium, with particular reference to those
one-dimensional models that can be solved exactly using a matrix
product method.  The most prominent physical phenomena exhibited by
these systems are phase transitions which in open systems are induced
by changes in the interactions with the boundaries, and in periodic
systems by the addition of particle species that have different
dynamics.  Additionally we have seen in a number of cases the
formation of shocks in these one-dimensional systems.

As we have also discussed, the matrix-product approach can be used to
determine the steady-state statistics of models with a variety
of microscopic dynamics.  The majority of models that have matrix
product states involve diffusing particles with hard-core interactions
and whose numbers are conserved, except possibly at boundary sites.
Under such conditions, certain generalisations to multi-species models
have been found.  Furthermore, there are a few cases in which
non-conserving particle reactions can occur that also admit a
convenient solution in terms of matrix products.

However, it has to be said that the dream scenario, of being able to 
systematically construct  any
nonequilibrium steady state  in matrix product form starting from the 
microscopic dynamics of the system, appears a remote goal.
Although the existence proofs discussed in Section~\ref{formal}
tell us that a matrix product formulation should nearly 
always be possible, we are in practice still feeling our way
with particular examples.

There remain a number of simple physical systems that exhibit
nontrivial out-of-equilibrium behaviour, but that have nevertheless
been unyielding to exact solution by matrix products, or any other
means.  It is perhaps appropriate here to highlight some of these
cases as challenges for future research.

\subsection{Particle- and site-wise disorder in the ASEP}

As was discussed in Section~\ref{pwdis}, it is possible to solve for
the steady state of the ASEP on the ring when particles have different
hop rates, as would occur when each particle is randomly assigned a
hop rate from some distribution.  As far as we aware, the
corresponding dynamics with open boundaries has not been solved apart
from the Karimipour's model of overtaking which we reviewed in
Section~\ref{sec:Karimipour}.  Furthermore, if the disordered hop
rates are associated with sites, rather than particles, even the model
on the ring has so far eluded a complete exact solution.  Indeed, even
if only a single site has a different hop rate to the rest, the steady
state does not appear to have a manageable form \cite{JL92,JL94}.
On the other hand, some nontrivial symmetries in the disordered case have
been identified \cite{TB98,KP00}.

One phenomenon that can emerge when disorder is present is phase
separation.  
One sees this most clearly for the periodic system with
particle-wise disorder, where as we noted
in Section~\ref{BEC},  a condensation of particles
behind the slowest  may occur \cite{Evans96}.  With site-wise disorder
\cite{TB98}, a
flattening of the macroscopic current-density relation $J(\rho)$ is
seen (recall that it is parabolic $J(\rho)=\rho(1-\rho)$in the ASEP
without disorder), and the densities at which transitions into a
maximal current phase are correspondingly lowered \cite{HS04b}; it has
been suggested that a linear portion in the $J(\rho)$ curve may also
provide a mechanism for phase separation \cite{Krug00}.  
Another interesting observation is that under
site-wise disorder   the location of the first-order transition
for the ASEP  becomes sample-dependent (i.e., the free-energy-like quantity for the nonequilibrium system is not self-averaging) \cite{ED04}.

\subsection{Driven $n$-mers with open boundaries}

In all the models described in this work, particles occupy a single
site of a lattice.  If one has extended objects on a ring, the
dynamics are unaffected.  With open boundaries, however, the situation
changes and even if there is a single particle species and all
particles are the same length, a matrix product solution for the
steady state has proved elusive.  Although it is quite easy to arrange
for cancellation of terms in the master equation corresponding to
particles joining and leaving domains (see Section~\ref{RRproof}), the
necessary cancellation with boundary terms is much harder to arrange:
it is not even clear what the most convenient choice of boundary
conditions should be (e.g., extended particles sliding on and off
site-by-site, or appearing and disappearing in their entirety, or a
mixture of the two).  A phenomenological diffusion equation for the
macroscopic density profile in the open system with extended objects
appearing and disappearing at both boundaries has been presented
\cite{SZL03}.  Furthermore, it was conjectured in that work to be the
equation that would result in the continuum limit (along the lines of
that taken in Section~\ref{sec:mfh}), if an exact matrix-product
solution of this model were to be found.  Whether this should turn out
to be the case or not, an application of the extremal current
principle (see \ref{sec:PD}) gives predictions for the phase diagram
that are in good agreement with simulation data: the same three phases
that emerge in the ASEP are seen, but with shifted transition points.
Meanwhile an approach to this problem based on a local-equilibrium
approximation \cite{LC03} has shown excellent agreement with Monte
Carlo results and have also been conjectured to be exact.  The case of
the open system with extended objects and a single spatial
inhomogeneity in the hop rates has also been considered \cite{SKL04}.

\subsection{Bridge model}

The bridge model \cite{EFGM95a,EFGM95b} comprises oppositely-charged
particles in the sense of Section~\ref{2species}.  Positive charges
enter at the left at rate $\alpha$, hop with unit rate to the right
and exit at the right boundary at rate $\beta$.  The dynamics of
negative charges is exactly the same, but in the opposite direction.
When two opposing particles meet they may exchange with some rate $q$.
Thus this model has both a parity and charge-conjugation symmetry.  In
certain parameter ranges, this symmetry is reflected in the steady
state that results: both positive and negative charges flow freely in
their preferred directions.  In particular, the limit
$\alpha\to\infty$ at $\beta=1$, for which an exact solution in terms
of matrix products exists, is included within this range.  On the
other hand, when $\beta$ is small
the model's symmetries are found to be broken.
This has been proved in the $\beta \to 0$ limit
\cite{GLEMSS95} but
there is no  known exact solution for general $\beta$.
In a finite system,
the current flips between the positive and negative direction, but as
the system size increases the time between flips increases
exponentially, and so in the thermodynamic limit only the state with
positive or negative current state is seen.  It is unclear whether the
structure of the matrix product solution allows for the description of
such symmetry-broken states.  
This may also be related to whether the boundary conditions correspond
to particle reservoirs with well-defined densities \cite{PS04}

\subsection{ABC model}

The ABC model \cite{EKKM98a,EKKM98b} is another deceptively simple
model that has yet to be completely solved.  This model has a ring
that is fully occupied by particles from three species, which are
labelled $A$, $B$ and $C$ and exhibit a symmetry under cyclic
permutation of the labels.  Specifically, the rates at which
neighbouring particles exchange are
\begin{equation}
AB \mathop{\rightleftharpoons}\limits_1^q BA \qquad
BC \mathop{\rightleftharpoons}\limits_1^q CB \qquad
CA \mathop{\rightleftharpoons}\limits_1^q AC \;.
\end{equation}
That is, if $q<1$, $A$ particles prefer to be to the left of $B$s,
$B$s to the left of $C$s and $C$s to the left of $A$s.  This model was
 discussed in Section~\ref{sec:ABC}  and was
shown to be exactly solvable when the numbers of $A$, $B$ and $C$
particles are all equal as a consequence of detailed balance being
satisfied.  Here a phase separation is seen when $q\ne1$. If $q=1$, on
the other hand, all configurations become equivalent and the
stationary distribution is uniform.  When there is even the slightest
imbalance and particle numbers \emph{and} hop rates, a small  current starts
to flow and nothing is known about the properties of the stationary
distribution.

The model is of particular importance because in the special case of
equal numbers of each species, it is one of few models along with the
SSEP and KMP model \cite{BGL05} where a free energy functional for the
density profile is known.

\subsection{KLS model}

We end by mentioning the original paradigm of a system driven out of
equilibrium, the model due to Katz, Lebowitz and Spohn
\cite{KLS83,KLS84,SZ95}.  This model has Ising-like interactions between
pairs of neighbouring spins and evolves by neighbouring spins
exchanging places (Kawasaki dynamics) but with a symmetry-breaking
external field that favours hops of up-spins along a particular axis.
When the system is open, or has periodic boundaries, a nonequilibrium
steady state ensues.  In the limit where spins become non-interacting
and there is only one spatial direction, the partially asymmetric
exclusion process (PASEP) described in Section \ref{pasep} is
recovered.  When the spins interact, an Ising-like steady state that
involves products of $2\times 2$ matrices (as described in Section
\ref{ising}) can be found, albeit only if certain conditions on the
hop rates are satisfied. (These conditions do admit nonequilibrium steady
states in which a current flows, however).  In two dimensions, one
has, of course, a finite critical temperature in the absence of a
driving force, and there has been particular interest in the nature of
the low-temperature ordered phase in the presence of an external drive
\cite{SZ95}.

\subsection{Fixed random sequence and shuffled dynamics}

We discussed in Section~\ref{discretetime} different updating schemes
for which matrix product steady states have been determined.  Here we
mention some other updating schemes which have not been solved so far.
First an ordered discrete time sequential updating scheme could have
the order as a randomly chosen sequence.  The forward and backward
schemes of Section~ref{discretetime} correspond to special sequences.  It would be of
interest to construct the a matrix product steady state for an
arbitrary fixed sequence.

Recently a {\em shuffled} updating scheme has been considered
\cite{WSS06,SW07}, in which a new random sequence is chosen at each
timestep.  It has been argued that this scheme is relevant to the
modelling of pedestrian dynamics.  The shuffled updating scheme
guarantees that each site or bond is updated exactly once in each
timestep but is more stochastic than an ordered sequential scheme.
Again it would be of interest to see if a matrix product could be used
to describe the steady state.


\section*{Acknowledgements}

We thank all the many colleagues with whom it has been a pleasure to
discuss nonequilibrium matrix product states over the years.  
In particular we
acknowledge our collaborators and other authors whose work we have
summarised.  For their useful discussions and insightful comments
during the preparation of this manuscript we would wish to thank
Bernard Derrida, Rosemary Harris, Des Johnston, Joachim Krug, Kirone
Mallick, Andreas Schadschneider, Gunther Sch\"utz and Robin
Stinchcombe.  RAB further acknowledges the Royal Society of Edinburgh
for the award of a  Research Fellowship.
\appendix

\renewcommand{\thesection}{\Alph{section}}

\section{Method of characteristics}
\label{char}

The continuity equation (\ref{cty2}) is a particularly simple case of
a first order quasi-linear differential equation for the density
\begin{equation}
a(x,t,\rho) \frac{\partial \rho}{\partial t} + 
b(x,t,\rho) \frac{\partial \rho}{\partial x}  =
c(x,t,\rho)
\label{cty3}
\end{equation}
The term on the right hand side would represent a source or sink of
particles and is relevant for example to the case where creation and
annihilation processes exist. Such an equation can generally be solved
by the method of characteristics \cite{Debnath97} which involves
identifying the characteristic curves along which information from the
boundary or initial condition propagates through the space-time domain.
The characteristic curves satisfy
\begin{equation}
\frac{\D t}{a} =
\frac{\D x}{b} = 
\frac{\D \rho}{c} 
\end{equation}
of which the two independent conditions may be written
\begin{equation}
\frac{\D x}{\D t} = \frac{b}{a}\qquad
\frac{\D \rho}{\D t} = \frac{c}{b}
\label{xrho}
\end{equation}
These equations
generally have two families of solution
\begin{equation}
\phi(x,t,\rho(x,t)) =C_1
\qquad \psi(x,t,\rho(x,t)) = C_2
\end{equation}
Then the solution to (\ref{cty3})
is of the form
\begin{equation}
F(\phi,\psi)=0
\end{equation}
where the function $F$ is fixed by the initial data.

For the case (\ref{cty2}), i.e. $c=0$, the characteristics in the
$x$--$t$ plane are straight lines with slope $v_g(\rho)$ and $\rho$ is
constant along these lines, as can be seen from (\ref{xrho}).  More
generally the characteristics are curves in the $x$--$t$ plane and the
density varies along the characteristic.  This can be seen by
integrating (\ref{xrho})
\begin{equation}
G(\rho)-G(\rho_0) =t \qquad \mbox{where}\quad
G(\rho)= \int^\rho \rmd \rho 
 \frac{b(\rho)}{c(\rho)}
\label{patden}
\end{equation}
which gives $\rho$ implicitly as a function of $t$, then
the trajectory of the characteristic is given by
\begin{equation}
x = x_0 + \int_0^t b(\rho(t')) \rmd t' \;.
\label{patpos}
\end{equation}
If we consider a patch of the initial density profile
of density $\rho_0$ at $x_0$, its density evolves according to
(\ref{patden}) and (\ref{patpos}) gives the position
of the patch.
These more complicated characteristics
can produce a richer for boundary induced phase
transitions \cite{PFF03,EJS03} including stationary shocks.

\section{Generating functions and asymptotics}
\label{gf}

In this appendix, we collect together general formul\ae\ for obtaining
both exact and asymptotic expressions for the coefficients appearing
in a generating function if the latter is known.  We refer the reader
to \cite{Wilf94} for proofs and more detailed discussion.  For
brevity, we introduce the notation $\coeff{x^n} f(x)$ to mean ``the
coefficient of $x^n$ in the (formal) power series $f(x)$''.

\paragraph{Lagrange Inversion Formula}  If a generating function $f(x)$
satisfies the functional relation
\begin{equation}
f(x) = x \Phi(f)
\end{equation}
where $\Phi(f)$ satisfies $\Phi(0) = 1$, one has an expression for $x$
as a function of $f$ which can be inverted to find $f$ as a function
of $x$.  This procedure then allows one to determine with relative
ease the coefficient of $x^n$ in the expansion of any \emph{arbitrary}
function $F(f)$.  The formula for doing so reads
\begin{equation}
\label{gf:lif}
\coeff{x^n} F(f) = \frac{1}{n} \coeff{f^{n-1}} \left[ F^\prime(f)
  \Phi(f)^n \right] \;.
\end{equation}

\paragraph{Asymptotics}  If a generating function $f(x)$ has a
singularity at a point $x_0$ in the complex plane, it contributes a
term $A x_0^{-n} n^{-\mu}$ to the coefficient $\coeff{x^n} f(x)$,
where $A$ and $\mu$ are constants to be determined.  For sufficiently
large $n$, this coefficient is dominated by the singularity closest to
the origin.  To be more precise, we say that two sequences $a_n$ and
$b_n$ are asymptotically equivalent, denoted $a_n \sim b_n$, if
\begin{equation}
\lim_{n \to \infty} \frac{a_n}{b_n} = 1 \;.
\end{equation}
Then, we have that $\coeff{x^n} f(x) \sim A x_0^{-n} n^{-\mu}$.

Let us decompose the generating function $f(x)$ into its regular and
singular parts around $x_0$, denoting the former by $f_r(x)$.  That
is,
\begin{equation}
f(x) = f_r(x) \left(1 - \frac{x}{x_0}\right)^{-\nu} \;.
\end{equation}
The cases that interest us are poles $\nu=1,2,\ldots$ and algebraic
singularities (noninteger $\nu$).  For the former case one finds
\begin{equation}
\label{gf:pole}
\coeff{x^n} f(x) \sim {\nu + n -1 \choose n} f_r(x_0) x_0^{-n} \;.
\end{equation}
For sufficiently large $n$, the binomial coefficient behaves as
\begin{equation}
{\nu + n-1 \choose n} \sim \frac{n^{\nu - 1}}{(\nu-1)!} \;.
\end{equation}
Replacing the factorial with a Gamma function gives the extension of
(\ref{gf:pole}) to algebraic singularities
\begin{equation}
\label{gf:sing}
\coeff{x^n} f(x) \sim \frac{f_r(x_0)}{\Gamma(\nu)} x_0^{-n} n^{\nu-1} \;.
\end{equation}
The prefactor $f_r(x_0)$ in this expression can be obtained by taking
the limit
\begin{equation}
f_r(x_0) = \lim_{x\to x_0} \left( 1 - \frac{x}{x_0} \right)^{\nu} f(x) \;.
\end{equation}

\section{Equivalence of the integral and sum representations of the
  ASEP normalisation}
\label{intsum}

In order to show that the integral representation of the normalisation
(\ref{asep:Z0}) is equivalent to the summation (\ref{open:Z2}),
 the simplest approach is to show that the
generating functions for $Z_N$ computed from the two expressions are
equal.  As we have seen in Section~\ref{gfa} the generating function
computed from the finite sum (\ref{open:Z2}) yields
(\ref{open:Zed}) which can be developed further to give
\begin{eqnarray}
\Zed(z) &\equiv&  \sum_{N=0}^\infty z^N Z_N\label{gendefa}\\
&=& \frac{\alpha \beta \left[ 2\alpha\beta - 2z - (\alpha+\beta
 -1)(1+ \sqrt{1-4z})\right]}{2 \left[z-\alpha(1-\alpha)\right]
 \left[z-\beta(1-\beta)\right]}\;.\label{open:Zed2}
\end{eqnarray}

We now turn to the integral representation
(\ref{asep:Z0}) which 
may be  recast as a contour integral by change of variable
$u= \rme^{i\theta}$
\begin{equation}
Z_N = \frac{(1-ab)}{4\pi i} \oint_{K} \frac{\rmd
  u}{u} \frac{(2 - u^2 - u^{-2}) (2 + u +
  u^{-1})^N}{(1-au)(1-au^{-1})(1-bu)(1-bu^{-1})}
\label{Zinta}
\end{equation}
where $K$ is the (positively oriented) unit circle
in the complex $u$ plane. 

We compute the generating function (\ref{gendefa})
using (\ref{Zinta}). Summing a geometric series (which converges for
$|z|<1/4$) we find
\begin{equation}
\fl \Zed(z) = \frac{(1-ab)}{4\pi i} \oint_{K} \frac{\rmd
  u}{u} \frac{(2 - u^2 - u^{-2})} {(1-au)(1-au^{-1})(1-bu)(1-bu^{-1})
(1-z(2+u+u^{-1}))}
\label{Finta}
\end{equation}
We now use the residue theorem to evaluate this integral. We take $a$
and $b<1$ (although we get the same final result for other ranges of
$a$ and $b$).  In this case the singularities within the unit circle
are simple poles at $u=a$, $u=b$ and $u=u_-$ where we define
\begin{equation}
\label{upm}
u_\pm = 1- \frac{1}{2z}(1 \pm \sqrt{1-4z})
\end{equation}
The residues from the three poles yield
\begin{eqnarray}
\Zed(z) &=& \frac{(1-a^2)}{2(a-b)} \frac{\alpha^2}{\left[z-\alpha(1-\alpha)\right]}
-\frac{(1-b^2)}{2(a-b)} \frac{\beta^2}{\left[z-\beta(1-\beta)\right]}\\
&&
-\frac{(1-ab)(2-u_-^2-u_-^{-2})}
{2z(1+a^2-a(u_-+u_-^{-1}))(1+b^2-b(u_-+u_-^{-1}))(u_- - u_+)}\;.
\end{eqnarray}
Some simple algebra then shows that this expression is equivalent
to (\ref{open:Zed2}).

\section{Matrix  representations}
\label{appreps}

Here, we collect together some representations of the various matrix
product algebras that have been discussed in this work.

\subsection{Representations of PASEP algebra}
We begin by recapping representations for the PASEP algebra
in the case  of  injection at the left boundary and extraction at the right boundary
\begin{eqnarray}
\label{pasep:DEa}
DE - qED &=& D + E \\
\beta D \ket{V} &=& \ket{V}
\label{pasep:EWb}\\
\alpha \bra{W}  E &=& \bra{W}  \;.
\label{pasep:EWa}
\end{eqnarray}
These representations  are generalisations of the three
representations for  the totally asymmetric case first given in
\cite{DEHP93}.

From Section~\ref{alsalam} we have
\begin{eqnarray}
D &=& \frac{1}{1-q} \left( \begin{array}{ccccc}
1+b & \sqrt{c_0} & 0 & 0 & \cdots\\
0 & 1+bq & \sqrt{c_1} & 0 &\\
0 & 0 & 1+bq^2 & \sqrt{c_2} & \\
0 & 0 & 0 & 1+bq^3 &  \\
\vdots & & & & \ddots
\end{array} \right) \\
E &=& \frac{1}{1-q} \left( \begin{array}{ccccc}
1+a & 0 & 0 & 0 & \cdots\\
\sqrt{c_0} & 1+aq & 0 & 0 & \\
0 & \sqrt{c_1} & 1+aq^2 & 0 &\\
0 & 0 & \sqrt{c_2} & 1+aq^3 &\\
\vdots & & & & \ddots
\end{array} \right)
\end{eqnarray}
where
\begin{equation}
\label{pasep:aba}
a = \frac{1-q}{\alpha} - 1 \;,\quad
b = \frac{1-q}{\beta} - 1 \;,\quad
c_n  = (1-q^{n+1}) (1-ab q^n) \;,
\label{abcndef}
\end{equation}
and  the boundary vectors are
\begin{equation}
\bra{W} = \bra{0} \quad\mbox{and}\quad
\ket{V} = \ket{0}\;.
\end{equation}
From this representation one sees that for  certain parameter curves, namely
\begin{equation}
1-ab q^n = 0\;,
\end{equation}
that the representations become finite dimensional since
$c_n =0$ and the upper left corner of the matrices  become
disconnected from  the rest. Curves of this type 
were first noted in the more general case $\gamma,\delta \neq0$
in \cite{ER96} and finite dimensional representations were catalogued  in \cite{MS97}.

In the limit $q\to 1$ (\ref{pasep:D1}) and 
(\ref{pasep:D1}) have well-defined limits
\begin{eqnarray}
D &=&  \left( \begin{array}{ccccc} 
1/\beta & \sqrt{g} & 0 & 0 & \cdots\\
0 & 1+ 1/\beta & \sqrt{2g} & 0 &\\
0 & 0 & 2+ 1/\beta & \sqrt{3g} & \\
0 & 0 & 0 & 3+ 1/\beta &  \\
\vdots & & & & \ddots
\end{array} \right) \\
E &=&  \left( \begin{array}{ccccc}
1/\alpha  & 0 & 0 & 0 & \cdots\\
\sqrt{g} & 1+1/\alpha & 0 & 0 & \\
0 & \sqrt{2g} & 2+1/\alpha & 0 &\\
0 & 0 & \sqrt{3g} & 3+1/\alpha &\\
\vdots & & & & \ddots
\end{array} \right)
\end{eqnarray}
where $\ds g = \frac{1}{\alpha} + \frac{1}{\beta}$.

A second representation of (\ref{pasep:DEa}--\ref{pasep:EWa}) given in Section~\ref{pasymp}
is
\begin{eqnarray}
\label{pasep:D2a}
D &=& \frac{1}{1-q} \left( \begin{array}{ccccc}
1 & \sqrt{1-q} & 0 & 0 & \cdots\\
0 & 1 & \sqrt{1-q^2} & 0 &\\
0 & 0 & 1 & \sqrt{1-q^3} & \\
0 & 0 & 0 & 1  &  \\
\vdots & & & & \ddots
\end{array} \right) \\
\label{pasep:E2b}
E &=& \frac{1}{1-q} \left( \begin{array}{ccccc}
1 & 0 & 0 & 0 & \cdots\\
\sqrt{1-q} & 1 & 0 & 0 & \\
0 & \sqrt{1-q^2} & 1 & 0 &\\
0 & 0 & \sqrt{1-q^3} & 1  &\\
\vdots & & & & \ddots
\end{array} \right)
\end{eqnarray}
with 
\begin{equation}
\braket{W}{n} = \kappa \frac{a^n}{\sqrt{(q;q)_n}} \quad\mbox{and}\quad
\braket{n}{V} = \kappa \frac{b^n}{\sqrt{(q;q)_n}}
\end{equation}
where the parameters $a$ and $b$ are  given in (\ref{abcndef}),
and $\kappa$ is a constant usually 
chosen so that $\braket{V}{W} = 1$. For these matrices,
the $q\to 1$ limit is not defined.

A third  representation, which we have not yet encountered, is
\begin{eqnarray}
D &=& \left( \begin{array}{ccccc}
1/\beta  & 1/\beta & 1/\beta & 1/\beta  & \cdots\\
0 & d^{(11)} & d^{(12)} &  d^{(13)} &  \cdots\\
0 & 0   & d^{(22)} &  d^{(23)} & \cdots\\
0 & 0 & 0 &  d^{(33)}  &  \\
\vdots & & & & \ddots
\end{array} \right) \nonumber \\
E &=&  \left( \begin{array}{ccccc}
0 & 0 & 0 & 0 & \cdots\\
1 & 0 & 0 & 0 & \\
0 & 1 & 0 & 0 &\\
0 & 0 & 1 & 0 &\\
\vdots & & & & \ddots
\end{array} \right)
\label{DmuEalg}\\
\braket{W}{n}  &=& (  1/\alpha)^n
\quad \braket{n}{V} = \delta_{n,0}
\label{DE3}
\end{eqnarray}
where
\begin{equation}
d^{(ij)}= 
\sum_{m=0}^{i-1}q^m {j-i+m \choose m  } +  \frac{q^i}{\beta} 
{j \choose i } \;.
\end{equation}
In the $q\to 1$ limit the elements of $D$ reduce  to
\begin{equation}
d^{(ij)}= 
{j \choose i-1  } +  \frac{1}{\beta} 
{j \choose i } \;.
\end{equation}

\subsection{Representations of two-species  and multispecies algebras}
We now give  representations  corresponding to the physically relevant and nontrivial algebras classified in Section~\ref{2species} and the multispecies
generalisations of Section~\ref{sec:multispecies}.

\paragraph{Solution AII and multispecies generalisation} 

This is a multispecies model with matrices $E$ and $D(v)$,
with an algebra
\begin{eqnarray}
D(v) E  &=& \frac{1}{v} D(v) +E \nonumber\\
D(v) D(v') &=&  \frac{v}{v-v'}D(v') - \frac{v'}{v-v'}D(v)
\quad v >v' \nonumber\\
D(v) \ket{V} &=& \frac{1}{\beta(v)} \ket{V}\nonumber\\
\bra{W} E  &=& \frac{1}{\alpha}  \;.
\end{eqnarray}
where  $\beta(v)= \frac{v -(1-\beta)}{v}$.
One possible representation is
\begin{eqnarray}
D(v) &=& \left( \begin{array}{ccccc}
1/\beta(v)  & 1/(\beta(v)v) &  1/(\beta(v) v^2) &  1/(\beta(v) v^3) & \cdots\\
0 & 1 & 1/v &  1/v^2 &   \cdots\\
0& 0 & 1 & 1/v &     \cdots\\
0& 0& 0 & 1 &        \cdots\\
\vdots & & & & \ddots
\end{array} \right) \nonumber\\
E &=&  \left( \begin{array}{ccccc}
0 & 0 & 0 & 0 & \cdots\\
1 & 0 & 0 & 0 & \\
0 & 1 & 0 & 0 &\\
0 & 0 & 1 & 0 &\\
\vdots & & & & \ddots
\end{array} \right) \nonumber\\
\braket{W}{n}  &=& (  1/\alpha)^n
\quad \braket{n}{V} = \delta_{n,0}
\label{Karimrep}
\end{eqnarray}
This is the generalisation of the third PASEP representation (\ref{DE3})
in the case $q=0$.

\paragraph{Solution BI and BII} 

With the algebra (\ref{BIalg}), 
\begin{eqnarray}
p_1 X_1 X_0 - q_1 X_0 X_1 &=& x_0 X_1 -x_1 X_0 \nonumber\\
p_2 X_2 X_0 - q_2 X_0 X_2 &=& x_0 X_2 \nonumber\\
p_2 X_1 X_2 - q_2 X_2 X_1 &=& -x_1 X_2 
\end{eqnarray}
one has a representation
\begin{eqnarray}
\label{BIrep1}
X_1  &=& -\frac{x_1}{(p_2-q_2)} \left( \begin{array}{ccccc}
1 & \sqrt{c_0} & 0 & 0 & \cdots\\
0 & 1 & \sqrt{c_1} & 0 &\\
0 & 0 & 1 & \sqrt{c_2} & \\
0 & 0 & 0 & 1  &  \\
\vdots & & & & \ddots
\end{array} \right) \\
X_2 &=&  \left( \begin{array}{ccccc}
1 & 0 & 0 & 0 & \cdots\\
0 & (q_2/p_2) & 0 & 0 & \\
0 & 0 & (q_2/p_2)^2 & 0 &\\
0 & 0 & 0 & (q_2/p_2)^3 &\\
\vdots & & & & \ddots
\end{array} \right)\\
X_0 &=& \frac{x_0}{(p_2-q_2)} \left( \begin{array}{ccccc}
1 & 0 & 0 & 0 & \cdots\\
\sqrt{c_0} & 1 & 0 & 0 & \\
0 & \sqrt{c_1} & 1 & 0 &\\
0 & 0 & \sqrt{c_2} & 1 &\\
\vdots & & & & \ddots
\end{array} \right)
\label{BIrep0}
\end{eqnarray}
where
\begin{equation}
c_n = (1-(q_1/p_1)^{n+1})  \;.
\end{equation}
As noted in the main text a representation for solution BII
is to use a presentation of the PASEP algebra and write $X_2
= \ket{V}\bra{W}$, i.e.  a projector.

\paragraph{Solution CII and multispecies generalisation}
This is the case of the ASEP with disordered hopping rates.
The matrix algebra (\ref{msalg1})
\begin{equation}
p_\mu D_\mu E - q_\mu E D_\mu = D_\mu
\end{equation}
can be represented via
\begin{eqnarray}
D_{\mu} &=& \left( \begin{array}{ccccc}
d_\mu^{(00)} & d_\mu^{(01)} &  d_\mu^{(02)} & d_\mu^{(03)} & \cdots\\
0 & d_\mu^{(11)} & d_\mu^{(12)} &  d_\mu^{(13)} &  \cdots\\
0 & 0   & d_\mu^{(22)} &  d_\mu^{(23)} & \cdots\\
0 & 0 & 0 &  d_\mu^{(33)}  &  \\
\vdots & & & & \ddots
\end{array} \right) \nonumber \\
E &=&  \left( \begin{array}{ccccc}
0 & 0 & 0 & 0 & \cdots\\
1 & 0 & 0 & 0 & \\
0 & 1 & 0 & 0 &\\
0 & 0 & 1 & 0 &\\
\vdots & & & & \ddots
\end{array} \right)
\end{eqnarray}
where
\begin{equation}
d_\mu^{(ij)}= 
{i+j \choose i }  \frac{q_\mu^{i}}{p_\mu^{i+j}}\;.
\end{equation}

\paragraph{Solution D and 
Deformed commutators}  
As noted in \ref{sec:closeg} the Solution class D operators can be constructed
from tensor products of operators  obeying deformed
commutator relations.
The deformed commutator is defined as
\begin{equation}
X_1 X_2 - r X_2 X_1 = 0
\end{equation}
where $r$ is a deformation parameter: $r=1$ corresponds to the usual
commutator $[X_1, X_2]=0$.  Deformed commutators appear in the
analysis of the PASEP, and in some of the two-species
models of Section~\ref{2species}.  One possible representation of
these matrices is
\begin{eqnarray}
\label{defcomrep}
X_1  &=&  \left( \begin{array}{ccccc}
0 & 1 & 0 & 0 & \cdots\\
0 & 0 & 1 & 0 &\\
0 & 0 & 0 & 1 & \\
0 & 0 & 0 & 0 &  \\
\vdots & & & & \ddots
\end{array} \right) 
\quad
X_2 = \left( \begin{array}{ccccc}
1  & 0 & 0 & 0 & \cdots\\
0 & r  & 0 & 0 & \\
0 & 0  & r^2 & 0 &\\
0 & 0 & 0  & r^3 &\\
\vdots & & & & \ddots
\end{array} \right) \;.
\end{eqnarray}
We note that both $X_1$ and any product of $X_1$ and $X_2$
including at least one $X_1$  are traceless.


\end{document}